\documentclass[fleqn]{2023SCGE}
\usepackage{amsmath}
\usepackage{amssymb}
\usepackage{xspace}
\usepackage{makecell}
\usepackage{graphicx}
\newcommand{\Ni}{$^{56}$Ni\xspace}

\usepackage{bib_macros}

\begin{document}

\ensubject{subject}

\ArticleType{Article}
\SpecialTopic{SPECIAL TOPIC: Astronomy}
\Year{2026}
\Month{September}
\Vol{xx}
\No{x}
\DOI{10.1360/s11433-026-3120-6}
\ArtNo{000000}
\ReceiveDate{June 8, 2026}
\AcceptDate{September 3, 2026}

\title{Supernova nucleosynthesis: a review}{Supernova nucleosynthesis: a review}

\author[1]{Shuai Zha}{{zhashuai@ynao.ac.cn}}%
\author[2]{Yudong Luo}{}
\author[1]{Zhanwen Han}{{zhanwenhan@ynao.ac.cn}}

\AuthorMark{Zha S}

\AuthorCitation{Zha S, et al}

\address[1]{International Centre of Supernovae (ICESUN), Yunnan Key Laboratory of Supernova Research, Yunnan Observatories, \\ Chinese Academy of Sciences (CAS), 650216 Kunming, People's Republic of China}
\address[2]{Center for Exotic Nuclear Studies, Institute for Basic Science, Daejeon 34126, Korea}


\abstract{Supernovae are major drivers of cosmic chemical evolution. They synthesize heavy elements and disperse them into the interstellar medium via their explosion shocks. Light elements are converted into heavier ones during both presupernova evolution and the explosive event itself. Supernova explosions generate nucleosynthesis environments rich in neutrons, protons, and neutrinos under unique thermodynamic conditions, which can enable the production of heavy elements beyond iron. Modern numerical simulations are constructing increasingly realistic explosion models of various supernova channels, providing more accurate nucleosynthesis conditions and chemical yields. In tandem with advances in large-scale spectroscopic surveys delivering precise, high-resolution stellar abundance data, as well as isotopic ratios from presolar grains and meteorites, our understanding of the supernova role in cosmic nucleosynthesis is poised to advance significantly. We review recent progress in modeling various supernova channels, with particular emphasis on nucleosynthesis yields derived from state-of-the-art simulations. We examine the roles of Type Ia, core-collapse, electron-capture, and pair-instability supernovae in producing intermediate-mass, iron-peak, trans-iron, and very heavy elements, as well as their characteristic chemical imprints. We also outline major theoretical uncertainties that affect yield predictions. We intend this review to serve as a timely reference for theoretical model development and a practical guide for interpreting observational abundance data.}



\maketitle


\begin{multicols}{2}

\section{Introduction}

Humans inhabit a world of striking diversity. At a fundamental level, everything we see is composed of molecules and atoms, which in turn consist of protons, neutrons, and electrons. Currently, there are 118 confirmed chemical elements, each defined by its unique proton number (also called atomic number), together with more than 3,000 isotopes that vary by neutron number \cite{ciaaw}. Yet, the infant Universe was remarkably simple. According to the standard model of Big Bang nucleosynthesis \cite{peebles_1993}, the primordial matter emerging from the first few minutes of cosmic evolution consisted almost exclusively of hydrogen (H) and helium (He), with only trace amounts of lithium (Li). This stark contrast raises a fundamental question: how did the primordial light elements synthesize into heavier species, and how did galaxies become enriched to their present-day chemical\footnote{``Chemistry" refers to the elemental abundance analysis in the astronomical convention.} abundances? Addressing the chemical enrichment of the Universe is one of the grand challenges in astrophysics, which demands a synergistic effort across the astronomical and nuclear communities.

Stars are the main sites for elemental transformation, dubbed \emph{nucleosynthesis}, and serve as the primary origin of elements heavier than He \cite{B2FH_1957,clayton_1968}. The fundamental obstacle to the fusion of light nuclei into heavier ones is the Coulomb repulsion between positively charged nuclei, and overcoming this barrier requires sufficiently high relative velocities between the daughter nuclei, which correspond to the high temperatures found in stellar interiors ($\gtrsim3\times10^6$\,K). The centers of stars function as gravitationally confined thermonuclear reactors, where successive fusion reactions convert H into progressively heavier elements until the iron (Fe) group. These newly synthesized elements would reside in stars, provided the stars remained gravitationally bound. However, numerous stars end their lives in catastrophic supernovae that eject chemically enriched material into the interstellar medium (ISM) \cite{arnett_1996}. Moreover, supernovae generate extreme physical conditions that enable additional nucleosynthesis processes, such as temperatures far exceeding those in quiescent stellar burning and neutron(proton)-rich environments. The resulting supernova ejecta eventually serve as the building blocks for the next generation of stars, planetary systems, and, ultimately, life.

Pioneered by Hoyle and Fowler \cite{hoyle_1960}, tremendous progress has been made in understanding the chemical contribution of various supernova channels \cite{thielemann_2018}. Consensus has been reached on key points: for example, core-collapse supernovae provide prompt enrichment in $\alpha$ elements (O, Mg, Si, Ca, etc.), whereas Type Ia supernovae dominate the Fe-group contribution with a significant delay after a burst of star formation (see, e.g., \cite{arcones_thielemann_2023,kobayashi_2025} and Table~\ref{tab:list} for a more comprehensive list). Nevertheless, key challenges remain, such as: How can supernova yields from emerging first-principle explosion models be integrated into galactic chemical evolution (GCE) models? Do core-collapse supernovae still contribute to heavy elements beyond iron synthesized via the rapid neutron-capture process, although neutron-star mergers are currently the preferred source? How do yields of various supernova channels depend on metallicity, binarity, and rotation?

\begin{table*}[ht]
    \centering
    \begin{threeparttable}
        \caption{\normalsize Elemental origins and main uncertainties \label{tab:list}}
        \begin{tabular}{cccc}
        \toprule
        Elements     &  Major source & Secondary source & Main uncertainties\\ \hline
        \makecell{$\alpha$-elements \\ (O-Ca)}     &  Core-collapse supernova & -- & \makecell{stellar evolution \\ explosion energy} \\ \hline
        \makecell{Fe-group \\ (Cr-Ni)}    & Type Ia supernova & Core-collapse supernova & \makecell{progenitor channel \\
        explosion mechanism} \\  \hline
        \makecell{$r$-process elements \\ (As-Bi)} & Neutron-star merger \tnote{1} & \makecell{Magnetorotational \\ supernova/Collapsar} & \makecell{event rates \\ explosion mechanism} \\ \hline
        $p$-nuclei & $\gamma$-process & $\nu p$-process & \makecell{synthesis condition \\ reaction rates} \\
        \bottomrule
        \end{tabular}
        \begin{tablenotes}
            \footnotesize    
            \item[1] The interpretation of the AT2017gfo spectra has not yet reached consensus regarding whether neutron-star mergers can account for the full range of $r$-process elements observed in the solar system \cite{pian_2017,smartt_2017,watson_2017}. 
        \end{tablenotes}
    \end{threeparttable}
    \label{tab:placeholder}
\end{table*}

To promote future theoretical developments, we review our current understanding of supernova explosion physics and their nucleosynthesis yields from various channels. We also outline the uncertainties in supernova yields introduced by stellar models, supernova explosion models, nuclear reaction rates, etc. Our review is organized as follows. Sections~\ref{sec:snphysics} and \ref{sec:reaction} provide an overview of the underlying physics of supernova explosion and nucleosynthesis processes, respectively. Sections~\ref{sec:ia}, \ref{sec:ccsn}, and \ref{sec:other} review the nucleosynthesis yields of Type Ia supernovae, core-collapse supernovae, and other supernova channels (electron-capture and pair-instability supernovae). We give our conclusions and outlook in Section~\ref{sec:final}.

\section{Supernova Explosion Physics \label{sec:snphysics}}

The theory of stellar evolution dictates that the fate of a star is primarily determined by its zero-age main-sequence mass ($M_{\rm ZAMS}$; \cite{kippenhahn_2013}). For $M_{\rm ZAMS} \lesssim 8\,M_{\odot}$, nuclear fusion in the stellar core goes no further than the production of O and Ne. The star subsequently ejects its outer envelope, leaving behind an inert, electron-degenerate core known as a white dwarf (WD). If a WD accretes sufficient mass from a companion star or undergoes a merger event, it may approach the Chandrasekhar mass limit ($\sim 1.4\,M_{\odot}$). Upon reaching this limit, the WD is expected to ignite thermonuclear fusion explosively under degenerate conditions, leading to a Type Ia supernova (SN Ia; \cite{liu_2023}). In contrast, for $M_{\rm ZAMS} \gtrsim 8\,M_{\odot}$ (up to $\sim140\,M_\odot$; see, e.g. \cite{ibeling_2013}), the stellar core evolves through successive burning stages, ultimately forming an Fe core. Because Fe-group nuclei have the highest binding energy per nucleon, further nuclear fusion is endothermic and cannot sustain the star against gravity. As the Fe core contracts, electron-capture processes reduce the degeneracy pressure support and trigger catastrophic collapse. The outcome is either the formation of a neutron star (NS) or a black hole (BH), often accompanied by a core-collapse supernova (CCSN; \cite{bethe_1990}). Note that the exact threshold mass depends on the metallicity, binarity, and rotation of the progenitor star. These two scenarios represent the major channels of supernova explosion, and other exceptional channels will be discussed in Section~\ref{sec:other}. With this foundation, we now present a brief overview of the key physical processes governing these supernova explosions.

\subsection{Type Ia Supernovae}

A comprehensive review of the in-depth physical mechanisms that govern thermonuclear explosions of WDs (SNe Ia) can be found in \cite{hillebrandt_2000}. In these events, nucleosynthesis plays a central role in driving the explosion and powering the supernova emission. The conversion of WD fuel (typically C and O) into \Ni releases nuclear binding energy, which overcomes the gravitational binding energy of WD and powers the explosion. Assuming that WD is composed of equal masses of C and O that burn into \Ni, the total released nuclear energy is:
\begin{equation}
\begin{aligned}
    &\frac{M_{^{56}\rm Ni}}{m_u}\times\Bigg[ \frac{B(^{56}{\rm Ni})}{56}-0.5\times\bigg(\frac{B(^{12}{\rm C})}{12}+\frac{B(^{16}{\rm O})}{16}\bigg)\Bigg] \\
     \simeq &\frac{M_{^{56}\rm Ni}}{M_\odot}\times\frac{1.99\times10^{33}\,{\rm g}}{1.66\times10^{-24}\,{\rm g}} \\
    & \times[8.64\,{\rm MeV}-0.5\times(7.68\,{\rm MeV}+7.98\,{\rm MeV})] \\
     \simeq  &\frac{M_{^{56}\rm Ni}}{M_\odot}\times 1.56\times10^{51}\,{\rm erg},
\end{aligned}
\end{equation}
where $M_{^{56}\rm Ni}$ is the produced \Ni mass, $m_u$ is the atomic mass unit, and $B(X)$ is the nuclear binding energy of $X$. This yields an energy scale consistent with typical supernova explosion energies, approximately $10^{51}$\,erg, which is defined as 1\,Bethe (short as B). The synthesized radioactive \Ni also powers the optical emission of the supernova \cite{arnett_1982}.


The thermonuclear combustion in SNe Ia shares many similarities with terrestrial combustion (see Chapter~14 of \cite{landau_1987}). There are various scenarios for how the fuel ignites and subsequently burns, and these different pathways lead to distinct explosion outcomes and nucleosynthesis yields. In the single-degenerate scenario, where a WD accretes material from a non-degenerate companion star \cite{whelan_1973,nomoto_1982,canal_1996,han_2004}, the fuel eventually ignites into a thermonuclear runaway, a state in which nuclear energy is released at a rate faster than it can be transported away from the reaction zone \cite{starrfield_2017}. Once a runaway occurs, reactive hydrodynamics takes over where nuclear combustion pushes the reactive flow into the unburned fuel \cite{hillebrandt_2000}. The combustion front can propagate either as a deflagration (subsonic, with propagation aided by electron conduction \cite{timmes_1992} and turbulent mixing \cite{niemeyer_1995}) or as a detonation (supersonic) \cite{arnett_1969}. Here, sub(super)sonic means that the propagation speed is slower (faster) than the speed of sound. In a deflagration, more intermediate-mass elements (IMEs) such as Si, S, and Ca are produced \cite{nomoto_1976}. In a detonation, most of the fuel is instead converted into Fe-group elements \cite{arnett_1971}. However, a pure deflagration has difficulty completely unbinding the WD. To reconcile the production of IMEs with a successful explosion, several other explosion scenarios have been proposed, such as:
\begin{itemize}
    \item Double detonation \cite{nomoto_1982b,fink_2007}: A helium shell detonation on the surface of a sub-Chandrasekhar mass WD triggers a secondary carbon detonation in the core.
    \item Pure turbulent deflagration (PTD) \cite{niemeyer_1995}: A subsonic burning front propagates outward, accelerated by turbulence.
    \item Deflagration-to-detonation transition (DDT) \cite{khokhlov_1997,ropke_2007,zhang_2026}: A subsonic deflagration pre-expands the WD, which later transits into a supersonic detonation.
    \item Gravitationally confined detonation (GCD) \cite{plewa_2004}: An off-center deflagration breaks the WD surface,  its ash flows converge at the opposite pole, and then the accumulated material ignites a detonation.
\end{itemize}
Numerical modeling of such reactive hydrodynamics remains both computationally challenging and time-consuming. For a recent status report on these complex mechanisms, see \cite{ropke_2017}.

Alternatively, in a double-degenerate system, two WDs in a close orbit can merge due to gravitational-wave radiation and produce a SN Ia \cite{webbink_1984,iben_1984}. This progenitor channel also involves the double-detonation mechanism \cite{fink_2007,guillochon_2010}. In this scenario, a violent merger first ignites a detonation in the helium shell of the primary (more massive) WD. The resulting shock then compresses and heats the CO core, triggering a secondary C detonation that ultimately explodes the WD. There is no consensus yet on which progenitor channel dominates the observed normal SN Ia events \cite{liu_2023}, and chemical evolution may offer an additional constraint on this critical issue.

\subsection{Core-collapse Supernovae}

\begin{figure*}
\centering
\includegraphics[width=\textwidth]{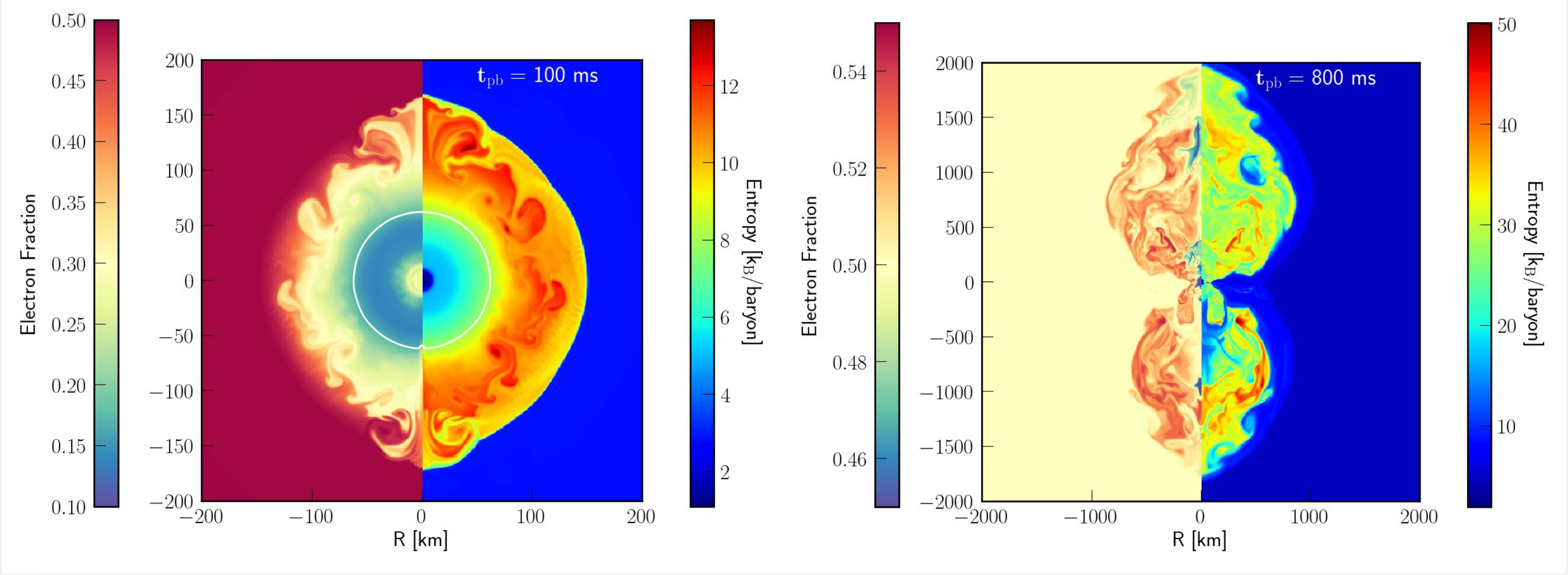}
\caption{Results from an unpublished two-dimensional simulation of a core-collapse supernova with a 15\,$M_\odot$ progenitor model. Shown are slices of electron fraction and entropy per baryon at 100~ms (left panel) and 800~ms (right panel) after core bounce. The supernova shock is visible as the boundary between cool (blue) and hot (yellow) matter, with hydrodynamic instabilities developing behind it. In the left panel, the white line indicates the density contour of $10^{11}$\,g\,cm$^{-3}$, defining the surface of the protoneutron star. \label{fig:ccsn2d}}
\end{figure*}

The association between the collapse of massive stars and supernova explosions was first proposed by Baade and Zwicky while investigating the origin of cosmic rays \cite{baade_1934}. They hypothesized that ``a super-nova represents the transition of an ordinary star into a \emph{neutron star}, consisting mainly of neutrons" at a time when NSs had not yet been observed. Such objects were later identified as rapidly rotating radio sources, or pulsars \cite{hewish_1968}. This association was firmly established with the detection of two dozen neutrinos from SN 1987A \cite{Kamiokande-II:1987idp,Bionta:1987qt,Aglietta:1987it}, which could only originate from a collapsed dense core at nuclear saturation density ($\rho_{\rm sat}\simeq2.7\times10^{14}\,{\rm g\,cm^{-3}}$, which gives the maximum nucleon packing per unit volume within a nucleus.)~\cite{bruenn_1987}. The collapse of the Fe core in massive stars releases a large amount of gravitational energy, approximately by the following formula 
\begin{equation}
\frac{GM_{\rm NS}^2}{R_{\rm NS}} \simeq 3.45\times10^{53}\,{\rm erg} \times \Big( \frac{M_{\rm NS}}{1.4\,M_\odot}\Big)^2 \Big(\frac{R_{\rm NS}}{15\,{\rm km}}  \Big)^{-1},
\end{equation}
where $M_{\rm NS}$ and $R_{\rm NS}$ are the mass and radius of the resulting NS. This energy is initially converted into thermal energy within the collapsed core, with the majority subsequently carried away by neutrinos as observed in SN 1987A \cite{arafune_1987,sato_1987}; however, even a 1\%  conversion of this energy reservoir would be sufficient to power CCSN explosions at 1\,B. Understanding how this conversion occurs has become a grand challenge in modern astrophysics, akin to a millennium problem \cite{science_2012}, and we briefly describe our current understanding in the following.

The dynamical collapse of the Fe core ensues once its mass approaches the effective Chandrasekhar mass, given by \cite{woosley_2002}:
\begin{equation}
M_{\rm Ch,eff} \simeq M_{\rm Ch} \Big[1+ \Big( \frac{s_e}{\pi Y_e} \Big)^2 \Big], \label{eq:mch}
\end{equation}
where $M_{\rm Ch}$ is the canonical Chandrasekhar mass for WDs \cite{chandrasekhar_1931}, $s_e$ is the electronic entropy, and $Y_e$ is the electron fraction (the number of electrons per nucleon), which is equivalent to the number of protons per nucleon. Collapse is accelerated by the depletion of pressure support due to electron captures. The central density ($\rho_{\rm c}$) rises rapidly from $\sim10^{10}$\,g\,cm$^{-3}$ to $\rho_{\rm sat}$ in less than 1\,s. Once $\rho_{\rm c}$ exceeds $\sim10^{12}$\,g\,cm$^{-3}$, even neutrinos become trapped in the core on timescales longer than the dynamical timescale and play a critical role in the subsequent evolution \cite{bruenn_1985}. As $\rho_{\rm c}$ surpasses $\rho_{\rm sat}$, the equation of state (EoS) stiffens due to strong repulsive nuclear forces. The stiff core overshoots its equilibrium density because of its large inertia and then rebounds, launching a shock wave into the overlying infalling material. This bounce shock, known as the prompt shock, was initially thought to produce a successful explosion, referred to as the prompt explosion mechanism \cite{bck_1985,baron_1987}. However, with theoretical advances in presupernova evolution, neutrino transport, and EoS, the prompt shock was found to lose kinetic energy due to neutrino emission and the dissociation of heavy nuclei into nucleons in postshock matter, rendering it unable to drive a successful explosion \cite{bruenn_1989a,bruenn_1989b,bethe_1990,janka_1993}.

The prompt shock soon transforms into an accretion shock and stalls at a radius of $\sim100$-200\,km from the center within a few tens of ms (the left panel of Figure~\ref{fig:ccsn2d}). After this phase, the most plausible mechanism to reenergize the shock and trigger a successful explosion is the delayed neutrino-driven mechanism \cite{arnett_1966,colgate_1966,bethe_1985}. In this framework, neutrinos diffused from the inner dense region (often called protoneutron star, PNS) heat the material below the shock through charged current interactions \cite{sudarshan_1958,feynman_1958}: 
\begin{equation} \label{eq:nu}
    \begin{aligned}
        \nu_e + n &\to p+e^-,  \\
        \bar{\nu}_e + p &\to n+e^+,
    \end{aligned}
\end{equation}
where $\nu_e$ and $\bar{\nu}_e$ are electron-flavor neutrino and anti-neutrino, respectively. Under favorable conditions, the resulting pressure in the neutrino-heating region can overcome the ram pressure of the infalling matter above the shock, leading to a runaway shock expansion (the right panel of Figure~\ref{fig:ccsn2d}). Numerous numerical simulations have explored this scenario using progressively more sophisticated tools. However, one-dimensional (1D) simulations assuming spherical symmetry have consistently failed to produce explosions with the most sophisticated neutrino transport treatments \cite{rampp_2000,mezzacappa_2001,liebendorfer_2001}, except for the lowest-mass progenitors (around 9\,$M_\odot$; \cite{kitaura_2006,melson_2015,radice_2017}). Over the past two decades, advances in multi-D modeling have demonstrated successful explosions under this mechanism, aided by the enhanced neutrino heating and turbulence stresses arising from hydrodynamic instabilities such as convection and standing accretion shock instability (see comprehensive reviews in \cite{janka_2016,burrows_vartanyan_2021,janka_2025} and a 2D example in Figure~\ref{fig:ccsn2d}). With these self-consistent explosion models at hand, one can perform more accurate nucleosynthesis calculations, especially for the innermost ejecta which produce Fe-group and trans-Fe elements \cite{hix_2017}.

For progenitor stars with rapid rotation and strong magnetic fields in the Fe core, a rare class of supernova explosions can be triggered by the magnetorotational (MR) mechanism \cite{leblanc_1970,meier_1976}. Strong magnetic fields, amplified through MR instabilities or dynamo actions, can tap into the rotational kinetic energy of the rapidly rotating collapsed core, launching bipolar jets at the rotational axis and driving hyper-energetic explosions up to $\sim$10\,B \cite{mosta_2014,kuroda_2020,obergaulinger_2020,powell_2023}. Other mechanisms exist but may require additional unknown physics or remain speculative without detailed numerical exploration. For instance, if a first-order hadron-quark phase transition occurs inside the PNS, the core may collapse once again and launch a second bounce shock, which can trigger a successful explosion \cite{sagert_2009,fischer_2018,zha_2020}. Whether this works strongly depends on the underlying EoS as well as the progenitor properties \cite{fischer_2011,zha_2021,huang_2025}. Another possibility is the jittering jet mechanism, which explodes the progenitor star with jets rapidly changing direction \cite{papish_2011} as inferred from certain phenomenological features in supernova remnants \cite{soker_2024}. However, it lacks rigorous and quantitative investigations. After all, some massive stars with highly compact Fe cores can fail to explode and their cores continue the collapse to form stellar-mass BHs in events known as failed supernovae \cite{macfadyen_1999,sumiyoshi_2007,kochanek_2008,oconnor_2011}.

\section{Nucleosynthesis Processes \label{sec:reaction}}

\subsection{Explosive nuclear burning}

\begin{figure*}
    \centering
    \includegraphics[width=\columnwidth]{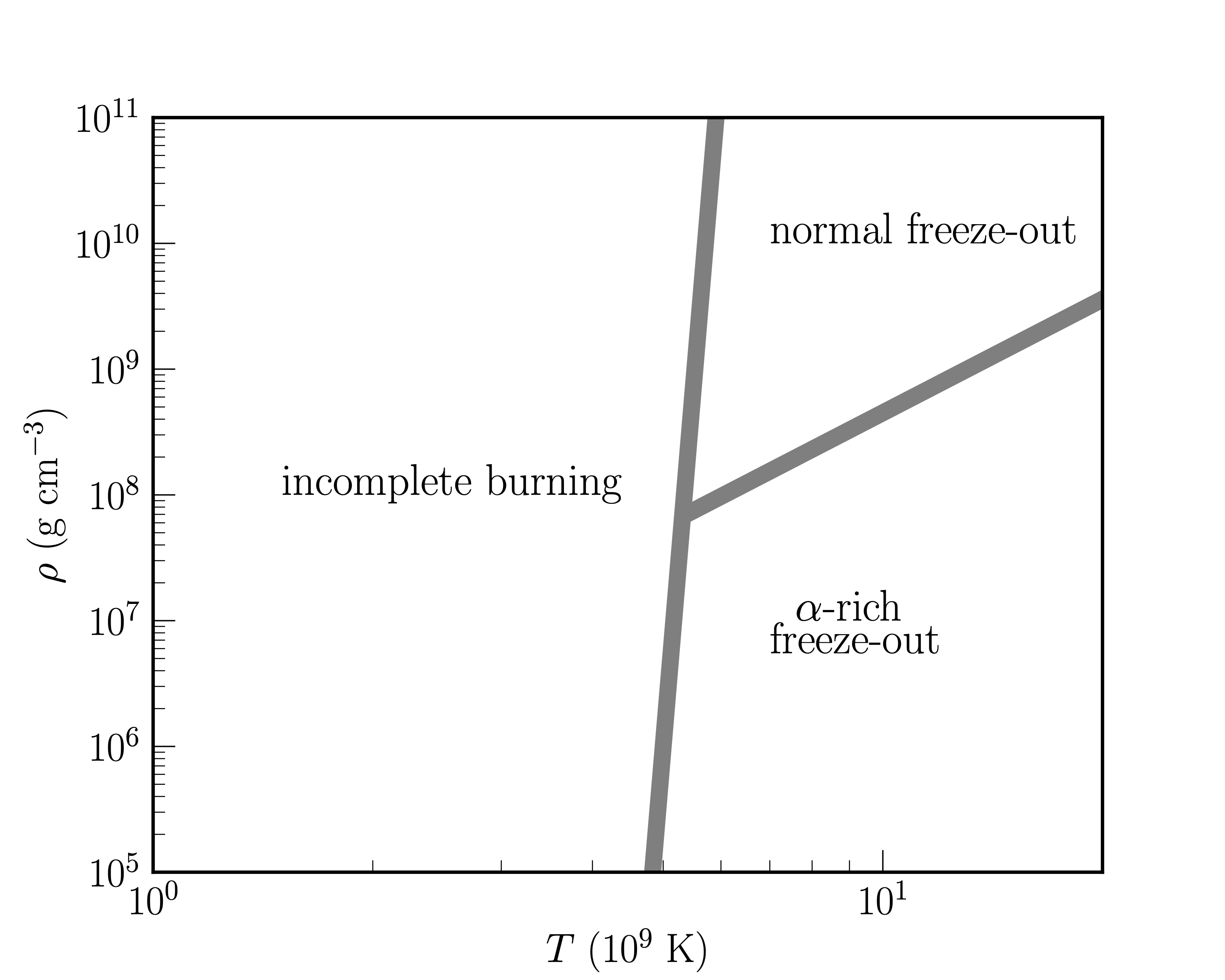}
    \caption{Three different regimes of explosive Si burning following Woosley et al. (1973) \cite{woosley_1973}. }
    \label{fig:burn}
\end{figure*}

During most of a star's life, while it remains in near-perfect hydrostatic and thermal equilibrium, nuclear burning proceeds slowly over timescales ranging from millions to billions of years \cite{kippenhahn_2013}. In contrast, during a supernova, a supersonic shock wave (or a detonation) propagates through the star and heats the material to extreme temperatures. This is followed by rapid expansion and cooling over a timescale of just seconds or less. As a result, nuclear burning occurs under vastly different conditions compared to quiescent stellar burning, giving rise to explosive nuclear nucleosynthesis \cite{arnett_1973}. 

As the temperature exceeds $\sim5\times10^9$\,K, the material achieves nuclear statistical equilibrium (NSE), in which nuclear reactions proceed so rapidly that forward and reverse processes reach a detailed balance. For a detailed discussion of the conditions required to achieve NSE, we refer the reader to \cite{woosley_1973,imshennik_1981,hix_1999,calder_2007}. In NSE, one can determine the abundances of all nuclei by their binding energies and the temperature and density through the so-called Saha equation \cite{clifford_1965}. Baryon-number and charge conservations fully determine the mass fractions of all nuclear species once the thermodynamic state, specified by the triplet $(T,\rho,Y_e)$, is given. Since neutrinos normally escape from the star without interaction due to their small cross section with nuclei, a true equilibrium involving weak interactions is not achieved. However, weak interactions must be carefully monitored because they further modify $Y_e$ which determines the composition of the Fe-peak elements (see e.g., \cite{frohlich_2006a}).

The nucleosynthesis outcome of explosive burning is controlled not only by the establishment of NSE, but also by the thermodynamic trajectory during freeze-out \cite{woosley_1973}, see Figure~\ref{fig:burn} for different regimes of explosive Si burning. When the density is sufficiently high and the abundance of free light particles at the onset of freeze-out is small, the system undergoes a particle-poor freeze-out (\emph{normal} freeze-out). Under these conditions, the composition remains close to the NSE or quasi-statistical-equilibrium distribution, and the ejecta are dominated by Fe-peak nuclei. The final isotopic abundances depend primarily on the neutron excess, entropy, and expansion timescale, but only moderately depart from the equilibrium distribution. At lower densities, the equilibrium abundances of light particles become appreciable, such as $\alpha$ particles. If the subsequent expansion and cooling are rapid compared with the timescales for He burning and $\alpha$ captures, these particles cannot be fully incorporated into Fe-peak nuclei before reactions freeze out. The resulting $\alpha$-rich freeze-out terminates the equilibrium in the presence of an excess of $\alpha$ particles and modifies the final abundance pattern relative to the normal freeze-out case. In environments with a small neutron excess, $^{56}\mathrm{Ni}$ generally remains the dominant product, while the production of species such as $^{44}\mathrm{Ti}$, $^{45}\mathrm{Sc}$, and $^{58,60}\mathrm{Ni}$ is significantly enhanced.

\subsection{$r$/$\gamma$-process and neutrino-induced processes}
In addition to explosive nuclear burning, supernova explosions can generate some exceptional conditions, such as high neutron fluxes and neutrino irradiation that enable distinct nucleosynthesis pathways, particularly for the synthesis of elements beyond the Fe peak. These relevant processes include:
\begin{enumerate}
    \item $r$-process: The rapid neutron-capture process ($r$-process) accounts for producing approximately half of the stable elements beyond Fe in the solar system \cite{cowan_2021}, such as Pt and Au. Under an intense flux of free neutrons, seed nuclei (primarily Fe-group elements) capture neutrons on timescales much shorter than the intervening $\beta$-decay half-lives. The resultant unstable, neutron-rich nuclei follow a path along the neutron-drip line and their $\beta$-decays occur only after the neutron flux subsides. This allows nucleosynthesis to bypass the instability gaps associated with short-lived nuclei and produce the characteristic abundance peaks regulated by closed nuclear shells at neutron numbers of 50, 82 and 126. In sufficiently neutron-rich environments, the $r$-process flow may reach the region where spontaneous, neutron-induced, and $\beta$-delayed fission terminate the flow and recycle material toward lower mass numbers. The corresponding fission rates, fragment-mass distributions, and neutron multiplicities influence the number of recycling cycles and the final abundance pattern, particularly in the second-peak and rare-earth regions \cite{panov_2004,Goriely_2013,horowitz_2019,kajino_2019}. Because these properties remain poorly constrained for the extremely neutron-rich nuclei involved, fission is one of the major nuclear physics uncertainties in $r$-process calculations.
    Currently, neutron-star merger is the preferred astronomical site for $r$-process nucleosynthesis \cite{eichler_1989,pian_2017}, while neutrino-driven winds in CCSNe \cite{woosley_1992b} and a rare class of CCSNe \cite{thielemann_2017} remain potential contributing sources. For a nuclear physics view of $r$-process, we refer the readers to Refs~\cite{horowitz_2019,kajino_2019}.
    \item $\gamma$-process: The 35 neutron-deficient stable isotopes between $^{74}$Se and $^{196}$Hg (known as proton-rich nuclei or $p$-nuclei; \cite{B2FH_1957}) cannot be synthesized via neutron-capture processes. Their production is also not feasible through proton-capture processes under stellar conditions, because the Coulomb barriers inhibit such reactions while the inverse photodisintegration reactions (i.e., ($\gamma,p$)\footnote{The symbol $X(a,b)Y$ denote that a reactant $X$ interacts with an incoming particle $a$, emits an outgoing particle $b$ , and produces the final nucleus $Y$.}) become dominant at elevated temperatures. The $\gamma$-process was proposed to make these $p$-nuclei via a sequence of photodisintegration reactions of pre-existing heavy seed nuclei, including ($\gamma,n$), or ($\gamma,p$) and ($\gamma,\alpha$) followed by subsequent $\beta$-decays at temperatures of $\sim2$-3$\times10^9$\,K \cite{rauscher_2013}.  The required physical conditions can be achieved during explosive O/Ne burning in CCSNe \cite{woosley_1978}, as well as in SNe Ia with nuclear seeds from preceding $s$-process nucleosynthesis \cite{howard_1991}.
    \item neutrino-induced processes: Neutrinos not only play a critical role in triggering successful CCSN explosions, but also influence nucleosynthesis by modifying the proton-to-neutron ratio in the ejecta closest to the PNS (see~Eq.~\ref{eq:nu}), which can affect the potential $r$-process. Beyond the transport effect, two main neutrino-induced nucleosynthesis mechanisms (the $\nu$-process and $\nu p$-process) have been proposed to operate under the intense neutrino flux emanating from the PNS in CCSNe, contributing to heavy-element synthesis \cite{kajino_2014,balasi_2015,gabriel_2017,wangxl_2023,fischer_2024}. In the $\nu$-process \cite{woosley_1990,haxton_1990,haxton_1991,hartmann_1991}, neutrinos induce nuclear reactions with abundant nuclei to produce $\nu$-isotopes such as $^7$Li, $^{11}$B, $^{19}$F, $^{92}$Nb, $^{98}$Tc, $^{138}$La and $^{180}$Ta \cite{Heger_2005, Hayakawa_2013, Hayakawa_2018, kobayashi_2011}. In proton-rich ejecta, the $\nu p$-process operates that electron anti-neutrinos ($\bar{\nu}_e$) transform protons into neutrons, then $\nu p$-process could occur from $^{56}$Ni with sequence of $(p,\gamma)$ and $(n, p)$ to bypass the $\beta^+$-waiting points (where nucleosynthesis flow stalls on the proton-rich side) to synthesized light $p$-nuclei such as $^{92,94}$Mo and $^{96,98}$Ru \cite{frohlich_2006c, wanajo_2011, pruet_2006} or even cross the $\beta^+$-decay stability line to produce neutron-rich high-mass nuclei for high entropy ejecta \cite{balantekin_2024}. More recently, a $\nu r$-process has been proposed, in which charge-current neutrino interactions produce light p-nuclei (e.g., Nb) on the neutron-rich side via the after neutron capture freeze-out \cite{xiong_2024}. 
\end{enumerate}

\subsection{Theoretical approach}
\begin{figure*}[ht]
\centering
\includegraphics[width=\textwidth]{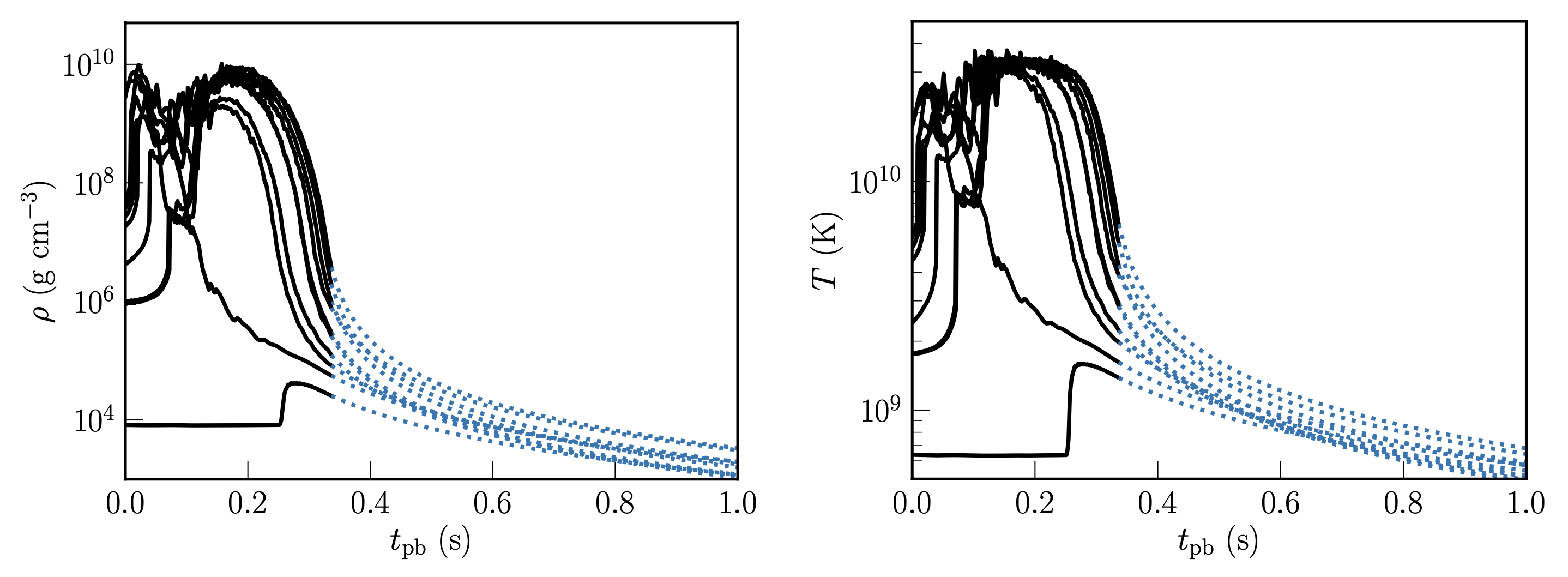}
\caption{Thermodynamic histories of selected tracers for the ejecta in a core-collapse supernova simulation. Here $t_{\rm pb}$ denotes the time with respect to the moment of core bounce. The black solid lines are directly interpolated from recorded snapshots, while the blue dotted lines represent extrapolation to late time assuming radiation-dominated expansion (i.e., constant $\rho/T^3$). \label{fig:traj}}
\end{figure*}

To compute the nucleosynthesis yields of various astronomical sites, one employs a nuclear reaction network that solves a coupled system of ordinary differential equations governing the time evolution of isotopic abundances (see e.g., Eq.~10 of Ref.~\cite{hix_2006}):
\begin{equation} \label{eq:network}
\begin{aligned}
    \dfrac{\mathrm{d}Y_i}{\mathrm{d}t} = & \sum\limits_{j} \mathcal{N}_j^i \lambda_j Y_j + \sum\limits_{j,k} \mathcal{N}_{j,k}^i \rho N_{\rm A} \langle j,k \rangle Y_j Y_k \\ 
    &+ \sum\limits_{j,k,l} \mathcal{N}_{j,k,l}^i \rho^2 N_{\rm A}^2 \langle j,k,l \rangle Y_j Y_k Y_l,
\end{aligned}
\end{equation}
where $Y_i$ is the nuclear abundance of $i$-th species, i.e., its number fraction relative to all nucleons; $\lambda_j$, $\langle j,k \rangle$ and $\langle j,k,l \rangle$ are the cross sections for reactions involving one-, two- and three-body reactions, respectively; $\rho$ is the rest-mass density and $N_{\rm A}$ is the Avogadro's constant. The numeric factors $\mathcal{N}$'s properly account for the numbers of nuclei created or destroyed in each reaction. 

Because a reaction network typically involves hundreds to thousands of nuclear species, Eq.~\ref{eq:network} has a very large and extremely stiff Jacobian matrix. The characteristic timescales of different processes span many orders of magnitude, from microseconds to millions of years. To maintain efficiency and accuracy, implicit integration methods are combined with various sparse matrix solvers \cite{hix_1999,timmes_1999}. Several open-source numerical tools are available to perform reliable nucleosynthesis calculations, such as \textsc{torch} \cite{timmes_1999}, \textsc{SkyNet} \cite{lippuner_2017}, and \textsc{WinNet} \cite{reichert_2023}. The reaction rates used in the network are typically evaluated either through parametric fitting formulae provided by libraries such as \textsc{REACLIB} \cite{reaclib} or via numerical interpolation from tabulated datasets (e.g., \cite{langanke_2001}).

Supernova explosion models provide the required physical inputs for the reaction network, namely the thermodynamic histories of the ejecta (more specifically, the time evolution of temperature and density). From Lagrangian hydrodynamic simulations in which fluid parcels are sampled, this information is straightforward to record. In Eulerian hydrodynamic simulations which use grids to sample the fluid, additional operations are required to extract this information. Either passive tracer particles are employed to record the thermodynamic history alongside the simulations \cite{seitenzahl_2010,harris_2017}, or high-cadence snapshots of the simulation results are stored, from which interpolation forward or backward in time provides the desired information \cite{wanajo_2018,sieverding_2023}. In addition, since multi-D models usually cannot run for a sufficient duration to allow the ejecta to expand and cool completely, one has to extrapolate the thermodynamic history to later times, often assuming radiation-dominated expansion (a constant $\rho/T^3$) \cite{arcones_2007}, explicitly as 
\begin{equation}
    \begin{aligned}
        \rho(t) & = \rho(t_{\rm e})[1+(t-t_{\rm e})/\tau]^{-2}, \\
        T(t) & = T(t_{\rm e})[1+(t-t_{\rm e})\tau]^{-2/3}, 
    \end{aligned}
\end{equation}
where $t_{\rm e}$ is the time when the simulations terminate, and one fit the last few ms of the trajectories to get the expansion timescale $\tau$. Figure~\ref{fig:traj} shows an example of the thermodynamic history of supernova ejecta extracted from simulations \cite{zha_2024}.

Finally, when presenting or discussing nucleosynthesis yields, it is common to use the nucleon fraction 
\begin{equation}
    X_i = A_i Y_i,
\end{equation}
where $A_i$ is the mass number of the nuclear species $i$, i.e., the total number of protons and neutrons in its nucleus; and $\sum_{i} X_i=1$. Note that the nucleon fraction is often treated interchangeably with the mass fraction (which is routinely employed in the astronomical community), but they are not strictly equivalent because of the nuclear binding energies. In practice, to interpret the nucleosynthesis contribution of a given astronomical site, one typically normalizes the $X_i$ with a reference abundance pattern (e.g., the solar composition $X_{i,\odot}$, see \cite{ander_grevesse_1989,asplund_2009,grevesse_2011}) and defines the production factor as:
\begin{equation}
    \mathrm{pr.~factor} \equiv X_i/X_{i,\odot}. \label{eq:pr}
\end{equation}
Values greater (smaller) than 1 indicate overproduction (underproduction) relative to solar. Another commonly used quantity is the logarithmic abundance ratio relative to the solar value:
\begin{equation}
    [A/B] = \log_{10}(X_A/X_B) - \log_{10}(X_{A,\odot}/X_{B,\odot}).
\end{equation}
A positive value of $[A/B]$ indicates that $A$ is more abundant relative to $B$ than in the solar composition, whereas a negative value indicates the opposite.

\section{Yields of Type Ia Supernovae \label{sec:ia}}

\begin{figure*}[ht]
    \centering
    \includegraphics[width=\textwidth]{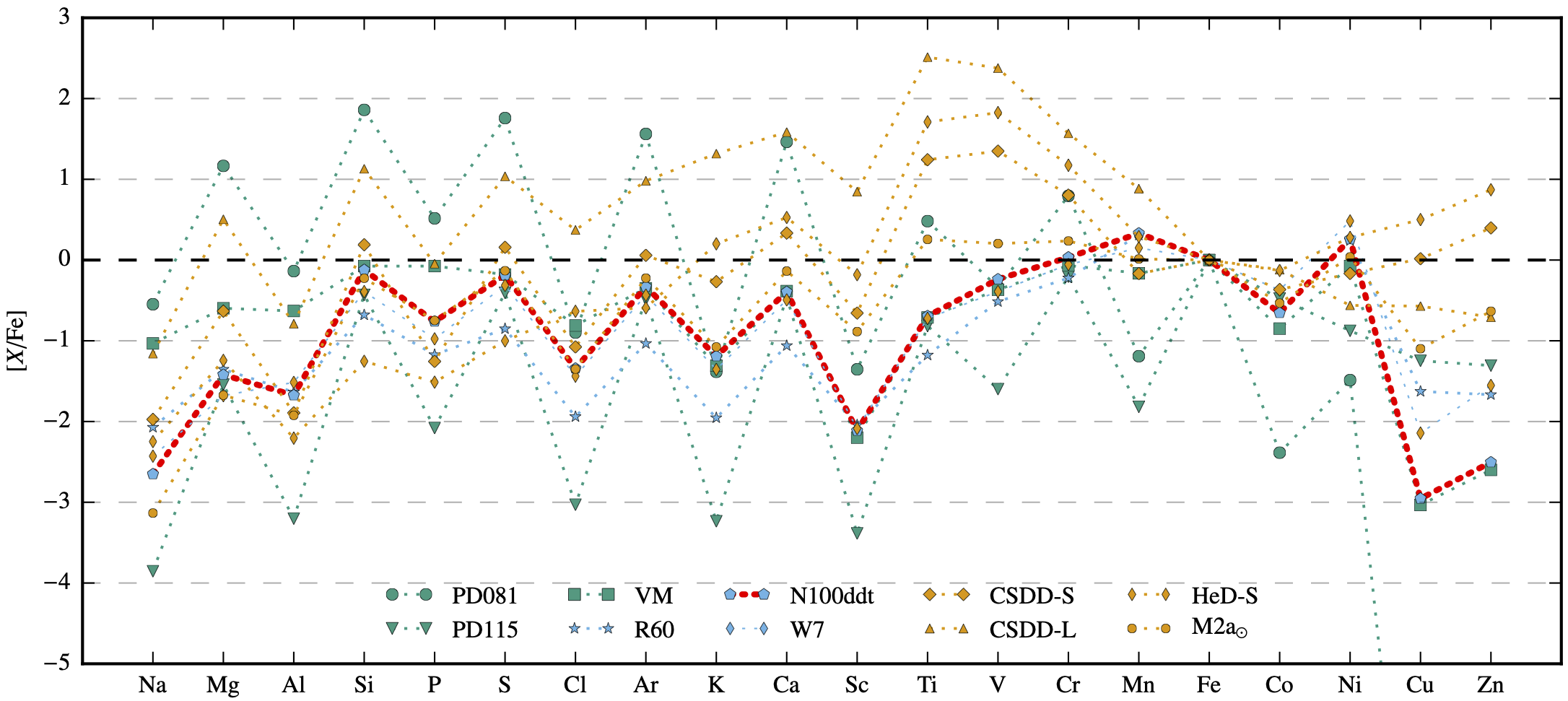}
    \caption{Comparison of elemental yield patterns from Na through Zn predicted by different Type Ia supernova explosion models. PD081, PD115 \cite{sim_2010} and VM \cite{pakmor_2012} are sub-$M_{\rm Ch}$ detonation models; R60 \cite{seitenzahl_2013} and W7 \cite{iwamoto_1999} are pure $M_{\rm Ch}$ deflagration models; N100ddt \cite{seitenzahl_2013} is a delayed-detonation model; M2a$_\odot$ \cite{gronow_2020}, CSDD-S and CSDD-L \cite{sim_2012} are double-detonation models;  HeD-S \cite{lach_2020} is a helium-detonation model. Reprinted from Lach et al. (2020) \cite{lach_2020} with permission from EDP Sciences.}
    \label{fig:ia_yields}
\end{figure*}

SNe Ia are believed to dominate cosmic Fe-group production on delayed timescales ($[\rm Fe/H]\gtrsim-1$), contributing $\sim70\%$ solar iron \cite{matteucci_1986}. The detailed yield pattern depends sensitively on both the progenitor channel and the explosion mechanism. In general, the yields are governed by three key conditions:
\begin{enumerate}
    \item the density for the occurrence of nuclear combustion, which controls electron captures and therefore the production of neutron-rich isotopes;
    \item the degree of pre-expansion before detonation happens, which sets the partition between Fe-peak nuclei and IMEs such as Si, S, and Ca;
    \item the neutron excess of the progenitor, determined primarily by the initial metallicity through $^{22}$Ne.
\end{enumerate}
Together, these factors link the nucleosynthesis yields of SNe Ia to different explosion channels and the progenitor metallicity. 

\subsection{Different explosion scenarios}

Early yield calculations of SNe Ia were based on idealized detonation and deflagration scenarios. For a prompt detonation in a near-$M_{\rm Ch}$ WD, the $^{12}$C-detonation burns fuels at high density and produces large masses of Fe-group elements and little IMEs \cite{arnett_1971}. The resulting ejecta are too Fe-rich to be consistent with the spectra of normal SNe Ia \cite{branch_1982}. By contrast, in the classical near-$M_{\rm Ch}$ deflagration scenario, subsonic burning allows substantial pre-expansion, which leads to the synthesis of more IMEs while still producing enough $^{56}$Ni to power the light curve \cite{nomoto_1984a,thielemann_1986,iwamoto_1999}. However, pure deflagration models generally underproduce the kinetic energies and stratification of the ejecta (e.g., leaving some unburned C/O at the center) inferred for normal Ia events \cite{gamezo_2004}. 

The deflagration-to-detonation transition (DDT) scenario, in which an initial deflagration phase pre-expands the WD before a subsequent detonation completes the burning, provides one of the most successful frameworks for reproducing both the observables and the nucleosynthesis yields (e.g., the red dotted line in Figure~\ref{fig:ia_yields}). The density at which the DDT occurs ($\rho_{\rm DDT}$) is a primary factor that determines the $^{56}$Ni mass, the partition between IMEs and Fe-peak elements, and the stratification of the ejecta. Tuning the transition conditions was shown to reproduce a broad range of luminosities while preserving the layered IME/Fe-peak structure characteristic of normal SNe Ia \cite{khokhlov_1991,seitenzahl_2013,seitenzahl_2017,leung_2018}. Multi-D simulations of the DDT scenario demonstrate that different ignition geometries and flame morphologies also lead to a range of \Ni masses and different distributions of stable Fe-peak isotopes and IMEs (see e.g., 2D models \cite{maeda_2012} and 3D models \cite{seitenzahl_2013}).

Sub-$M_{\rm Ch}$ channels have become increasingly prominent because they may account for part of the observed diversity and event rates of SNe Ia. These include double detonations involving a He shell and a C/O core, as well as violent mergers \cite{sim_2010,pakmor_2012}. A key nucleosynthesis distinction is that sub-$M_{\rm Ch}$ detonations occur at a lower central density, significantly reducing electron captures and therefore neutron richness. As a result, they synthesize less neutron-rich stable Fe-peak material, such as $^{58}$Ni, $^{54}$Fe, $^{55}$Mn, than near-$M_{\rm Ch}$ events at the same $^{56}$Ni yield \cite{leung_2020a}. In particular, the Mn/Fe ratio is a useful discriminant between near-$M_{\rm Ch}$ and sub-$M_{\rm Ch}$ populations. 

Systematic comparisons have been made to confront both chemical evolution models and remnant observations across different explosion scenarios \cite{truran_2012,mori_2018,leung_2018,lach_2020,leung_2020a,leung_2020b}. Figure~\ref{fig:ia_yields} shows a collection of elemental yield patterns (Na through Zn) from various explosion models \cite{lach_2020}, including pure $M_{\rm Ch}$ deflagration, sub-$M_{\rm Ch}$ detonation, delayed detonation, double detonation, and He detonation. Due to large variations in modeling, a conclusive determination of the explosion mechanism is not yet possible from the nucleosynthesis viewpoint. Nevertheless, the diversity in SNe Ia observations may instead point to a need for multiple explosion scenarios. For a more comprehensive overview of explosion scenarios and the corresponding nucleosynthesis yields, see \cite{seitenzahl_2017}. 

\subsection{Metallicity dependence}

Metallicity affects the SNe Ia yields mainly through the neutron excess of the progenitor, usually parameterized by the mass fraction of $^{22}$Ne that is produced from the star’s initial CNO content $^{14}$N during stellar evolution. Recent systematic studies have quantified these metallicity-dependent trends across a range of explosion models \cite{leung_2020a,gronow_2020,keegans_2023}. Yields start to change when the $^{22}$Ne mass fraction exceeds $10^{-4}$, corresponding to $\sim1/100$ solar metallicity, with abundance differences varying by a factor of a few \cite{keegans_2023}. The isotopic pattern is more sensitive to metallicity, though the impact is subtle and element-dependent \cite{keegans_2023}. It remains unclear how these metallicity-dependent yields can be observationally constrained and how they can affect GCE models.

\subsection{Beyond iron}
Although SNe Ia are best known for Fe-peak and IME production, they also contribute to rarer isotopes beyond the Fe group under certain conditions. $\gamma$-process nucleosynthesis can occur in SN Ia models when pre-existing $s$-process seeds (in accreted material or in the WD) are photodisintegrated in hot layers with an appropriate temperature, producing $p$-nuclei such as $^{92,94}$Mo and $^{96,98}$Ru \cite{howard_1991, kusakabe_2005, Kusakabe_2010, travaglio_2011}. As expected, the presupernova abundance of $s$-nuclei in the WD has a large impact on the $\gamma$-process. Depending on the amount of $s$-process seeds, the contribution of SNe Ia to the solar $p$-process composition is estimated to exceed 50\% \cite{travaglio_2011}. More broadly, there is growing interest in trans-Fe production in thermonuclear supernova environments \cite{battino_2025}. For example, during a double WD merger event, $^{22}$Ne$(\alpha,n)$$^{25}$Mg reaction releases free neutrons,  enabling weak $s$-process nucleosynthesis that peaks at Kr, with a production factor over $\sim1000$ \cite{battino_2025}.

\subsection{Nuclear-physics uncertainties}
Nuclear-physics uncertainties can propagate into the SN Ia yields in the following aspects. First, uncertainties arise from the thermonuclear reaction rates and their temperature dependence in the relevant burning regimes \cite{bravo_2012,parikh_2013}. The main nucleosynthesis products are not very sensitive to individual rate variations, with the $^{12}$C + $^{12}$ C fusion reaction having the greatest impact \cite{parikh_2013}. Second, weak rates affect the production of neutron-rich isotopes and the stable/unstable partition of Fe-peak material \cite{mori_2016}. Dense plasmas of the WD require accounting for Coulomb screening, which can modify electron-capture rates and reduce the production of neutron-rich isotopes such as $^{50}$Ti, $^{54}$Cr, and $^{58}$Fe \cite{mori_2020}. Last, certain reaction rates may be important for the synthesis of rare isotopes. For example, the $^{91}$Zr(p,$\gamma$)$^{92m}$Nb cross section has been highlighted as important for assessing SN Ia contributions to $^{92}$Nb \cite{gyurky_2021}.

\subsection{Observational constraints}
Theoretical predictions of the SN Ia yields are increasingly confronted with both direct and indirect observational constraints. Nebular-phase spectra probe the composition and ionization state of the inner ejecta and have long been used to infer Fe-group abundances and stratification in SNe Ia \cite{ruiz-lapuente_1992,liu_1997,tucker_2020}, which is an especially exciting frontier with the advent of JWST observations \cite{kwok_2023,kwok_2025}. Furthermore, observations of supernova remnants (e.g., Kepler's supernova remnant \cite{sato_2020}) can reveal element distributions and Fe richness. At low metallicity, chemically peculiar stars can serve as fossil records of individual enrichment events, and iron-rich metal-poor stars have been proposed as probes of early SN Ia contributions and their explosion scenarios \cite{reeves_2023,reggiani_2023}. Together, these constraints can provide a more quantitative mapping between SN Ia explosion scenarios (near-$M_{\rm Ch}$ vs. sub-$M_{\rm Ch}$), their metallicity-dependent yields, and the integrated chemical evolution signatures observed in galaxies.

\section{Yields of Core-collapse Supernovae \label{sec:ccsn}}

As a massive star evolves towards core collapse, it develops an ``onion-shell" structure in which successive stages of hydrostatic burning produce concentric layers composed of increasingly heavier elements toward the center: an H-rich envelope, He shell, C/O layer, O/Ne/Mg layer, Si/S-rich shell, and an Fe-group core. During these quiescent phases, mixing processes such as convection and dredge-up can transport nuclear-processed material to the outermost layer, which can be lost through stellar winds \cite{thielemann_1985,nomoto_1986,weaver_1993,limongi_2000,hirschi_2017,higgins_2024}. When the Fe core becomes gravitationally unstable and collapses, a PNS forms and a runaway shock is launched at radii of order $\sim 10^2$\,km. As the shock propagates outward through the pre-existing composition stratification, it heats and compresses the overlying shells. The innermost shocked material can reach NSE, and its final composition is strongly modified by neutrino interactions (cf. Eq.~\ref{eq:nu}), while matter further out that is shock-heated to sufficiently high peak temperatures undergoes explosive burning. With these distinct nucleosynthesis regimes in mind, i.e., neutrino-processed inner ejecta versus shock-driven explosive burning in the outer layers, we discuss the nucleosynthesis yields of CCSN ejecta across different explosion scenarios.

\subsection{Neutrino-driven explosion}

The delayed neutrino-driven mechanism is currently the most plausible engine for successful CCSN explosions \cite{janka_2016,burrows_vartanyan_2021}. Systematic studies of nucleosynthesis in the innermost, neutrino-processed CCSN ejecta began with early neutrino-driven explosion calculations \cite{arnett_1970,hillebrandt_1986}. Progress was long limited by the fact that spherically symmetric (1D) models with self-consistent neutrino transport typically failed to explode robustly. Therefore, the PNS mass cut, thermodynamic histories, neutrino irradiation of the innermost ejecta, and therefore their nucleosynthesis yields were not determined in first-principles calculations. This situation has now begun to change with the advent of multi-D explosion models over the last two decades.

\subsubsection{Multi-dimensional explosion models}

\begin{figure*}
    \centering
    \includegraphics[width=\columnwidth]{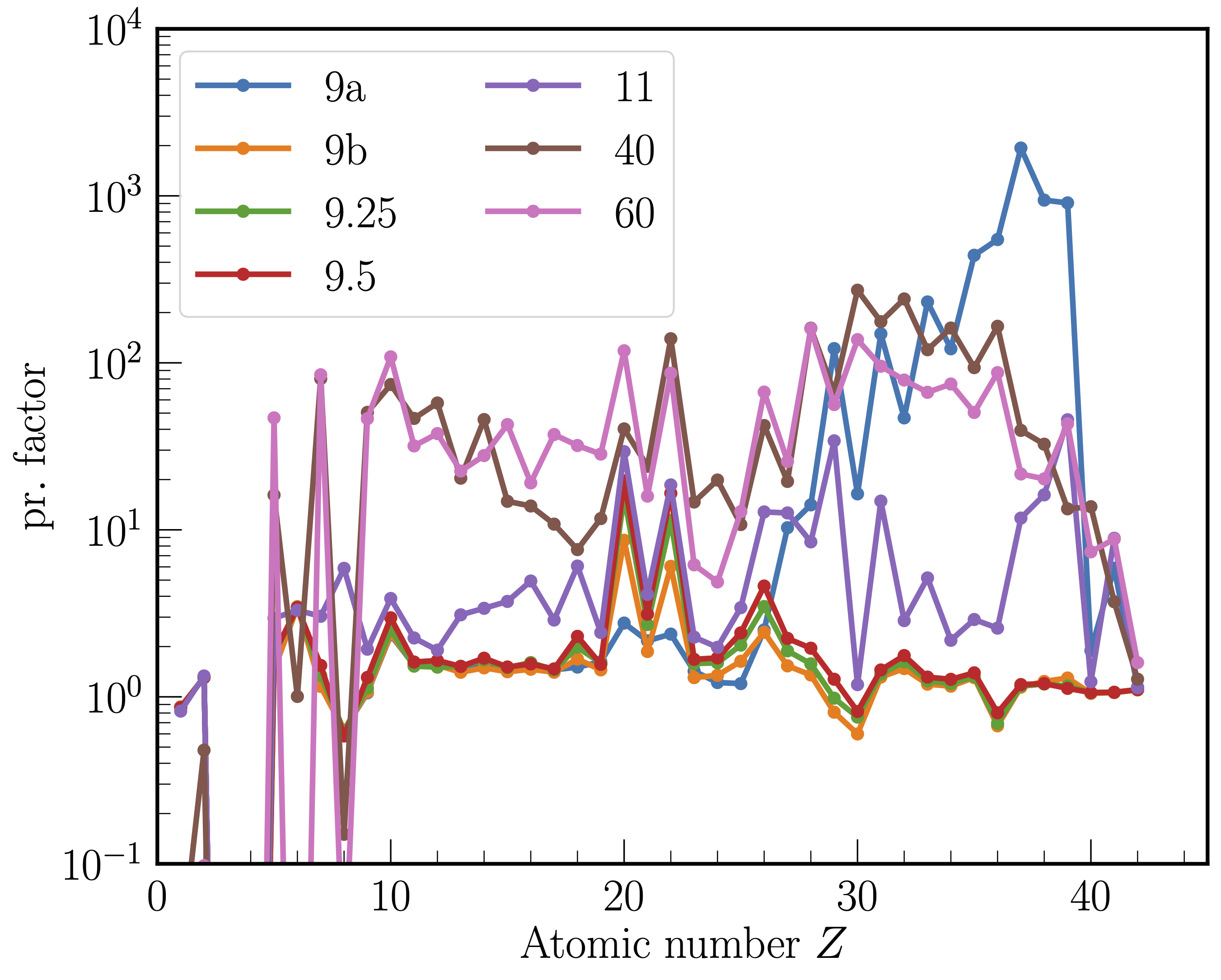}
    \caption{Elemental yield patterns for the 3D core-collapse supernova models of Wang et al. (2024) \cite{wang_2024a}. Each curve corresponds to a different model, with the label indicating the zero-age main-sequence mass, $M_{\rm ZAMS}$. Models 9a and 9b use the same 9\,$M_\odot$ progenitor, but 9a includes an aspherical perturbation added to the progenitor model, which leads to more neutron-rich ejecta and the production of heavier elements towards $Z=40$. The nucleosynthesis trend with respect to $M_{\rm ZAMS}$ is non-monotonic, due to the mass-dependent nature of the explosion. Replotted with data provided by Tianshu Wang.}
    \label{fig:wang}
\end{figure*}

Quantitatively useful yield predictions have become feasible with multi-D neutrino-driven explosion models \cite{pruet_2005,arcones_2011,fujimoto_2011,bliss_2018,wanajo_2018}. In these models, non-radial instabilities and turbulence assist shock revival and shape the conditions in the neutrino-driven outflows. The thermodynamic history and $Y_e$ (or neutron richness) of the innermost ejecta can be directly extracted from such models (see Figure~\ref{fig:traj}). An important note is that turbulence and mixing differ qualitatively between 2D axisymmetric and 3D: in 3D, a turbulent cascade transfers large flows down to small scales, whereas in 2D the cascade is reversed \cite{couch_2013,dolence_2013}. Therefore, nucleosynthesis inferred from 2D calculations can be biased. During the last decade, fully 3D neutrino hydrodynamic simulations have matured to the point where they can be used to compute nucleosynthesis of the innermost ejecta with substantially improved physical fidelity \cite{Wongwathanarat_2013,sieverding_2020,sieverding_2023b,wang_2024a,wang_2024b}.

Wang and Burrows \cite{wang_2024a} presented CCSN yields based on the currently largest suite of fully 3D and long-term CCSN simulations using the \textsc{Fornax} code, which follows progenitors of $\sim$9–60\,$M_\odot$ to several seconds after core bounce. Because the innermost ejecta arise from inherently multi-D, time-dependent flows in the gain region and outflows, their electron fraction $Y_e$ and thermodynamic trajectories exhibit strong spatial and temporal variability, rather than the nearly monotonic behavior in 1D parameterized models. Across most of their models, the neutrino-processed ejecta span moderately proton-rich to mildly neutron-rich conditions, with typical $Y_e \simeq 0.45$–0.60. The publicly released yields (Figure~\ref{fig:wang}) show efficient production of intermediate-mass and light trans-iron nuclei (roughly Si through Ge), largely through explosive burning and charged-particle–dominated freeze-out. They do not find a robust main $r$-process in these neutrino-driven ejecta, which is consistent with the lack of sufficiently neutron-rich conditions. Instead, some of their models exhibit a clear weak $r$-process, producing nuclei up to the $^{90}$Zr peak. In particular, their lowest-mass model with prescribed perturbation (9a) exhibits ejecta with $Y_e$ down to $\sim$0.35 that produce abundant trans-Fe elements, which may be an important chemical signature of CCSNe at the lower mass end ($\sim9\,M_\odot$).

\subsubsection{Surrogate 1-dimensional models}
Despite these advances, it remains computationally impractical to compute a large, systematic grid of fully self-consistent 3D CCSN models spanning many progenitors while also exploring key uncertainties (neutrino transport, microphysics, progenitor structure) and enabling broad comparison to observational constraints. Consequently, many nucleosynthesis studies (especially GCE models) still rely on reduced-physics explosion prescriptions in 1D. For example, the classic ``thermal bomb” method deposits a prescribed energy near an imposed mass cut for the PNS to launch a shock and then follows explosive burning (e.g., \cite{hashimoto_1989,limongi_2003,umeda_2017,imasheva_2023}). Although valuable for examining nucleosynthesis in the outer shells, such approaches do not model neutrino heating and charged-current interactions, and therefore cannot reliably predict the $Y_e$ of the innermost ejecta. This is a critical limitation because $Y_e$ strongly regulates Fe-group and trans-Fe production.

To bridge the gap between tractability and realism, intermediate approaches have been developed that include approximate neutrino transport and/or calibrated effective physics. For example, the \textsc{PUSH} framework \cite{push_1} uses the isotropic diffusion source approximation (IDSA; \cite{idsa}) for neutrino transport and triggers explosions by enhancing the heating associated with heavy-flavor neutrinos. Its free parameters are calibrated to reproduce key properties of SN 1987A, and yield sets have been published for solar-metallicity progenitors \cite{push_3}, low-metallicity progenitors \cite{push_4}, and different nuclear EoSs \cite{push_5}. In a complementary direction, 1D CCSN simulations have been augmented with an effective treatment of multi-D convection and turbulent pressure which is calibrated against 2D/3D results \cite{couch_2020}, enabling emerging ``1D+” nucleosynthesis studies \cite{boccioli_2024,Boccioli_2025,Boccioli_2026}. Ultimately, the predictive power of these reduced-cost models depends on how well they reproduce the thermodynamic histories and the $Y_e$ distributions of the innermost ejecta found in full 3D explosion models.

\subsubsection{Binarity, rotation and metallicity}
The discussion above implicitly targets CCSNe from single, non-rotating, solar-metallicity progenitors. Real massive-star populations deviate from this idealization in at least three ways: binarity, rotation, and metallicity. Each can alter the presupernova structure and therefore the explosion outcome and yields.

\emph{Binarity}. More than half of massive stars sit in binary or multiple systems, interact with companions, and experience mass transfer and/or envelope stripping \cite{sana_2012,chen_2024}. Such interactions can produce hydrogen-deficient progenitors that explode as Type Ib/Ic supernovae, and their nucleosynthesis yields have been explored with parametrized explosions \cite{shigeyama_1990,yoshida_2017}. More recently, binary interaction has been emphasized to change the stellar evolution pathways and nucleosynthesis yields in ways that cannot be captured by treating the system as two independent single stars \cite{farmer_2023}. Detailed comparisons show systematic differences in presupernova core structure between single and binary-stripped models \cite{laplace_2021}, and these structural differences can translate into different patterns of explodability, remnant masses, and fallback behavior against $M_{\rm ZAMS}$ \cite{vartanyan_2021}. It remains unknown how binarity would affect the contribution of CCSNe to GCE.

\emph{Rotation}. Through rotation-induced mixing and angular-momentum transport \cite{maeder_2000}, rotation can modify the chemical stratification and compactness of the pre-collapse core. Using parameterized explosion prescriptions applied to rotating progenitor models, yield sets have been computed for low-metallicity and solar-metallicity cases \cite{limongi_2018,roberti_2024a} and for super-solar metallicity cases \cite{roberti_2024b}. During quiescent evolution, rotating massive stars tend to synthesize a larger amount of heavy nuclei (up to Pb) than their non-rotating counterparts, and they are likely to explode as Wolf-Rayet stars rather than red supergiants. Magnetic fields add further complexity that can qualitatively change the explosive dynamics (MR-driven CCSN) and nucleosynthesis, and their coupled impact with rotation on CCSN nucleosynthesis is discussed in Section~\ref{ssec:MR}.

\emph{Metallicity}. Line-driven winds are weaker at reduced metallicity \cite{smith_2014}, so stars typically lose less mass and angular momentum over their lifetimes. For a fixed $M_{\rm ZAMS}$, low-metallicity progenitors often reach collapse with different core–envelope configurations and may exhibit systematically different explodability \cite{heger_2003}. They also tend to have a lower neutron excess (with $Y_e \simeq 0.5$), which suppresses the production of odd-$Z$ nuclei relative to even-$Z$ nuclei and therefore strengthens the so-called even–odd effect in the yields (isotopes with even-$Z$ elements are significantly more abundant than the odd-$Z$ elements) \cite{heger_2010}. Several groups have explored CCSN nucleosynthesis from low metallicity to metal-free (Population III) progenitors, using parameterized explosion models to provide broad yield grids \cite{jura_1986,tominaga_2007,yoshida_2008,heger_2010,limongi_2012,izutani_2010,bessell_2015}. Finally, very massive stars at low metallicity can explode in a distinct channel, which will be discussed in Section~\ref{ssec:PISN}.

Overall, the exploration of these effects has relied on parameterized studies with 1D explosion models. These models and yield grids remain valuable for mapping trends with progenitor mass and composition, especially for nucleosynthesis in the outer layers, but not for the innermost ejecta.

\subsubsection{Uncertainties in reaction rates}
CCSN yield predictions are sensitive to uncertainties of nuclear reaction rates in two ways: (1) through stellar evolution, where reaction rates change the presupernova core structure and composition that set the stage for the explosion, and (2) through explosive nucleosynthesis, where rates directly control burning and freeze-out in shocked material.

On the stellar-evolution side, Fields et al. \cite{fields_2018} conducted a systematic Monte Carlo survey of reaction-rate uncertainties and their impact on presupernova properties. They found that a relatively small set of rates (about 8), prominently including the triple-$\alpha$ and $^{12}$C($\alpha,\gamma$)$^{16}$O reactions, dominate the uncertainties in quantities such as core masses, composition profiles, and presupernova compactness. They further emphasized that the resulting variations can be comparable to differences introduced by numerical/modeling choices (such as mass resolution and network size), implying that nuclear and numerical uncertainties must be treated on a similar footing when assessing presupernova structure.

Several studies have then examined how rate uncertainties propagate into explosive yields \cite{tur_2010,kikuchi_2015,subedi_2020,jin_2020,jayatissa_2022,nishimura_2026}. For instance, varying the triple-$\alpha$ and $^{12}$C($\alpha,\gamma$)$^{16}$O rates within $\sim$2$\sigma$ can change the predicted production of long-lived radioisotopes such as $^{26}$Al and $^{60}$Fe by factors of $\gtrsim 5$, whereas $^{44}$Ti is comparatively less sensitive \cite{tur_2010}. 
Moreover, the de-excitation of the Hoyle state in $^{12}$C by neutrons, protons or alpha particles also affect the $\nu$p-process\cite{jin_2020, Sasaki_2023}.
For the explosion phase, 11 reactions (dominated by ($\alpha,p$) and ($p,\gamma$) channels) were identified to strongly affect the synthesis of $^{44}$Ti and $^{56}$Ni \cite{subedi_2020}. In another extensive study focused on $\gamma$-ray emitters \cite{hermansen_2020}, a much larger set (141) of rates can produce significant yield variations (see their Table 3 for the largest effects), demonstrating that the relevant rate set expands rapidly once many isotopes and burning regimes are considered.

It should be noted that the importance of reaction-rate uncertainties depends on the relevant thermodynamic regime. In the innermost ejecta that reach NSE, abundances are set primarily by bulk conditions such as entropy, expansion timescale, and $Y_e$ rather than by individual charged-particle rates. Using the \textsc{PUSH} approach, Nishimura et al. \cite{nishimura_2026} found that only a small number of rates affect the outcome of many Fe-group yields (see their Table 4-7). 

\subsubsection{Neutrino-induced processes and flavor oscillation}
Despite the small cross section for neutrino-nucleus interactions, the intense flux of neutrinos with $\sim10^{53}$\,erg\,s$^{-1}$ can induce significant nuclear reactions. The primary mechanism involves inelastic neutral-current scattering, where heavy-flavor neutrinos ($\nu_\mu$, $\nu_\tau$) with characteristic temperatures excite nuclei to particle-unbound levels through giant resonance transitions, leading to spallation reactions that evaporate neutrons, protons, or $\alpha$-particles \cite{woosley_1990}. These products then react with the surrounding medium to synthesize rare isotopes. The $\nu$-process successfully explains the solar abundances of several odd-Z nuclei: $^{11}$B is produced primarily via $^{12}$C($\nu$,$\nu'$n)$^{11}$C in the C/O-shell and via $^4$He($\nu$,$\nu'$p)$^3$H($\alpha$,$\gamma$)$^7$Li($\alpha$,$\gamma$)$^{11}$B and $^4$He($\nu$,$\nu'$n)$^3$He($\alpha$,$\gamma$)$^7$Be($\alpha$,$\gamma$)$^{11}$C($\beta^+$)$^{11}$B in He-layer, $^{19}$F is synthesized through $^{20}$Ne($\nu$,$\nu'$p)$^{19}$F in the Ne-shell, and $^7$Li via $^4$He($\nu$,$\nu'$p)$^3$H($\alpha$,$\gamma$)$^7$Li in He/C shells \cite{Yoshida_2005,Kusakabe_2019}. $\nu$-process also accounts for the rare isotopes $^{138}$La and $^{180}$Ta through ($\nu$,$\nu'$n) and ($\nu_e$,$e^-$) reactions on seeds nuclei in the O/Ne shell \cite{Heger_2005}. The earlier calculations \cite{Heger_2005,Hayakawa_2013,Hayakawa_2018,kobayashi_2011} are subject to significant uncertainties associated with the assumed neutrino energy spectra, since they employed energy-independent neutrino-induced reaction cross sections. The resulting nucleosynthesis yields can differ substantially when realistic energy-dependent cross sections are combined with neutrino spectral change through propagation, arising from collective or Mikheyev-Smirnov-Wolfenstein flavor-oscillation effects. This has been demonstrated by Ref. \cite{yao_2025} for $^{138}$La and by Ref. \cite{Yoshida_2005} for $^7$Li and $^{11}$B. The $\nu$p-process, which may occur in hot dense proton-rich matter with strong neutrino flux, can produce intermediate-mass $p$-nuclei, particularly $^{92,94}$Mo and $^{96,98}$Ru \cite{pruet_2005,wanajo_2006,frohlich_2006b}. The $\nu p$-process could happen since $\bar{\nu}_e$ captures on free protons continuously generate neutrons, enabling $(n,p)$ reactions to bypass electron-capture bottlenecks such as $^{64}$Ge and facilitate the production of heavy proton-rich nuclei.  More elaborate processes, $\nu r$ process \cite{xiong_2024} and $\nu i$ process \cite{wangxl_2023}, can produce $p$-nuclei and lanthanides up to $A \sim 200$, though their synthesis conditions in CCSNe are more uncertain. 

A further source of uncertainty for nucleosynthesis in the innermost ejecta is the flavor evolution of neutrinos. Neutrinos are produced and transported in flavor states (i.e., as electron-flavors $\nu_e$, $\bar\nu_e$, muon- and tau-flavors $\nu_x$ with $x=\mu,\tau$ and their antiparticles), but they propagate as mixtures of mass eigenstates and can oscillate between flavors \cite{fukuda_1998}. This matters for nucleosynthesis because $Y_e$ is controlled mainly by charged-current reactions of electron-flavor neutrinos on free nucleons, see Eq.~\ref{eq:nu}.  Early studies showed that neutrino oscillations can modify the conditions for heavy-element production (especially the $r$-process) and allow nucleosynthesis constraints on neutrino mixing parameters \cite{qian_1993,qian_1995}. Subsequent work has shown that neutrino–neutrino forward scattering makes the flavor evolution nonlinear and can produce collective oscillation phenomena that may alter the $\nu_e/\bar\nu_e$ spectra experienced by outflows, modifying both the $r$-process and the $\nu p$-process in proton-rich ejecta \cite{pllumbi_2015,wu_2015,balantekin_2024, sasaki_2017, Sasaki_2021}. More recently, fast flavor conversion has become a highly debated topic and is important for both the explosion mechanism and nucleosynthesis \cite{tamborra_2021,fujimoto_2023}. Finally, because flavor conversion depends on the neutrino mass ordering and on mixing parameters, CCSNe nucleosynthesis and neutrino signals can serve as potential probes of the mass hierarchy \cite{mathews_2012}, a fundamental problem in particle physics \cite{wolenstein_1978}.

\subsubsection{$\gamma$-process and $p$-nuclei production}
In CCSNe, $\gamma$-process occurs during explosive O/Ne-shell burning. The seed nuclei originate either from pre-existing $s$- and $r$-process isotopes in the protostellar cloud or are produced in situ during the weak $s$-process in He-core burning and convective C-shell burning phases~\cite{Travaglio_2018, Rayet_1995, Rayet_1990, Prantzos_1990, Rauscher_2002}. In general, light $p$-nuclei are produced in layers with high temperature, whereas heavy $p$-nuclei form in low-temperature regions. For a peak temperature $T_{\rm p}\lesssim 2.6\times10^9$\,K, heavy $p$-nuclei with neutron number $N>82$, such as $^{180}$Ta, $^{190}$Pt and $^{184}$Os, are synthesized via $(\gamma,n)$, $(\gamma,p)$ and $(\gamma,\alpha)$ reactions. In layers with intermediate peak temperatures ($2.6\times10^9\,\mathrm{K}\lesssim T_{\rm p}\lesssim3.0\times10^9$\,K), seed nuclei with $50<N<82$ are destroyed by photodisintegration, allowing the production of intermediate-mass $p$-nuclei (also with $50<N<82$). In the innermost layers, where the peak temperature exceeds $T_{\rm p}\gtrsim3.0\times10^9$\,K, light $p$-nuclei ($N<$50) are produced directly via the $(\gamma,n)$ reaction. Most $p$-nuclei in the mass ranges 124 $\leq$ A $\leq$ 150 and 168 $\leq$ A $\leq$ 200 can be produced in CCSNe, while significant deficiencies remain in the A $\leq$ 124 and  A$\sim$160 mass regions, where solar abundances are severely underproduced~\cite{Rauscher_2002,Sensitivity2006,Hayakawa_2008,rauscher_2013}. For these mass regions, neutrino-induced processes are proposed as the main production mechanism, as we discussed in the previous section.

Of particular note, $\gamma$-process relies on extensive reaction-network calculations involving about 10,000 reactions across nearly 2,000 stable and unstable nuclei. Consequently, predicted abundances are highly sensitive to uncertainties in nuclear properties and reaction rates. Despite experimental efforts to constrain key reactions~\cite{2020PhRvC.102d5811S,2020PhLB..80735575F,2021PhRvL.127k2701L,2021ApJ...915...78C,2022EPJWC.26011030V,2023EPJWC.27911004W,2023PhRvC.107c5803W,2023PhRvC.107c5808T}, many relevant nuclei remain inaccessible due to instability or extremely low natural isotopic abundances. Therefore, most reaction rates still rely on statistical Hauser–Feshbach (HF) calculations~\cite{Hauser1952The,HauserFeshbach}. Codes like TALYS~\cite{2012Modern}, combined with various phenomenological and microscopic models, have become essential for calculating cross sections across a wide range of nuclei~\cite{Utsunomiya2003Cross,2011PhRvC..84a5802D,2006PhRvC..74b5805G,2014PhRvC..90c5806N,2009PhRvC..80c5804G,2020PhLB..80535431W}. However, the nuclear inputs for HF calculations remain poorly constrained in the $\gamma$-process mass region, highlighting the need for dedicated statistical-model tests.

\subsection{Hypernovae and jet-induced nucleosynthesis \label{ssec:MR}}

Observations show that some CCSNe have extremely high explosion energies ($\sim$10\,B) inferred from broad spectral lines and light-curve fitting. These events are sometimes associated with long gamma-ray bursts and termed hypernovae \cite{iwamoto_1998, woosley_1999}. The neutrino-driven mechanism cannot achieve the high explosion energies characteristic of hypernovae. Such events are rare ($\sim1\%$ of all CCSNe) in the local universe, but they could serve as major $r$-process sites in the early universe, when neutron-star mergers are not yet viable. From a GCE model, the occurrence rate of hypernovae in the early Universe may be quite high (about half of all CCSNe) to explain the Fe-peak ratios in metal-poor stars \cite{grimmett_2020}. In the following, we review two contemporary candidate explosion scenarios for hypernovae and their associated jet-induced nucleosynthesis.

\subsubsection{Magnetorotational supernovae}
\begin{figure*}[ht]
    \centering
    \includegraphics[width=1.0\linewidth]{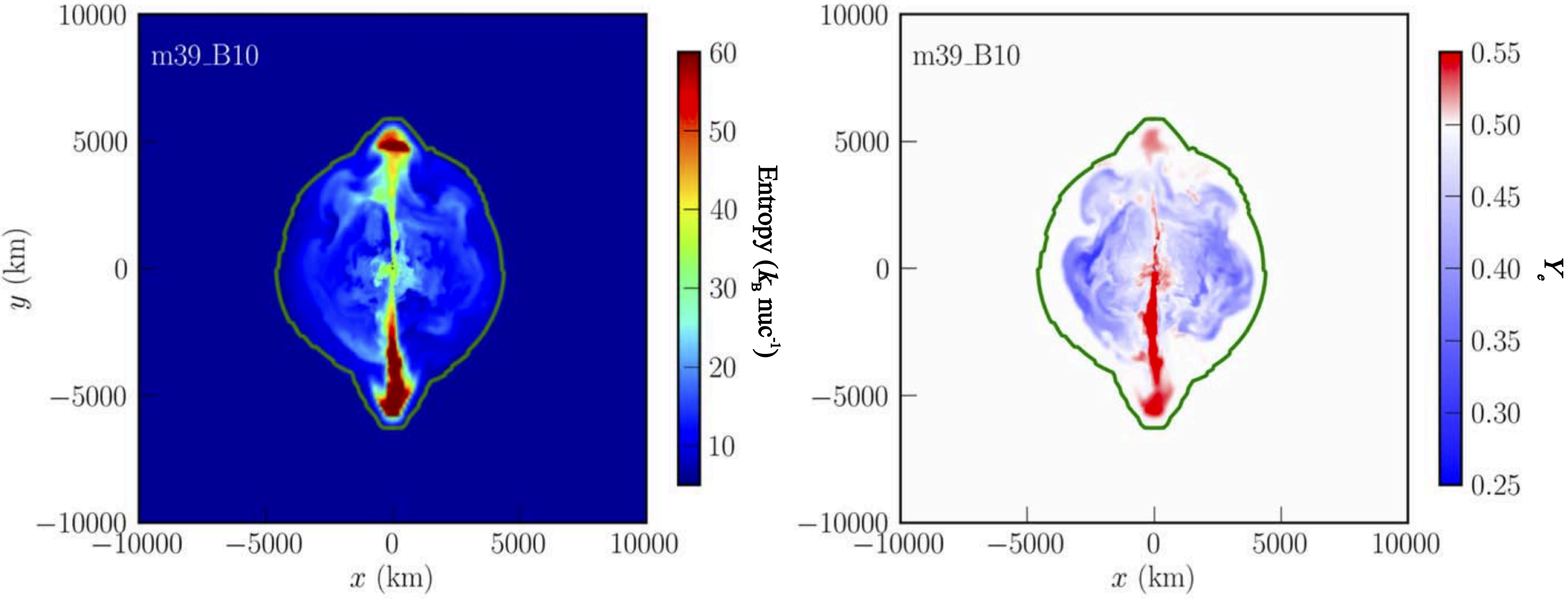}
    \caption{Explosion configuration in a magnetorotational supernova simulation. Two-dimensional slices of entropy (left) and electron fraction (right) are shown in the $x$-$y$ plane, with the $y$-axis aligned with the rotation axis. Jets are launched along this axis. Reprinted from Zha et al. (2024) \cite{zha_2024}.}
    \label{fig:jet}
\end{figure*}

Magnetorotational supernovae (MRSNe) arise when rapid rotation and strong magnetic fields in the pre-supernova Fe core substantially alter the collapse and explosion dynamics relative to the delayed neutrino-driven paradigm. Differential rotation can amplify seed fields by winding and the MR instabilities, enabling magnetic stresses to extract rotational energy and launch bipolar outflows, see Figure~\ref{fig:jet} for an example of MRSNe with jet launching \cite{powell_2023,zha_2024}. From a nucleosynthesis perspective, the key difference from neutrino-driven explosions is that a fraction of the innermost ejecta can be expelled in fast, low-$Y_e$ channels (often jet-like), thus enabling $r$-process \cite{trivedi_1978}. 

\begin{figure*}[ht]
    \centering
    \includegraphics[width=\textwidth]{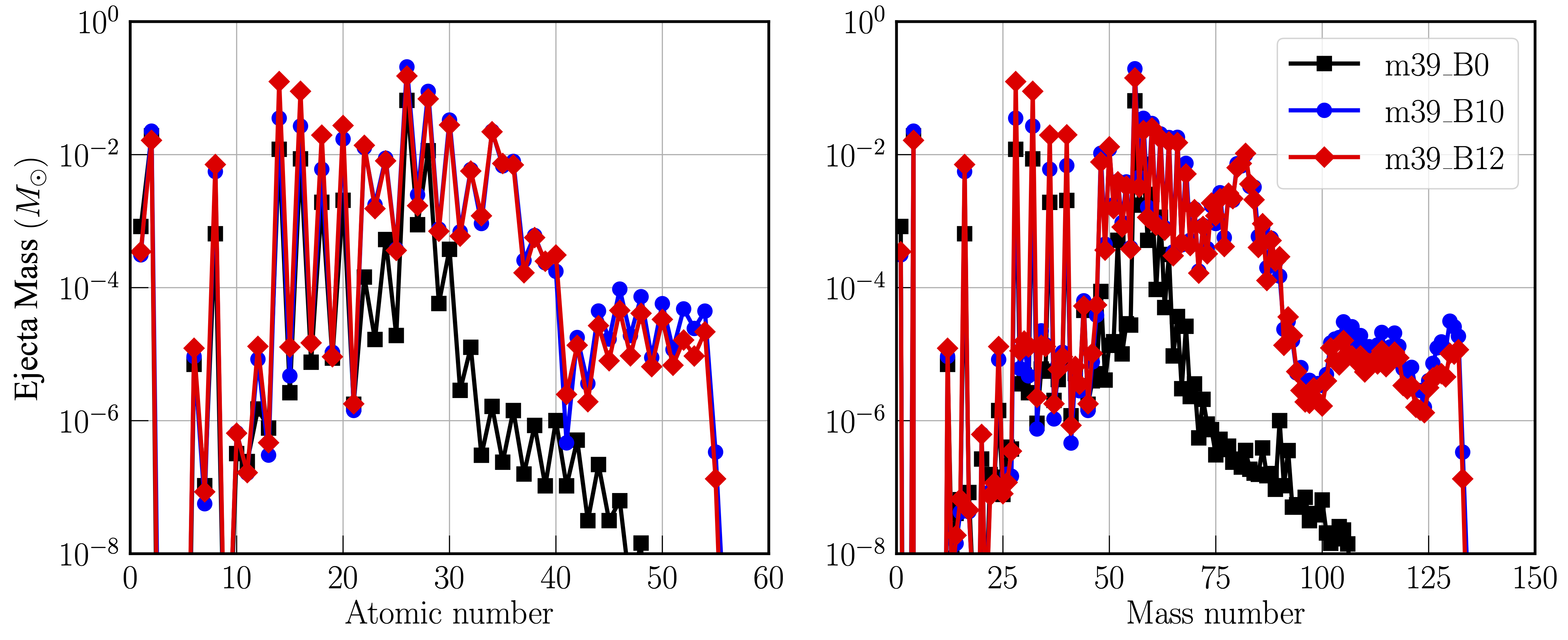}
    \caption{Yield patterns as a function of atomic number $Z$ (left panel) and mass number $A$ (right panel) for the nonmagnetic model (m39\_B0) and the magnetic models (m39\_B10 and m39\_B12). In the magnetic models, nucleosynthesis of heavy elements proceeds up to $A\sim130$. Reprinted from Zha et al. (2024) \cite{zha_2024}.}
    \label{fig:mhd_yields}
\end{figure*}

Multi-D simulations have explored the nucleosynthesis yields of MRSNe and their sensitivity to progenitor conditions. With 2D simulations coupled with a neutrino leakage scheme, Nishimura et al. \cite{nishimura_2015} demonstrated that jet-like ejecta in rapidly rotating, strongly magnetized cores can achieve neutron-rich conditions compatible with a strong $r$-process, compatible with the pattern found in an extremely metal-poor star \cite{yong_2021}. They found that the outcome is sensitive to the assumed rotation and magnetic-field strength, and successful $r$-process occurs in models with a prompt magnetic jet (within 30\,ms after core bounce) but not a delayed jet. Later, it was found that 3D non-axisymmetric effects \cite{mosta_2018,powell_2023,reichert_2023b,liubov_2026} and magnetic-field configuration and misalignment \cite{halevi_2018,reichert_2024} can significantly affect the emergence, stability, and composition of jets. In general, 3D simulations produce weaker, proton-rich jets, while the bulk ejecta is mildly neutron-rich with the minimum $Y_e$ of $\sim0.25$ (see the right panel of Figure~\ref{fig:jet}). As a result, MRSNe can only contribute to elemental synthesis up to the second $r$-process peak ($A\sim130$, see Figure~\ref{fig:mhd_yields}). Clearly, more efforts are needed to justify whether MRSNe can contribute significantly to the main Galactic $r$-process inventory.

\subsubsection{Collapsar}
In the collapsar model, rapidly rotating massive stars undergo core collapse to a BH while the still infalling outer envelope circularizes into an accretion disk and is partially expelled in winds \cite{macfadyen_1999}. The angular momentum profile of the progenitor is important for whether a centrifugally supported disk can form around the BH, or the matter is instead swallowed in a quasi-radial inflow. The composition of the ejected matter in collapsars is mainly determined by $Y_e$ and the entropy of the accretion disk, which depend on the competition between electron/positron captures in the hot and dense disk and neutrino absorption in the surrounding region. Early work \cite{fujimoto_2004,fujimoto_2006,fujimoto_2007} showed that collapsar disks can drive neutron-rich outflows capable of producing heavy elements, and that the yields are sensitive to disk accretion rate, viscosity, and neutrino cooling. More recently, detailed disk-wind calculations find that with BH accretion rates suitable for explaining long GRBs, the disk outflow can produce robust $r$-process yields up to $A\sim195$ \cite{siegel_2019}.
In slowly expanding collapsar ejecta, neutrons released by fission recycling after $r$-process freeze-out may drive secondary $i$- and $s$-process nucleosynthesis and modify the rare-earth abundances ratios such as Tm/Eu and Lu/Eu \cite{He_2024,He_2026}.
Under certain event-rate assumptions, the integrated yields from collapsars can exceed those of neutron-star mergers, accounting for $\gtrsim80$\% of the cosmic $r$-process inventory. However, collapsar models with $\alpha$-viscosity and detailed neutrino transport fail to drive a robust $r$-process \cite{just_2022}, suggesting that more genuinely magnetohydrodynamic effects are required. A similar conclusion is reached in Ref.~\cite{coleman_2024}. Therefore, whether collapsars are significant $r$-process contributors remains highly debated.

\subsubsection{A note on $r$-process nucleosynthesis}
Although observations of the kilonova AT2017gfo provide strong evidence that neutron-star mergers (NSMs) produce neutron-capture elements, it remains premature to regard them as the unique confirmed site of the entire $r$-process. Early identifications of Te I and Cs I in the optical-to-near-infrared spectra were subsequently replaced by the more physically plausible identification of Sr II \cite{pian_2017,smartt_2017,watson_2017}. The latter represents compelling spectroscopic evidence for the production of a first-peak $r$-process element, but its interpretation still depends on simplified treatments of the ejecta temperature, geometry, ionization state, and radiative transfer. Moreover, the production of light $r$-process nuclei such as Sr, Y, and Zr may require contributions from both the dynamical ejecta and accretion-disk winds, whose relative masses remain uncertain \cite{thielemann_2017_2}. GCE studies provide an additional constraint: the characteristic delay times of neutron-star mergers may make it difficult for this channel alone to explain the presence of $r$-process elements in extremely metal-poor stars. This difficulty is particularly relevant to the remarkably uniform abundance pattern observed among heavy $r$-process elements, especially from the second peak upward in old metal-poor halo stars. 

\begin{figure*}[ht]
    \centering
    \includegraphics[width=\columnwidth]{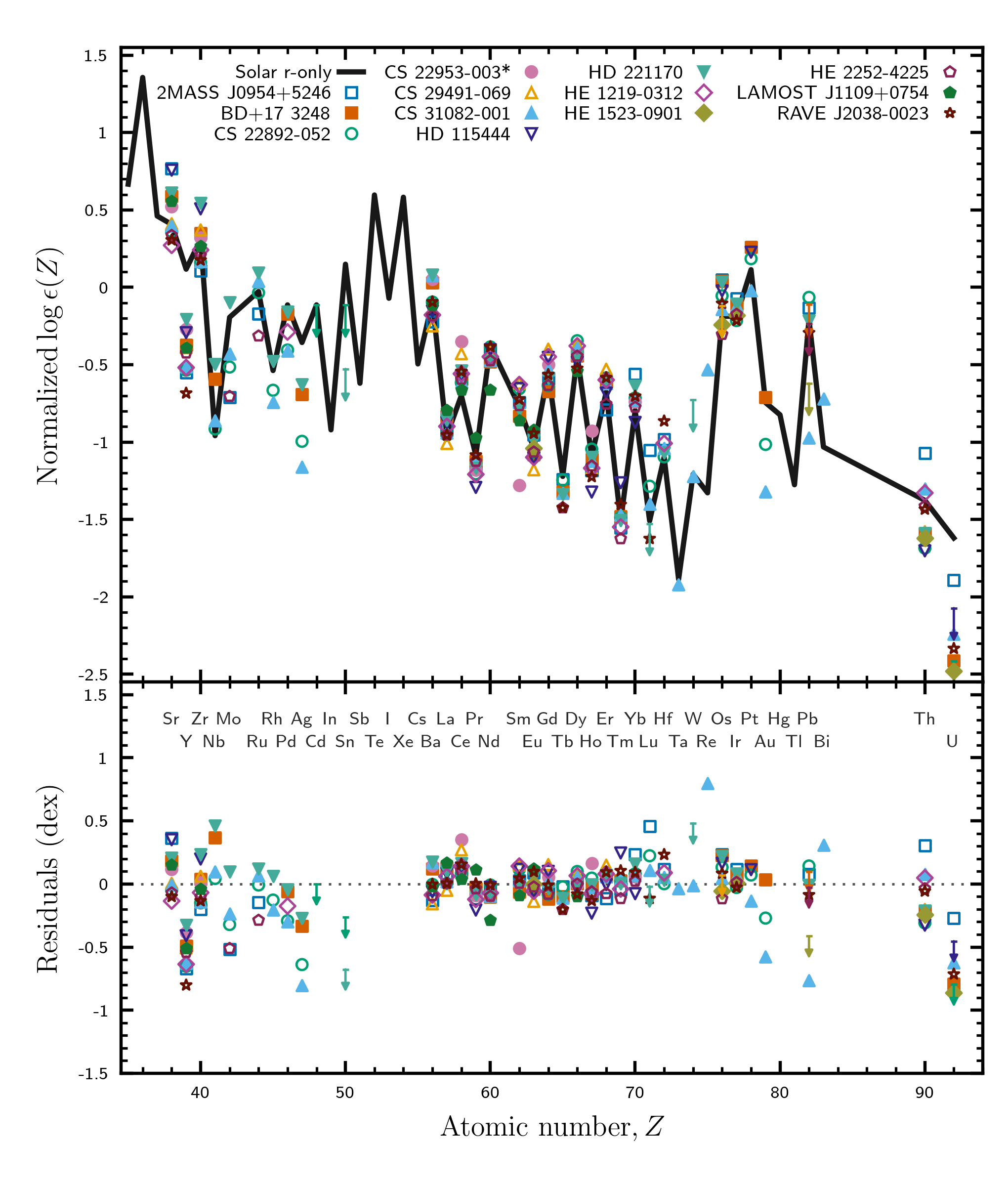}
    \caption{The upper panel shows normalized $r$-process-element abundances of $r$-II stars overlaid with the scaled solar $r$-process-only pattern \cite{burris2000} (black curve). 
The lower panel shows the residuals relative to the best-fitting solar $r$-process-only pattern.
For each star, the solar-system $r$-process-only pattern from Ref. \cite{burris2000} was shifted vertically to match the measured heavy-element abundances. 
The vertical offset was determined by a weighted least-squares fit to the measured Ba–Dy abundances \cite{ji2016}. The data sets are taken from: CS~22892--052
\cite{sneden2003}; HD~115444 \cite{westin2000}; BD$+17$~3248
\cite{cowan2002}; CS~31082--001 \cite{siquiera-mello2013}; HD~221170
\cite{ivans2006}; HE~1523--0901 \cite{frebel2007}; CS~29491--069 and
HE~1219--0312 \cite{hayek2009}; CS~22953--003 \cite{francois2007};
HE~2252--4225 \cite{mashonkina2014}; LAMOST~J110901.22$+$075441.8
\cite{li2015}; RAVE~J203843.2$-$002333 \cite{placco2017}; and
2MASS~J09544277$+$5246414 \cite{holmbeck2018}.}
    \label{fig:rprocess_patt}
\end{figure*}

Figure \ref{fig:rprocess_patt} shows the neutron-capture abundances in 13 $r$-II stars (i.e., $r$-process-rich stars with [Eu/Fe]$>$+1.0 and [Ba/Eu]$<$0 ) and their mean abundance differences to the scaled solar system $r$-process values. The observed rare-earth elements with $56<Z<70$ in $r$-process-enhanced stars show the similar relative pattern, while the lighter and heavier elements ($Z\sim40$ and $Z>70$) have larger scattering.
The solar-system $r$-process residual reflects the cumulative enrichment of
the interstellar medium over many stellar generations, whereas metal-poor
$r$-II stars preserve nucleosynthetic signatures from much earlier stages of
Galactic evolution. 

Their close agreement therefore indicates that the $r$-process yields are
remarkably robust, or universal, at least for elements at and beyond the
second $r$-process peak. This universality constrains the nucleosynthetic
conditions but does not by itself distinguish between NSMs and massive-star
sources. Although NSMs can synthesize heavy $r$-process nuclei up to Th and U,
several GCE calculations find that their delayed contribution \cite{lorimer_2008,dimple_2022} alone cannot
reproduce the observed Eu abundances at the lowest metallicities
\cite{shibagaki_2016,kobayashi_2020,yamazaki_2022}. These results motivate an
additional prompt $r$-process channel associated with massive stars, such as
magneto-rotational supernovae or collapsars. However, this conclusion depends
on the adopted delay-time distribution and treatment of chemical mixing.
Inhomogeneous models show that NSM ejecta may still be incorporated into very
metal-poor stars when chemically primitive gas survives in low-mass Galactic
progenitors \cite{shen_2015}. Taken together, the current evidence favors a
mixed scenario in which prompt massive-star events contribute substantially
to the earliest enrichment, while NSMs remain an important $r$-process source
and become increasingly significant at later times.

\subsubsection{Jet-induced nucleosynthesis}

Despite the different central engines in MRSNe and collapsars, they may share similar nucleosynthesis outcomes induced by jets that propagate through the outer stellar envelope above the NS/BH. The shock-heated, jet-processed material can synthesize IMEs and $^{56}$Ni depending on peak temperatures/densities. Parameterized and semi-analytic jet models have long been used to connect energetics, asphericity, and yields \cite{tominaga_2009}, and more recent studies have revisited jet-induced yields with improved treatments of the outflow thermodynamics and composition \cite{leung_2023,leung_2024}. In particular, Grimmett et al. \cite{grimmett_2021} employed an analytic relation between the jet energy flux and the kinetic energy from the differential rotation of the PNS. The main findings so far are that jet-driven models can produce the \Ni mass required for Type Ic-BL supernovae at hypernova energies \cite{grimmett_2021} and reproduce the abundance patterns observed in some extremely metal-poor galaxies \cite{leung_2024}.

\subsection{Other explosion scenarios}
Because a definitive answer for the CCSN explosion mechanism remains elusive, alternative scenarios have been proposed along with their potential nucleosynthesis signatures. In the so-called jittering-jet mechanism, even without extreme progenitor rotation, stochastic angular-momentum accretion onto the PNS can produce a sequence of short-lived jet episodes with varying axes, whose cumulative effect can unbind the star \cite{papish_2011,soker_2024}. Jet-driven dynamics can strongly enhance asphericity, produce high-entropy channels, and dredge up inner material, thus imprinting distinct abundance patterns including $r$-process elements \cite{papish_2012}. In binary-evolution contexts, jets may also operate during a common-envelope phase. In particular, neutron-rich outflows associated with a common-envelope jet supernova have been proposed as a potential environment for robust  $r$-process nucleosynthesis \cite{jin_2024}. 

Another nonstandard mechanism is the phase-transition-driven supernova, in which a strong first-order phase transition (hadron-quark deconfinement) in the PNS triggers a secondary collapse and launches an additional shock. Under suitable conditions, this secondary shock can power the explosion and can lead to distinctive neutrino and gravitational-wave signatures \cite{sagert_2009,fischer_2018,zha_2020}. From the nucleosynthesis perspective, the altered thermodynamic history and the potential for rapid, shock-driven ejection from deep layers can enable $r$-process beyond the third peak and into the actinides \cite{nishimura_2012,fischer_2020}.

\section{Yields of Other Supernova Channels \label{sec:other}}
\subsection{Electron-capture supernovae}

\begin{figure*}
    \centering
    \includegraphics[width=\columnwidth]{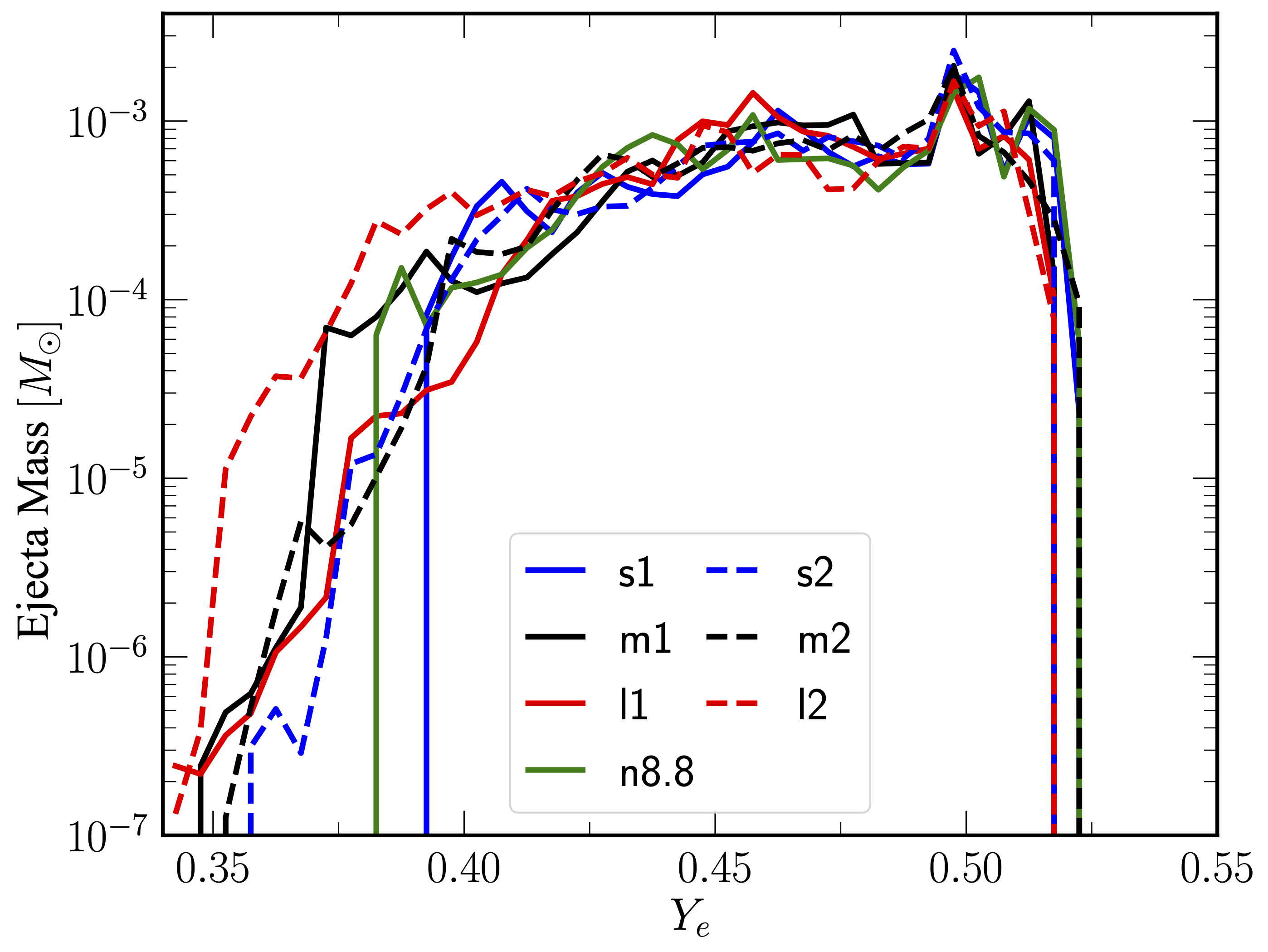}
    \caption{Distribution of the electron fraction $Y_e$ in the neutrino-driven ejecta of electron-capture supernova models \cite{zha_2022}. Each curve corresponds to a different progenitor model from Zha et al.~\cite{zha_2019}, while the n8.8 model is the canonical $8.8\,M_\odot$ progenitor model taken from Nomoto \cite{nomoto_1987}. }
    \label{fig:ecsn}
\end{figure*}

Electron-capture supernovae (ECSNe) are proposed to originate from stars near the lower-mass threshold for CCSNe (i.e., $M_{\rm ZAMS}$=$\sim$8-10\,$M_\odot$), whose evolution produces an electron-degenerate ONeMg core rather than an Fe core \cite{nomoto_1984b}. In this channel, the core approaches the effective Chandrasekhar mass ($M_{\rm Ch,eff}$; see Eq.~\ref{eq:mch}) and becomes unstable when electron captures on nuclei such as $^{24}$Mg and $^{20}$Ne reduce the electron fraction $Y_e$ and degeneracy pressure support, triggering collapse. Note that the electron capture processes are exothermic and will ignite an oxygen-burning flame during electron capture on $^{20}$Ne, which may instead lead to a partial thermonuclear explosion instead of core collapse \cite{nomoto_1991,jones_2016}. Whether ECSNe are collapse or thermonuclear explosions is still under debate \cite{kirsebom_2019,zha_2019,holas_2026}. For the collapse case, because the progenitor core features a steep density gradient at the core–envelope interface, the post-bounce accretion rate drops rapidly, the ram pressure on the bounce shock is relatively low, and neutrino heating can more easily drive a successful, albeit weak explosion \cite{kitaura_2006}.

Early self-consistent simulations established the basic picture that ECSNe can explode via the neutrino-driven mechanism even in the 1D case, but with relatively low explosion energies ($\sim0.1$\,B) and small synthesized $^{56}$Ni masses compared with standard CCSN explosions \cite{kitaura_2006}. More recent multi-D studies have examined how turbulence, asymmetries, and more detailed microphysics affect the outcome. In particular, Zha et al. have explored the pre-collapse oxygen flame stage using 2D reactive hydrodynamic simulations \cite{zha_2019} and simulated the subsequent collapse and explosion phase with detailed neutrino-transport hydrodynamic simulations with their 2D progenitor models \cite{zha_2022}. Nevertheless, much of the explosion properties do not differ significantly from the previous studies with the canonical $8.8\,M_\odot$ progenitor models from Nomoto \cite{nomoto_1987}.

The nucleosynthesis of ECSNe reflects their distinctive core-envelope structure and prompt explosion nature. Because the explosion is relatively weak and the ejecta mass from deep metal-rich layers is modest, ECSNe contribute little to $\alpha$-elements and Fe-group elements compared with more energetic CCSNe of more massive stars. Their chemical signature might be in accordance with the Crab explosion \cite{nomoto_1982c}. On the other hand, their potential importance lies in the production of specific isotopes in the trans-Fe region under neutron-rich conditions. In ECSNe, early ejecta can become mildly neutron-rich due to their early explosion (with $Y_e$ down to $\sim0.35$, see Figure~\ref{fig:ecsn}; \cite{zha_2022}), enabling a charged-particle freeze-out that favors certain neutron-rich species rather than a full main $r$-process.

Several studies have quantified these yield patterns and emphasized their sensitivity to the $Y_e$ distribution of the innermost neutrino-processed ejecta. Wanajo et al. \cite{wanajo_2009} showed that with the neutron richness achieved in the earliest neutrino-driven ejecta, ECSN-like conditions can efficiently synthesize elements around the first trans-Fe peak (including neutron-rich isotopes often difficult to produce in ordinary CCSN models). Later, they emphasized the potential of ECSNe on contributing to weak $r$-process elements \cite{wanajo_2011b}, $^{60}$Fe \cite{wanajo_2013a} and $^{48}$Ca \cite{wanajo_2013b} with their 2D explosion models. Owing to the similar $Y_e$ distribution, the ECSN yields share similar characteristics with those from the lowest-mass CCSN models, e.g., model 9a in Figure~\ref{fig:wang} in particular. The potential role of ECSNe on GCE is still hindered by the large uncertainties in the occurrence rate relative to CCSNe, from a few percent to $\sim20$\% \cite{poelarends_2008,doherty_2017}.

\subsection{Pair-instability supernovae \label{ssec:PISN}}
\begin{figure*}[t]
    \centering
    \includegraphics[width=\columnwidth]{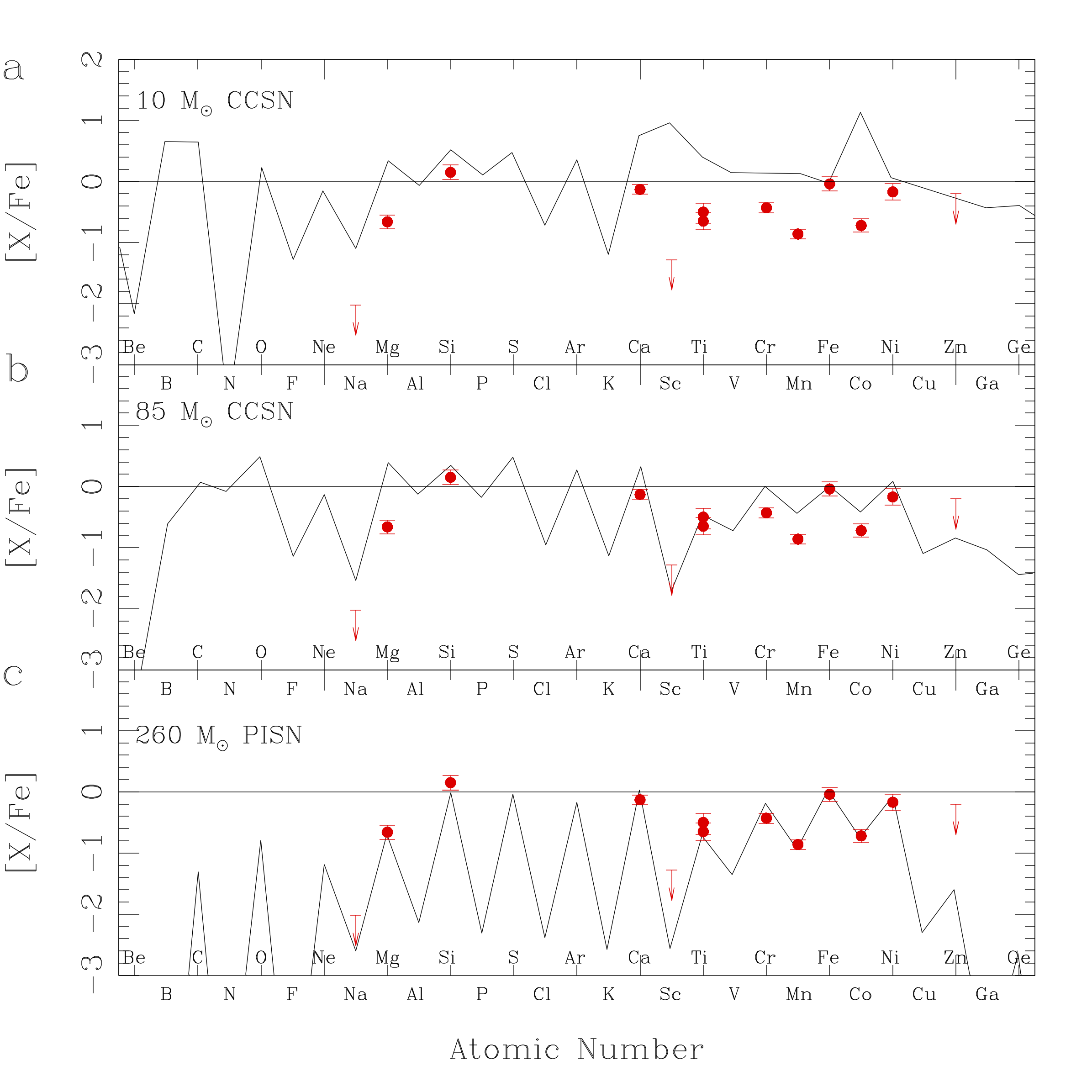}
    \caption{Abundance pattern of the Galactic halo star LAMOST J1010+2358 (red points) compared with nucleosynthesis yield predictions from Heger et al. \cite{heger_2010}: CCSN models with $M_{\rm ZAMS}=$10\,$M_\odot$ (top) and $M_{\rm ZAMS}=$85\,$M_\odot$ (middle), and a PISN model with $M_{\rm ZAMS}=$260\,$M_\odot$ (bottom). Adapted from Xing et al. (2023) \cite{xing_2023}, used under CC BY 4.0. }
    \label{fig:pisn}
\end{figure*}

Pair-instability supernovae (PISNe) occur in very massive stars whose cores reach conditions for thermal photons to produce electron–positron ($e^-e^+$) pairs \cite{eleid_1983,ober_1983}. This $e^-e^+$ pair creation reduces radiation pressure, lowers the effective adiabatic index of EoS below $4/3$, and causes the core to contract dynamically. The resulting compressional heating can ignite explosive O burning and later Si burning, releasing enough nuclear energy to reverse the collapse. The outcome depends primarily on the He-core mass: lower-mass cores undergo pulsational pair instability, ejecting shells without total disruption \cite{woosley_2007}, whereas higher-mass cores are completely disrupted in a single thermonuclear explosion. Although not specific to PISNe, early studies of very luminous, vibrationally unstable stars \cite{appenzeller_1970} helped to establish the framework for dynamical instabilities in such extreme stellar interiors. Metallicity is a very important factor because it regulates mass loss due to stellar wind and the ability of a star to retain the large He core required for pair instability. Therefore, PISNe are expected mainly from low-metallicity or Population III stars \cite{umeda_2005,heger_2010,kozyreva_2014}. For metal-free progenitors, the corresponding ZAMS mass range is approximately $140$-$260\,M_\odot$\cite{heger_2010}.

Because PISNe are thermonuclear explosions, their nucleosynthesis is governed mainly by the peak temperatures and densities reached during explosive O- and Si-burning, together with the pre-explosion core stratification \cite{heger_2010}. PISN yields show several characteristic features: strong production from O through the Fe-group, a pronounced odd-even effect—especially at low metallicity (due to a small neutron excess), and little production of elements heavier than the Fe-group \cite{heger_2010}. A low neutron excess forces nucleosynthesis through nuclei with equal numbers of protons and neutrons, for which the nuclear pairing force is strongest. In this regime, even-$Z$ nuclei are significantly more tightly bound than their odd-$Z$ neighbors. During freeze‑out from NSE, this binding energy difference is exponentially amplified, producing significantly more abundant even-$Z$ elements than odd-$Z$ elements (the so-called odd-even effect).  Consistent with this picture, Kozyreva et al. \cite{kozyreva_2014} found that the odd-even effect weakens as progenitor metallicity (and thus neutron excess) increases. The most massive PISN events can produce substantial $^{56}$Ni, which makes them ideal candidates for radioactive-powered superluminous supernovae \cite{gal-yam_2009,kasen_2011}. 

More recent studies have examined how progenitor properties and nuclear physics modify this classical behavior. Takahashi et al. \cite{takahashi_2018} studied PISN yields and their implications for early chemical enrichment with updated stellar and explosion models, including rotating progenitors. They found that rotation has little effect on the overall abundance pattern except for nitrogen, which is particularly sensitive to rotational mixing. In addition, uncertainties in key reaction rates can affect both the explosion energetics and the final abundance pattern. For example, a larger rate of the $^{12}$C($\alpha,\gamma$)$^{16}$O reaction can lead to higher $^{56}$Ni yields and explosion energies \cite{kawashimo_2024}. A notable limitation of current models is that multi-D effects remain unexplored in PISN explosion models and yields.

Because PISNe produce distinctive abundance patterns and are expected to occur in metal-free or metal-poor environments, their chemical imprint in extremely metal-poor stars has been a long-standing target of observational searches \cite{tumlinson_2006,karlsson_2013,bennassuti_2017}. Umeda and Nomoto \cite{umeda_2005} emphasized how to confront the distinctive yield patterns of PISNe with extremely metal-poor stars, and highlighted that the broad elemental signature (such as [C/Fe], [Zn/Fe], and [Co/Fe]) provides a discriminant from ordinary CCSNe. However, unambiguous identifications have been challenging to date, because early chemical enrichment is stochastic and other explosion channels (or mixing/fallback effects) can partially mimic some PISN features. Xing et al. \cite{xing_2023} recently reported a Galactic halo star whose abundance pattern exhibits an unusually pronounced even–odd signature and is best matched by the nucleosynthesis yields of a 260-$M_\odot$ PISN model from Heger et al. \cite{heger_2010}~(see Figure~\ref{fig:pisn}). Further confirmed stellar objects would not only provide the most direct evidence that PISNe enriched the early Galaxy, but also greatly enhance our knowledge about the initial mass function of the first stars \cite{ishigaki_2018}.

\section{Conclusions and outlook \label{sec:final}}

Supernovae represent catastrophic deaths of stars and play a critical role in synthesizing and dispersing heavy elements throughout the universe. In this review, we have presented the explosion mechanisms proposed for various supernova channels, the current state of their numerical modeling, and their characteristic nucleosynthesis yields. Thanks to advances in high-performance computing and the development of sophisticated magnetohydrodynamic codes with realistic neutrino transport and reaction networks, increasingly accurate supernova explosion models are now being constructed. The nucleosynthesis yields of different supernova channels, including 3D neutrino-driven and magnetorotational CCSN models, are continuously derived by extracting the nucleosynthesis conditions from these simulations. In particular, new insights into the synthesis of heavy elements beyond iron through $r$-process have emerged from 3D models of magnetorotational CCSNe. These developments have greatly advanced our understanding of the role that supernovae play in cosmic chemical evolution. In return, observational constraints on nucleosynthesis have helped refine our knowledge of supernova progenitor evolution and explosion mechanisms.

Despite this progress, significant challenges remain. Multi-D supernova simulations remain computationally too expensive to explore large parameter spaces that include progenitor binarity, rotation, and metallicity. Consequently, these models remain largely indicative and cannot yet be directly integrated into GCE frameworks, which still rely on effective 1D explosion models. Emerging 1D+ models represent an important improvement and warrant further investigation in GCE models. Artificial intelligence techniques may enhance effective yield prediction by leveraging information derived from multi-D explosion models. We further note that the bulk of current nucleosynthesis calculations with multi-D models are post-processed and do not self-consistently account for the dynamical impact of nuclear burning. Continued developments in coupling reduced networks (but more than just an $\alpha$-network) with hydrodynamic simulations are essential for achieving more robust and reliable nucleosynthesis predictions \cite{harris_2017,navo_2023,domingo_2024}.

Furthermore, the synthesis of heavy elements beyond iron via the $r$-, $\gamma$-, and neutrino-induced processes in supernovae remains poorly constrained, owing to insufficient characterization of the requisite physical conditions and persistent uncertainties in nuclear and particle physics. Nuclear experiments such as JUNA \cite{JUNA}, LUNA \cite{LUNA}, FRIB \cite{FRIB}, and HIAF \cite{HIAF} will significantly reduce the uncertainties in reaction rates often used in stellar evolution and explosive nucleosynthesis. 

We are entering an era of abundant observational data, with stellar abundance measurements from large-scale spectroscopic surveys such as LAMOST \cite{LAMOST}, GALAH \cite{GALAH}, APOGEE \cite{APOGEE}, and 4MOST \cite{4MOST} poised to deliver many discoveries. Presolar grains and meteorites preserve microscopic samples of stardust that formed in the outflows of ancient supernovae. Their isotopic anomalies may carry direct nucleosynthesis fingerprints of specific supernovae \cite{banerjee_2016,Jose_2016,yao_2025,Iizuka_2025,pal_2025,nguyen_2025}. These are therefore opportune times to advance supernova models toward more accurate and comprehensive nucleosynthesis yields, thereby sharpening our understanding of the origin of the elements.

\Acknowledgements{We thank Frank Timmes, Tianshu Wang, Qianfan Xing, and Jonas Lippuner for providing the research data and tools used in this work. We also thank Ming-chung Chu, Toshikata Kajino, Shingchi Leung, Bernhard M\"uller, Ken'ichi Nomoto, and Evan O'Connor for their continued collaboration. We have used \textsc{DeepSeek} extensively to improve the text and presentation. This work was supported by the National Natural Science Foundation of China (NSFC, Nos. 12288102, 12393811, 12473031), the Strategic Priority Research Program of the Chinese Academy of Sciences (grant Nos. XDB1160301, XDB1160300, XDB1160000), the CAS Project for Young Scientists in Basic Research (YSBR-148), the National Key R\&D Program of China (Nos. 2021YFA1600403 and 2021YFA1600400), the International Centre of Supernovae, Yunnan Key Laboratory (No. 202505AV340004), the Yunnan Revitalization Talent Support Program--Young Talent project, and the Yunnan Fundamental Research Projects (grant NOs. 202501AS070078, 202401BC070007).}

\InterestConflict{The authors declare that they have no conflict of interest.}



\bibliography{sn_nucleosynthesis}{}

@article{shibagaki_2016,
  author        = {Shibagaki, S. and Kajino, T. and Mathews, G. J. and Chiba,
                   S. and Nishimura, S. and Lorusso, G.},
  title         = {Relative Contributions of the Weak, Main, and
                   Fission-Recycling {$r$}-Process},
  journal       = {The Astrophysical Journal},
  year          = {2016},
  volume        = {816},
  number        = {2},
  eid           = {79},
  pages         = {79},
  doi           = {10.3847/0004-637X/816/2/79},
  eprint        = {1505.02257},
  archiveprefix = {arXiv}
}

@article{kobayashi_2020,
  author        = {Kobayashi, Chiaki and Karakas, Amanda I. and Lugaro, Maria},
  title         = {The Origin of Elements from Carbon to Uranium},
  journal       = {The Astrophysical Journal},
  year          = {2020},
  volume        = {900},
  number        = {2},
  eid           = {179},
  pages         = {179},
  doi           = {10.3847/1538-4357/abae65},
  eprint        = {2008.04660},
  archiveprefix = {arXiv}
}

@article{yamazaki_2022,
  author        = {Yamazaki, Yuta and He, Zhenyu and Kajino, Toshitaka and
                   Mathews, Grant J. and Famiano, Michael A. and Tang,
                   Xiaodong and Shi, Jianrong},
  title         = {Possibility to Identify the Contributions from Collapsars,
                   Supernovae, and Neutron Star Mergers from the Evolution of
                   the {$r$}-Process Mass Abundance Distribution},
  journal       = {The Astrophysical Journal},
  year          = {2022},
  volume        = {933},
  number        = {1},
  eid           = {112},
  pages         = {112},
  doi           = {10.3847/1538-4357/ac721c}
}

@article{shen_2015,
  author        = {Shen, Sijing and Cooke, Ryan and Ramirez-Ruiz, Enrico and
                   Madau, Piero and Mayer, Lucio and Guedes, Javiera},
  title         = {The History of {$r$}-Process Enrichment in the Milky Way},
  journal       = {The Astrophysical Journal},
  year          = {2015},
  volume        = {807},
  number        = {2},
  eid           = {115},
  pages         = {115},
  doi           = {10.1088/0004-637X/807/2/115},
  eprint        = {1407.3796},
  archiveprefix = {arXiv}
}

@article{thielemann_2017_2,
  author        = {Thielemann, F.-K. and Eichler, M. and Panov, I. V. and
                   Wehmeyer, B.},
  title         = {Neutron Star Mergers and Nucleosynthesis of Heavy Elements},
  journal       = {Annual Review of Nuclear and Particle Science},
  year          = {2017},
  volume        = {67},
  pages         = {253--274},
  doi           = {10.1146/annurev-nucl-101916-123246},
  eprint        = {1710.02142},
  archiveprefix = {arXiv}
}

@article{burris2000,
  author        = {Burris, Debra L. and Pilachowski, Catherine A. and Armandroff, Taft E. and Sneden, Christopher and Cowan, John J. and Roe, Henry},
  title         = {Neutron-Capture Elements in the Early Galaxy: Insights from a Large Sample of Metal-Poor Giants},
  journal       = {The Astrophysical Journal},
  year          = {2000},
  volume        = {544},
  pages         = {302--319},
  doi           = {10.1086/317172},
  eprint        = {astro-ph/0005188},
  archiveprefix = {arXiv}
}

@article{westin2000,
  author        = {Westin, Jenny and Sneden, Christopher and Gustafsson, Bengt and Cowan, John J.},
  title         = {The {$r$}-Process Enriched Low-Metallicity Giant {HD 115444}},
  journal       = {The Astrophysical Journal},
  year          = {2000},
  volume        = {530},
  pages         = {783--799},
  doi           = {10.1086/308407},
  eprint        = {astro-ph/9910376},
  archiveprefix = {arXiv}
}

@article{cowan2002,
  author        = {Cowan, John J. and Sneden, Christopher and Burles, Scott and Ivans, Inese I. and Beers, Timothy C. and Truran, James W. and Lawler, James E. and Primas, Francesca and Fuller, George M. and Pfeiffer, Bernd and Kratz, Karl-Ludwig},
  title         = {The Chemical Composition and Age of the Metal-Poor Halo Star {BD +17$^{\circ}$ 3248}},
  journal       = {The Astrophysical Journal},
  year          = {2002},
  volume        = {572},
  pages         = {861--879},
  doi           = {10.1086/340347},
  eprint        = {astro-ph/0202429},
  archiveprefix = {arXiv}
}

@article{sneden2003,
  author        = {Sneden, Christopher and Cowan, John J. and Lawler, James E. and Ivans, Inese I. and Burles, Scott and Beers, Timothy C. and Primas, Francesca and Hill, Vanessa and Truran, James W. and Fuller, George M. and Pfeiffer, Bernd and Kratz, Karl-Ludwig},
  title         = {The Extremely Metal-Poor, Neutron-Capture-Rich Star {CS 22892-052}: A Comprehensive Abundance Analysis},
  journal       = {The Astrophysical Journal},
  year          = {2003},
  volume        = {591},
  pages         = {936--953},
  doi           = {10.1086/375491},
  eprint        = {astro-ph/0303542},
  archiveprefix = {arXiv}
}

@article{ivans2006,
  author        = {Ivans, Inese I. and Simmerer, Jennifer and Sneden, Christopher and Lawler, James E. and Cowan, John J. and Gallino, Roberto and Bisterzo, Sara},
  title         = {Near-Ultraviolet Observations of {HD 221170}: New Insights into the Nature of {$r$}-Process-Rich Stars},
  journal       = {The Astrophysical Journal},
  year          = {2006},
  volume        = {645},
  pages         = {613--633},
  doi           = {10.1086/504069},
  eprint        = {astro-ph/0604180},
  archiveprefix = {arXiv}
}

@article{francois2007,
  author        = {Fran{\c{c}}ois, P. and Depagne, E. and Hill, V. and Spite, M. and Spite, F. and Plez, B. and Beers, T. C. and Andersen, J. and James, G. and Barbuy, B. and Cayrel, R. and Bonifacio, P. and Molaro, P. and Nordstr{\"o}m, B. and Primas, F.},
  title         = {First Stars {VIII}. Enrichment of the Neutron-Capture Elements in the Early Galaxy},
  journal       = {Astronomy \& Astrophysics},
  year          = {2007},
  volume        = {476},
  pages         = {935--950},
  doi           = {10.1051/0004-6361:20077706},
  eprint        = {0709.3454},
  archiveprefix = {arXiv}
}

@article{frebel2007,
  author        = {Frebel, Anna and Christlieb, Norbert and Norris, John E. and Thom, Christopher and Beers, Timothy C. and Rhee, Jaehyon},
  title         = {Discovery of {HE 1523-0901}, a Strongly {$r$}-Process-Enhanced Metal-Poor Star with Detected Uranium},
  journal       = {The Astrophysical Journal Letters},
  year          = {2007},
  volume        = {660},
  pages         = {L117--L120},
  doi           = {10.1086/518122},
  eprint        = {astro-ph/0703414},
  archiveprefix = {arXiv}
}

@article{hayek2009,
  author        = {Hayek, W. and Wiesendahl, U. and Christlieb, N. and Eriksson, K. and Korn, A. J. and Barklem, P. S. and Hill, V. and Beers, T. C. and Farouqi, K. and Pfeiffer, B. and Kratz, K.-L.},
  title         = {The {Hamburg/ESO R}-Process Enhanced Star Survey ({HERES}). {IV}. Detailed Abundance Analysis and Age Dating of the Strongly {$r$}-Process-Enhanced Stars {CS 29491-069} and {HE 1219-0312}},
  journal       = {Astronomy \& Astrophysics},
  year          = {2009},
  volume        = {504},
  pages         = {511--524},
  doi           = {10.1051/0004-6361/200811121},
  eprint        = {0910.0707},
  archiveprefix = {arXiv}
}

@article{siquiera-mello2013,
  author        = {Siqueira Mello, C. and Spite, M. and Barbuy, B. and Spite, F. and Caffau, E. and Hill, V. and Wanajo, S. and Primas, F. and Plez, B. and Cayrel, R. and Andersen, J. and Nordstr{\"o}m, B. and Sneden, C. and Beers, T. C. and Bonifacio, P. and Fran{\c{c}}ois, P. and Molaro, P.},
  title         = {First Stars {XVI}. {STIS/HST} Abundances of Heavy Elements in the Uranium-Rich Star {CS 31082-001}},
  journal       = {Astronomy \& Astrophysics},
  year          = {2013},
  volume        = {550},
  pages         = {A122},
  doi           = {10.1051/0004-6361/201219949},
  eprint        = {1212.0211},
  archiveprefix = {arXiv}
}

@article{mashonkina2014,
  author        = {Mashonkina, L. and Christlieb, N. and Eriksson, K.},
  title         = {The {Hamburg/ESO R}-Process Enhanced Star Survey ({HERES}). {X}. {HE 2252-4225}, One More {$r$}-Process-Enhanced and Actinide-Boost Halo Star},
  journal       = {Astronomy \& Astrophysics},
  year          = {2014},
  volume        = {569},
  pages         = {A43},
  doi           = {10.1051/0004-6361/201424017},
  eprint        = {1407.5379},
  archiveprefix = {arXiv}
}

@article{li2015,
  author        = {Li, Haining and Aoki, Wako and Honda, Satoshi and Zhao, Gang and Christlieb, Norbert and Suda, Takuma},
  title         = {Discovery of a Strongly {$r$}-Process-Enhanced Extremely Metal-Poor Star {LAMOST J110901.22+075441.8}},
  journal       = {Research in Astronomy and Astrophysics},
  year          = {2015},
  volume        = {15},
  pages         = {1264--1274},
  eprint        = {1506.05200},
  archiveprefix = {arXiv}
}

@article{placco2017,
  author        = {Placco, Vinicius M. and Holmbeck, Erika M. and Frebel, Anna and Beers, Timothy C. and Surman, Rebecca A. and Ji, Alexander P. and Ezzeddine, Rana and Points, Sean D. and Kaleida, Catherine C. and Hansen, Terese T. and Sakari, Charli M. and Casey, Andrew R.},
  title         = {{RAVE J203843.2-002333}: The First Highly {$r$}-Process-Enhanced Star Identified in the {RAVE} Survey},
  journal       = {The Astrophysical Journal},
  year          = {2017},
  volume        = {844},
  pages         = {18},
  doi           = {10.3847/1538-4357/aa78ef},
  eprint        = {1706.02934},
  archiveprefix = {arXiv}
}

@article{holmbeck2018,
  author        = {Holmbeck, Erika M. and Beers, Timothy C. and Roederer, Ian U. and Placco, Vinicius M. and Hansen, Terese T. and Sakari, Charli M. and Sneden, Christopher and Liu, Chao and Lee, Young Sun and Cowan, John J. and Frebel, Anna},
  title         = {The {R}-Process Alliance: {2MASS J09544277+5246414}, the Most Actinide-Enhanced {R-II} Star Known},
  journal       = {The Astrophysical Journal Letters},
  year          = {2018},
  volume        = {859},
  pages         = {L24},
  doi           = {10.3847/2041-8213/aac722},
  eprint        = {1805.11925},
  archiveprefix = {arXiv}
}

@article{ji2016,
  author        = {Ji, Alexander P. and Frebel, Anna and Simon, Joshua D. and Chiti, Anirudh},
  title         = {Complete Element Abundances of Nine Stars in the {$r$}-Process Galaxy {Reticulum II}},
  journal       = {The Astrophysical Journal},
  year          = {2016},
  volume        = {830},
  pages         = {93},
  doi           = {10.3847/0004-637X/830/2/93},
  eprint        = {1607.07447},
  archiveprefix = {arXiv}
}

@ARTICLE{He_2026,
       author = {{He}, Zhenyu and {Lin}, Yangming and {Luo}, Yudong and {Kajino}, Toshitaka and {Li}, Haining},
        title = "{Rare Earth Element Nucleosynthesis in Collapsars: Sensitivity to (n, {\ensuremath{\gamma}}) Reaction Rates of Unstable Isotopes during Secondary i- and s-processes}",
      journal = {\apj},
         year = 2026,
        month = apr,
       volume = {1000},
       number = {2},
          eid = {177},
        pages = {177},
          doi = {10.3847/1538-4357/ae4739},
       adsurl = {https://ui.adsabs.harvard.edu/abs/2026ApJ..1000..177H}
}

@ARTICLE{He_2024,
       author = {{He}, Zhenyu and {Kajino}, Toshitaka and {Kusakabe}, Motohiko and {Zhou}, Shan-Gui and {Koura}, Hiroyuki and {Chiba}, Satoshi and {Li}, Haining and {Lin}, Yangming},
        title = "{Possibility of Secondary i- and s-processes Following r-process in the Collapsar Jet}",
      journal = {\apjl},
         year = 2024,
        month = may,
       volume = {966},
       number = {2},
          eid = {L37},
        pages = {L37},
          doi = {10.3847/2041-8213/ad444c},
       adsurl = {https://ui.adsabs.harvard.edu/abs/2024ApJ...966L..37H}
}

@article{Goriely_2013,
    author = "Goriely, S. and Sida, J. -L. and Lema{\^\i}tre, J. -F. and Panebianco, S. and Dubray, N. and Hilaire, S. and Bauswein, A. and Janka, H. -Thomas",
    title = "{New fission fragment distributions and r-process origin of the rare-earth elements}",
    eprint = "1311.5897",
    archivePrefix = "arXiv",
    primaryClass = "astro-ph.SR",
    doi = "10.1103/PhysRevLett.111.242502",
    journal = "Phys. Rev. Lett.",
    volume = "111",
    number = "24",
    pages = "242502",
    year = "2013"
}

@article{Sasaki_2021,
    author = "Sasaki, Hirokazu and Yamazaki, Yuta and Kajino, Toshitaka and Kusakabe, Motohiko and Hayakawa, Takehito and Cheoun, Myung-Ki and Ko, Heamin and Mathews, Grant J.",
    title = "{Impact of Hypernova {\ensuremath{\nu}}p-process Nucleosynthesis on the Galactic Chemical Evolution of Mo and Ru}",
    eprint = "2106.01679",
    archivePrefix = "arXiv",
    primaryClass = "astro-ph.GA",
    doi = "10.3847/1538-4357/ac34f8",
    journal = "Astrophys. J.",
    volume = "924",
    number = "1",
    pages = "29",
    year = "2022"
}

@article{Sasaki_2023,
    author = "Sasaki, Hirokazu and Yamazaki, Yuta and Kajino, Toshitaka and Mathews, Grant J.",
    title = "{Effects of Hoyle state de-excitation on {\ensuremath{\nu}}p{\textendash}process nucleosynthesis and Galactic chemical evolution}",
    eprint = "2307.02785",
    archivePrefix = "arXiv",
    primaryClass = "astro-ph.HE",
    reportNumber = "LA-UR-23-27158",
    doi = "10.1016/j.physletb.2024.138581",
    journal = "Phys. Lett. B",
    volume = "851",
    pages = "138581",
    year = "2024"
}

@article{Sasaki_2017,
    author = "Sasaki, H. and Kajino, T. and Takiwaki, T. and Hayakawa, T. and Balantekin, A. B. and Pehlivan, Y.",
    title = "{Possible effects of collective neutrino oscillations in three-flavor multiangle simulations of supernova $\nu p$ processes}",
    eprint = "1707.09111",
    archivePrefix = "arXiv",
    primaryClass = "astro-ph.HE",
    doi = "10.1103/PhysRevD.96.043013",
    journal = "Phys. Rev. D",
    volume = "96",
    number = "4",
    pages = "043013",
    year = "2017"
}

@misc{ciaaw,
  author = {International Union of Pure and Applied Chemistry},
  title = {Commission on Isotopic and Atomic Weights},
  year = {2024},
  howpublished = {\url{https://ciaaw.org}},
}

@ARTICLE{hoyle_1960,
       author = {{Hoyle}, F. and {Fowler}, William A.},
        title = "{Nucleosynthesis in Supernovae.}",
      journal = {\apj},
         year = 1960,
        month = nov,
       volume = {132},
        pages = {565},
          doi = {10.1086/146963},
       adsurl = {https://ui.adsabs.harvard.edu/abs/1960ApJ...132..565H}
}

@BOOK{arnett_1996,
       author = {{Arnett}, David},
        title = "{Supernovae and Nucleosynthesis: An Investigation of the History of Matter from the Big Bang to the Present}",
         year = 1996,
       adsurl = {https://ui.adsabs.harvard.edu/abs/1996snih.book.....A}
}

@ARTICLE{ander_grevesse_1989,
       author = {{Anders}, E. and {Grevesse}, N.},
        title = "{Abundances of the elements: Meteoritic and solar}",
      journal = {\gca},
         year = 1989,
        month = jan,
       volume = {53},
       number = {1},
        pages = {197-214},
          doi = {10.1016/0016-7037(89)90286-X},
       adsurl = {https://ui.adsabs.harvard.edu/abs/1989GeCoA..53..197A}
}

@INCOLLECTION{grevesse_2011,
       author = {{Grevesse}, Nicolas and {Asplund}, Martin and {Sauval}, A. Jacques and {Scott}, Pat},
        title = "{The New Solar Composition and the Solar Metallicity}",
    booktitle = {The Sun, the Solar Wind, and the Heliosphere},
         year = 2011,
       editor = {{Miralles}, Mari Paz and {S{\'a}nchez Almeida}, Jorge},
       volume = {4},
        pages = {51},
       adsurl = {https://ui.adsabs.harvard.edu/abs/2011sswh.book...51G}
}

@ARTICLE{battino_2025,
       author = {{Battino}, U. and {Keegans}, J.~D. and {Allen}, M. and {R{\"o}pke}, F.~K. and {Herwig}, F. and {Best}, A. and {Hirschi}, R. and {Piersanti}, L. and {Straniero}, O. and {Sim}, S.~A. and {Travaglio}, C. and {Denissenkov}, P.~A.},
        title = "{Trans-Fe elements from type Ia supernovae: I. Heavy element nucleosynthesis during the formation of near-Chandrasekhar white dwarfs}",
      journal = {\aap},
         year = 2025,
        month = nov,
       volume = {703},
          eid = {A44},
        pages = {A44},
          doi = {10.1051/0004-6361/202555745},
archivePrefix = {arXiv},
       eprint = {2508.15079},
 primaryClass = {astro-ph.SR},
       adsurl = {https://ui.adsabs.harvard.edu/abs/2025A&A...703A..44B}
}

@ARTICLE{yao_2025,
       author = {{Yao}, Xingqun and {Luo}, Yudong and {Kajino}, Toshitaka and {Hayakawa}, Seiya and {Yamaguchi}, Hidetoshi and {Tang}, Xiaodong and {Gao}, Bingshui and {Liu}, Fulong},
        title = "{Implication of radioactive nuclear reaction $^{11}$C({\ensuremath{\alpha}}, p)$^{14}$N in Supernova {\ensuremath{\nu}}-process nucleosynthesis}",
      journal = {Chinese Physics C},
         year = 2025,
        month = aug,
       volume = {49},
       number = {8},
          eid = {084003},
        pages = {084003},
          doi = {10.1088/1674-1137/add680},
       adsurl = {https://ui.adsabs.harvard.edu/abs/2025ChPhC..49h4003Y}
}

@ARTICLE{leung_2024,
       author = {{Leung}, Shing-Chi and {Nomoto}, Ken'ichi},
        title = "{Hydrodynamics and Nucleosynthesis of Jet-driven Supernovae. II. Comparisons with Abundances of Extremely Metal-poor Galaxies and Constraints on Supernova Progenitors}",
      journal = {\apj},
         year = 2024,
        month = oct,
       volume = {974},
       number = {2},
          eid = {310},
        pages = {310},
          doi = {10.3847/1538-4357/ad6ddb},
archivePrefix = {arXiv},
       eprint = {2312.17226},
 primaryClass = {astro-ph.HE},
       adsurl = {https://ui.adsabs.harvard.edu/abs/2024ApJ...974..310L}
}

@ARTICLE{jin_2024,
       author = {{Jin}, Shilun and {Soker}, Noam},
        title = "{Robust r-process Nucleosynthesis beyond Lanthanides in the Common Envelop Jet Supernovae}",
      journal = {\apj},
         year = 2024,
        month = aug,
       volume = {971},
       number = {2},
          eid = {189},
        pages = {189},
          doi = {10.3847/1538-4357/ad5f8e},
archivePrefix = {arXiv},
       eprint = {2310.08907},
 primaryClass = {astro-ph.HE},
       adsurl = {https://ui.adsabs.harvard.edu/abs/2024ApJ...971..189J}
}

@ARTICLE{kawashimo_2024,
       author = {{Kawashimo}, Hiroki and {Sawada}, Ryo and {Suwa}, Yudai and {Moriya}, Takashi J. and {Tanikawa}, Ataru and {Tominaga}, Nozomu},
        title = "{Impacts of the $^{12}$C({\ensuremath{\alpha}}, {\ensuremath{\gamma}})$^{16}$O reaction rate on $^{56}$Ni nucleosynthesis in pair-instability supernovae}",
      journal = {\mnras},
         year = 2024,
        month = jun,
       volume = {531},
       number = {2},
        pages = {2786-2801},
          doi = {10.1093/mnras/stae1280},
archivePrefix = {arXiv},
       eprint = {2306.01682},
 primaryClass = {astro-ph.SR},
       adsurl = {https://ui.adsabs.harvard.edu/abs/2024MNRAS.531.2786K}
}

@ARTICLE{balantekin_2024,
       author = {{Balantekin}, A. Baha and {Cervia}, Michael J. and {Patwardhan}, Amol V. and {Surman}, Rebecca and {Wang}, Xilu},
        title = "{Collective Neutrino Oscillations and Heavy-element Nucleosynthesis in Supernovae: Exploring Potential Effects of Many-body Neutrino Correlations}",
      journal = {\apj},
         year = 2024,
        month = jun,
       volume = {967},
       number = {2},
          eid = {146},
        pages = {146},
          doi = {10.3847/1538-4357/ad393d},
archivePrefix = {arXiv},
       eprint = {2311.02562},
 primaryClass = {astro-ph.HE},
       adsurl = {https://ui.adsabs.harvard.edu/abs/2024ApJ...967..146B}
}

@ARTICLE{roberti_2024b,
       author = {{Roberti}, Lorenzo and {Limongi}, Marco and {Chieffi}, Alessandro},
        title = "{Presupernova Evolution and Explosive Nucleosynthesis of Rotating Massive Stars. II. The Supersolar Models at [Fe/H] = 0.3}",
      journal = {\apjs},
         year = 2024,
        month = may,
       volume = {272},
       number = {1},
          eid = {15},
        pages = {15},
          doi = {10.3847/1538-4365/ad391d},
archivePrefix = {arXiv},
       eprint = {2403.18945},
 primaryClass = {astro-ph.SR},
       adsurl = {https://ui.adsabs.harvard.edu/abs/2024ApJS..272...15R}
}

@ARTICLE{reichert_2024,
       author = {{Reichert}, Moritz and {Bugli}, Matteo and {Guilet}, J{\'e}r{\^o}me and {Obergaulinger}, Martin and {Aloy}, Miguel {\'A}ngel and {Arcones}, Almudena},
        title = "{Nucleosynthesis in magnetorotational supernovae: impact of the magnetic field configuration}",
      journal = {\mnras},
         year = 2024,
        month = apr,
       volume = {529},
       number = {4},
        pages = {3197-3209},
          doi = {10.1093/mnras/stae561},
archivePrefix = {arXiv},
       eprint = {2401.14402},
 primaryClass = {astro-ph.HE},
       adsurl = {https://ui.adsabs.harvard.edu/abs/2024MNRAS.529.3197R}
}

@ARTICLE{roberti_2024a,
       author = {{Roberti}, Lorenzo and {Limongi}, Marco and {Chieffi}, Alessandro},
        title = "{Zero and Extremely Low-metallicity Rotating Massive Stars: Evolution, Explosion, and Nucleosynthesis Up to the Heaviest Nuclei}",
      journal = {\apjs},
         year = 2024,
        month = feb,
       volume = {270},
       number = {2},
          eid = {28},
        pages = {28},
          doi = {10.3847/1538-4365/ad1686},
archivePrefix = {arXiv},
       eprint = {2312.02942},
 primaryClass = {astro-ph.SR},
       adsurl = {https://ui.adsabs.harvard.edu/abs/2024ApJS..270...28R}
}

@ARTICLE{boccioli_2024,
       author = {{Boccioli}, Luca and {Roberti}, Lorenzo},
        title = "{The Physics of Core-Collapse Supernovae: Explosion Mechanism and Explosive Nucleosynthesis}",
      journal = {Universe},
         year = 2024,
        month = mar,
       volume = {10},
       number = {3},
          eid = {148},
        pages = {148},
          doi = {10.3390/universe10030148},
archivePrefix = {arXiv},
       eprint = {2403.12942},
 primaryClass = {astro-ph.SR},
       adsurl = {https://ui.adsabs.harvard.edu/abs/2024Univ...10..148B}
}

@ARTICLE{reeves_2023,
       author = {{Reeves}, Zachary and {Schlaufman}, Kevin C. and {Reggiani}, Henrique},
        title = "{The Dependence of Iron-rich Metal-poor Star Occurrence on Galactic Environment Supports an Origin in Thermonuclear Supernova Nucleosynthesis}",
      journal = {\aj},
         year = 2023,
        month = sep,
       volume = {166},
       number = {3},
          eid = {127},
        pages = {127},
          doi = {10.3847/1538-3881/ace68d},
archivePrefix = {arXiv},
       eprint = {2307.05669},
 primaryClass = {astro-ph.GA},
       adsurl = {https://ui.adsabs.harvard.edu/abs/2023AJ....166..127R}
}

@ARTICLE{navo_2023,
       author = {{Nav{\'o}}, Gerard and {Reichert}, Moritz and {Obergaulinger}, Martin and {Arcones}, Almudena},
        title = "{Core-collapse Supernova Simulations with Reduced Nucleosynthesis Networks}",
      journal = {\apj},
         year = 2023,
        month = jul,
       volume = {951},
       number = {2},
          eid = {112},
        pages = {112},
          doi = {10.3847/1538-4357/acd640},
archivePrefix = {arXiv},
       eprint = {2210.11848},
 primaryClass = {astro-ph.HE},
       adsurl = {https://ui.adsabs.harvard.edu/abs/2023ApJ...951..112N}
}

@ARTICLE{leung_2023,
       author = {{Leung}, Shing-Chi and {Nomoto}, Ken'ichi and {Suzuki}, Tomoharu},
        title = "{Hydrodynamics and Nucleosynthesis of Jet-driven Supernovae. I. Parameter Study of the Dependence on Jet Energetics}",
      journal = {\apj},
         year = 2023,
        month = may,
       volume = {948},
       number = {2},
          eid = {80},
        pages = {80},
          doi = {10.3847/1538-4357/acbdf5},
archivePrefix = {arXiv},
       eprint = {2304.14935},
 primaryClass = {astro-ph.HE},
       adsurl = {https://ui.adsabs.harvard.edu/abs/2023ApJ...948...80L}
}

@ARTICLE{fujimoto_2023,
       author = {{Fujimoto}, Shin-ichiro and {Nagakura}, Hiroki},
        title = "{Explosive nucleosynthesis with fast neutrino-flavour conversion in core-collapse supernovae}",
      journal = {\mnras},
         year = 2023,
        month = feb,
       volume = {519},
       number = {2},
        pages = {2623-2629},
          doi = {10.1093/mnras/stac3763},
archivePrefix = {arXiv},
       eprint = {2210.02106},
 primaryClass = {astro-ph.HE},
       adsurl = {https://ui.adsabs.harvard.edu/abs/2023MNRAS.519.2623F}
}

@ARTICLE{reichert_2023b,
       author = {{Reichert}, M. and {Obergaulinger}, M. and {Aloy}, M. {\'A}. and {Gabler}, M. and {Arcones}, A. and {Thielemann}, F.~K.},
        title = "{Magnetorotational supernovae: a nucleosynthetic analysis of sophisticated 3D models}",
      journal = {\mnras},
         year = 2023,
        month = jan,
       volume = {518},
       number = {1},
        pages = {1557-1583},
          doi = {10.1093/mnras/stac3185},
archivePrefix = {arXiv},
       eprint = {2206.11914},
 primaryClass = {astro-ph.HE},
       adsurl = {https://ui.adsabs.harvard.edu/abs/2023MNRAS.518.1557R}
}

@ARTICLE{jayatissa_2022,
       author = {{Jayatissa}, H. and {Avila}, M.~L. and {Rehm}, K.~E. and {Talwar}, R. and {Mohr}, P. and {Auranen}, K. and {Chen}, J. and {Gorelov}, D.~A. and {Hoffman}, C.~R. and {Jiang}, C.~L. and {Kay}, B.~P. and {Kuvin}, S.~A. and {Santiago-Gonzalez}, D.},
        title = "{First direct measurement of the $^{13}$N({\ensuremath{\alpha}} ,p )$^{16}$O reaction relevant for core-collapse supernovae nucleosynthesis}",
      journal = {\prc},
         year = 2022,
        month = apr,
       volume = {105},
       number = {4},
          eid = {L042802},
        pages = {L042802},
          doi = {10.1103/PhysRevC.105.L042802},
archivePrefix = {arXiv},
       eprint = {2201.10016},
 primaryClass = {nucl-ex},
       adsurl = {https://ui.adsabs.harvard.edu/abs/2022PhRvC.105d2802J}
}

@ARTICLE{higgins_2024,
       author = {{Higgins}, Erin R. and {Vink}, Jorick S. and {Hirschi}, Raphael and {Laird}, Alison M. and {Sander}, Andreas A.~C.},
        title = "{New Wolf-Rayet wind yields and nucleosynthesis of Helium stars}",
      journal = {\mnras},
         year = 2024,
        month = sep,
       volume = {533},
       number = {1},
        pages = {1095-1110},
          doi = {10.1093/mnras/stae1853},
archivePrefix = {arXiv},
       eprint = {2407.07983},
 primaryClass = {astro-ph.SR},
       adsurl = {https://ui.adsabs.harvard.edu/abs/2024MNRAS.533.1095H}
}

@ARTICLE{nomoto_1986,
       author = {{Nomoto}, Kenichi and {Hashimoto}, Masa-Aki},
        title = "{Late stages of massive star evolution and nucleosynthesis.}",
      journal = {Progress in Particle and Nuclear Physics},
         year = 1986,
        month = jan,
       volume = {17},
        pages = {267-285},
          doi = {10.1016/0146-6410(86)90021-9},
       adsurl = {https://ui.adsabs.harvard.edu/abs/1986PrPNP..17..267N}
}

@ARTICLE{weaver_1993,
       author = {{Weaver}, Thomas A. and {Woosley}, S.~E.},
        title = "{Nucleosynthesis in massive stars and the $^{12}$C({\ensuremath{\alpha}}, {\ensuremath{\gamma}})$^{16}$O reaction rate}",
      journal = {\physrep},
         year = 1993,
        month = may,
       volume = {227},
       number = {1-5},
        pages = {65-96},
          doi = {10.1016/0370-1573(93)90058-L},
       adsurl = {https://ui.adsabs.harvard.edu/abs/1993PhR...227...65W}
}

@ARTICLE{limongi_2000,
       author = {{Limongi}, Marco and {Straniero}, Oscar and {Chieffi}, Alessandro},
        title = "{Massive Stars in the Range 13-25 M$_{solar}$: Evolution and Nucleosynthesis. II. The Solar Metallicity Models}",
      journal = {\apjs},
         year = 2000,
        month = aug,
       volume = {129},
       number = {2},
        pages = {625-664},
          doi = {10.1086/313424},
archivePrefix = {arXiv},
       eprint = {astro-ph/0003401},
 primaryClass = {astro-ph},
       adsurl = {https://ui.adsabs.harvard.edu/abs/2000ApJS..129..625L}
}

@ARTICLE{couch_2013,
       author = {{Couch}, Sean M.},
        title = "{On the Impact of Three Dimensions in Simulations of Neutrino-driven Core-collapse Supernova Explosions}",
      journal = {\apj},
         year = 2013,
        month = sep,
       volume = {775},
       number = {1},
          eid = {35},
        pages = {35},
          doi = {10.1088/0004-637X/775/1/35},
archivePrefix = {arXiv},
       eprint = {1212.0010},
 primaryClass = {astro-ph.HE},
       adsurl = {https://ui.adsabs.harvard.edu/abs/2013ApJ...775...35C}
}

@ARTICLE{dolence_2013,
       author = {{Dolence}, Joshua C. and {Burrows}, Adam and {Murphy}, Jeremiah W. and {Nordhaus}, Jason},
        title = "{Dimensional Dependence of the Hydrodynamics of Core-collapse Supernovae}",
      journal = {\apj},
         year = 2013,
        month = mar,
       volume = {765},
       number = {2},
          eid = {110},
        pages = {110},
          doi = {10.1088/0004-637X/765/2/110},
archivePrefix = {arXiv},
       eprint = {1210.5241},
 primaryClass = {astro-ph.SR},
       adsurl = {https://ui.adsabs.harvard.edu/abs/2013ApJ...765..110D}
}

@ARTICLE{idsa,
       author = {{Liebend{\"o}rfer}, M. and {Whitehouse}, S.~C. and {Fischer}, T.},
        title = "{The Isotropic Diffusion Source Approximation for Supernova Neutrino Transport}",
      journal = {\apj},
         year = 2009,
        month = jun,
       volume = {698},
       number = {2},
        pages = {1174-1190},
          doi = {10.1088/0004-637X/698/2/1174},
archivePrefix = {arXiv},
       eprint = {0711.2929},
 primaryClass = {astro-ph},
       adsurl = {https://ui.adsabs.harvard.edu/abs/2009ApJ...698.1174L}
}

@ARTICLE{couch_2020,
       author = {{Couch}, Sean M. and {Warren}, MacKenzie L. and {O'Connor}, Evan P.},
        title = "{Simulating Turbulence-aided Neutrino-driven Core-collapse Supernova Explosions in One Dimension}",
      journal = {\apj},
         year = 2020,
        month = feb,
       volume = {890},
       number = {2},
          eid = {127},
        pages = {127},
          doi = {10.3847/1538-4357/ab609e},
archivePrefix = {arXiv},
       eprint = {1902.01340},
 primaryClass = {astro-ph.HE},
       adsurl = {https://ui.adsabs.harvard.edu/abs/2020ApJ...890..127C}
}

@ARTICLE{sieverding_2023b,
       author = {{Sieverding}, Andre and {Kresse}, Daniel and {Janka}, Hans-Thomas},
        title = "{Production of $^{44}$Ti and Iron-group Nuclei in the Ejecta of 3D Neutrino-driven Supernovae}",
      journal = {\apjl},
         year = 2023,
        month = nov,
       volume = {957},
       number = {2},
          eid = {L25},
        pages = {L25},
          doi = {10.3847/2041-8213/ad045b},
archivePrefix = {arXiv},
       eprint = {2308.09659},
 primaryClass = {astro-ph.HE},
       adsurl = {https://ui.adsabs.harvard.edu/abs/2023ApJ...957L..25S}
}

@ARTICLE{jin_2020,
       author = {{Jin}, Shilun and {Roberts}, Luke F. and {Austin}, Sam M. and {Schatz}, Hendrik},
        title = "{Enhanced triple-{\ensuremath{\alpha}} reaction reduces proton-rich nucleosynthesis in supernovae}",
      journal = {\nat},
         year = 2020,
        month = dec,
       volume = {588},
       number = {7836},
        pages = {57-60},
          doi = {10.1038/s41586-020-2948-7},
       adsurl = {https://ui.adsabs.harvard.edu/abs/2020Natur.588...57J}
}

@ARTICLE{sieverding_2020,
       author = {{Sieverding}, A. and {M{\"u}ller}, B. and {Qian}, Y.-Z.},
        title = "{Nucleosynthesis of an 11.8 M$_{\odot}$ Supernova with 3D Simulation of the Inner Ejecta: Overall Yields and Implications for Short-lived Radionuclides in the Early Solar System}",
      journal = {\apj},
         year = 2020,
        month = dec,
       volume = {904},
       number = {2},
          eid = {163},
        pages = {163},
          doi = {10.3847/1538-4357/abc61b},
archivePrefix = {arXiv},
       eprint = {2008.12831},
 primaryClass = {astro-ph.HE},
       adsurl = {https://ui.adsabs.harvard.edu/abs/2020ApJ...904..163S}
}

@ARTICLE{subedi_2020,
       author = {{Subedi}, Shiv K. and {Meisel}, Zach and {Merz}, Grant},
        title = "{Sensitivity of $^{44}$Ti and $^{56}$Ni Production in Core-collapse Supernova Shock-driven Nucleosynthesis to Nuclear Reaction Rate Variations}",
      journal = {\apj},
         year = 2020,
        month = jul,
       volume = {898},
       number = {1},
          eid = {5},
        pages = {5},
          doi = {10.3847/1538-4357/ab9745},
archivePrefix = {arXiv},
       eprint = {2005.14702},
 primaryClass = {astro-ph.HE},
       adsurl = {https://ui.adsabs.harvard.edu/abs/2020ApJ...898....5S}
}

@ARTICLE{sato_2020,
       author = {{Sato}, Toshiki and {Bravo}, Eduardo and {Badenes}, Carles and {Hughes}, John P. and {Williams}, Brian J. and {Yamaguchi}, Hiroya},
        title = "{A Nucleosynthetic Origin for the Southwestern Fe-rich Structure in Kepler's Supernova Remnant}",
      journal = {\apj},
         year = 2020,
        month = feb,
       volume = {890},
       number = {2},
          eid = {104},
        pages = {104},
          doi = {10.3847/1538-4357/ab6aa2},
archivePrefix = {arXiv},
       eprint = {2001.02662},
 primaryClass = {astro-ph.HE},
       adsurl = {https://ui.adsabs.harvard.edu/abs/2020ApJ...890..104S}
}

@ARTICLE{tominaga_2007,
       author = {{Tominaga}, Nozomu and {Umeda}, Hideyuki and {Nomoto}, Ken'ichi},
        title = "{Supernova Nucleosynthesis in Population III 13-50 M$_{solar}$ Stars and Abundance Patterns of Extremely Metal-poor Stars}",
      journal = {\apj},
         year = 2007,
        month = may,
       volume = {660},
       number = {1},
        pages = {516-540},
          doi = {10.1086/513063},
archivePrefix = {arXiv},
       eprint = {astro-ph/0701381},
 primaryClass = {astro-ph},
       adsurl = {https://ui.adsabs.harvard.edu/abs/2007ApJ...660..516T}
}

@ARTICLE{yoshida_2008,
       author = {{Yoshida}, Takashi and {Umeda}, Hideyuki and {Nomoto}, Ken'ichi},
        title = "{{\ensuremath{\nu}}-Process Nucleosynthesis in Population III Core-Collapse Supernovae}",
      journal = {\apj},
         year = 2008,
        month = jan,
       volume = {672},
       number = {2},
        pages = {1043-1053},
          doi = {10.1086/523833},
archivePrefix = {arXiv},
       eprint = {0710.0251},
 primaryClass = {astro-ph},
       adsurl = {https://ui.adsabs.harvard.edu/abs/2008ApJ...672.1043Y}
}

@ARTICLE{imasheva_2023,
       author = {{Imasheva}, Liliya and {Janka}, Hans-Thomas and {Weiss}, Achim},
        title = "{Parametrizations of thermal bomb explosions for core-collapse supernovae and $^{56}$Ni production}",
      journal = {\mnras},
         year = 2023,
        month = jan,
       volume = {518},
       number = {2},
        pages = {1818-1839},
          doi = {10.1093/mnras/stac3239},
archivePrefix = {arXiv},
       eprint = {2209.10989},
 primaryClass = {astro-ph.HE},
       adsurl = {https://ui.adsabs.harvard.edu/abs/2023MNRAS.518.1818I}
}

@ARTICLE{tominaga_2009,
       author = {{Tominaga}, Nozomu},
        title = "{Aspherical Properties of Hydrodynamics and Nucleosynthesis in Jet-Induced Supernovae}",
      journal = {\apj},
         year = 2009,
        month = jan,
       volume = {690},
       number = {1},
        pages = {526-536},
          doi = {10.1088/0004-637X/690/1/526},
archivePrefix = {arXiv},
       eprint = {0711.4815},
 primaryClass = {astro-ph},
       adsurl = {https://ui.adsabs.harvard.edu/abs/2009ApJ...690..526T}
}

@ARTICLE{farmer_2023,
       author = {{Farmer}, R. and {Laplace}, E. and {Ma}, Jing-ze and {de Mink}, S.~E. and {Justham}, S.},
        title = "{Nucleosynthesis of Binary-stripped Stars}",
      journal = {\apj},
         year = 2023,
        month = may,
       volume = {948},
       number = {2},
          eid = {111},
        pages = {111},
          doi = {10.3847/1538-4357/acc315},
archivePrefix = {arXiv},
       eprint = {2303.04520},
 primaryClass = {astro-ph.SR},
       adsurl = {https://ui.adsabs.harvard.edu/abs/2023ApJ...948..111F}
}

@ARTICLE{mosta_2018,
       author = {{M{\"o}sta}, Philipp and {Roberts}, Luke F. and {Halevi}, Goni and {Ott}, Christian D. and {Lippuner}, Jonas and {Haas}, Roland and {Schnetter}, Erik},
        title = "{r-process Nucleosynthesis from Three-dimensional Magnetorotational Core-collapse Supernovae}",
      journal = {\apj},
         year = 2018,
        month = sep,
       volume = {864},
       number = {2},
          eid = {171},
        pages = {171},
          doi = {10.3847/1538-4357/aad6ec},
archivePrefix = {arXiv},
       eprint = {1712.09370},
 primaryClass = {astro-ph.HE},
       adsurl = {https://ui.adsabs.harvard.edu/abs/2018ApJ...864..171M}
}

@ARTICLE{gyurky_2021,
       author = {{Gy{\"u}rky}, Gy and {Hal{\'a}sz}, Z. and {Kiss}, G.~G. and {Sz{\"u}cs}, T. and {Husz{\'a}nk}, R. and {T{\"o}r{\"o}k}, Zs and {F{\"u}l{\"o}p}, Zs and {Rauscher}, T. and {Travaglio}, C.},
        title = "{Measurement of the $^{91}$Zr(p,{\ensuremath{\gamma}})$^{92m}$Nb cross section motivated by type Ia supernova nucleosynthesis}",
      journal = {Journal of Physics G Nuclear Physics},
         year = 2021,
        month = oct,
       volume = {48},
       number = {10},
          eid = {105202},
        pages = {105202},
          doi = {10.1088/1361-6471/ac2132},
archivePrefix = {arXiv},
       eprint = {2108.10006},
 primaryClass = {nucl-ex},
       adsurl = {https://ui.adsabs.harvard.edu/abs/2021JPhG...48j5202G}
}

@ARTICLE{asplund_2009,
       author = {{Asplund}, Martin and {Grevesse}, Nicolas and {Sauval}, A. Jacques and {Scott}, Pat},
        title = "{The Chemical Composition of the Sun}",
      journal = {\araa},
         year = 2009,
        month = sep,
       volume = {47},
       number = {1},
        pages = {481-522},
          doi = {10.1146/annurev.astro.46.060407.145222},
archivePrefix = {arXiv},
       eprint = {0909.0948},
 primaryClass = {astro-ph.SR},
       adsurl = {https://ui.adsabs.harvard.edu/abs/2009ARA&A..47..481A}
}

@ARTICLE{B2FH_1957,
       author = {{Burbidge}, E. Margaret and {Burbidge}, G.~R. and {Fowler}, William A. and {Hoyle}, F.},
        title = "{Synthesis of the Elements in Stars}",
      journal = {Reviews of Modern Physics},
         year = 1957,
        month = oct,
       volume = {29},
       number = {4},
        pages = {547-650},
          doi = {10.1103/RevModPhys.29.547},
       adsurl = {https://ui.adsabs.harvard.edu/abs/1957RvMP...29..547B}
}

@BOOK{clayton_1968,
       author = {{Clayton}, Donald D.},
        title = "{Principles of stellar evolution and nucleosynthesis}",
         year = 1968,
       adsurl = {https://ui.adsabs.harvard.edu/abs/1968psen.book.....C}
}

@BOOK{peebles_1993,
       author = {{Peebles}, P.~J.~E.},
        title = "{Principles of Physical Cosmology}",
         year = 1993,
          doi = {10.1515/9780691206721},
       adsurl = {https://ui.adsabs.harvard.edu/abs/1993ppc..book.....P}
}

@ARTICLE{woosley_2007,
       author = {{Woosley}, S.~E. and {Blinnikov}, S. and {Heger}, Alexander},
        title = "{Pulsational pair instability as an explanation for the most luminous supernovae}",
      journal = {\nat},
         year = 2007,
        month = nov,
       volume = {450},
       number = {7168},
        pages = {390-392},
          doi = {10.1038/nature06333},
archivePrefix = {arXiv},
       eprint = {0710.3314},
 primaryClass = {astro-ph},
       adsurl = {https://ui.adsabs.harvard.edu/abs/2007Natur.450..390W}
}

@ARTICLE{push_1,
       author = {{Perego}, A. and {Hempel}, M. and {Fr{\"o}hlich}, C. and {Ebinger}, K. and {Eichler}, M. and {Casanova}, J. and {Liebend{\"o}rfer}, M. and {Thielemann}, F.-K.},
        title = "{PUSHing Core-collapse Supernovae to Explosions in Spherical Symmetry I: the Model and the Case of SN 1987A}",
      journal = {\apj},
         year = 2015,
        month = jun,
       volume = {806},
       number = {2},
          eid = {275},
        pages = {275},
          doi = {10.1088/0004-637X/806/2/275},
archivePrefix = {arXiv},
       eprint = {1501.02845},
 primaryClass = {astro-ph.SR},
       adsurl = {https://ui.adsabs.harvard.edu/abs/2015ApJ...806..275P}
}

@ARTICLE{push_3,
       author = {{Curtis}, Sanjana and {Ebinger}, Kevin and {Fr{\"o}hlich}, Carla and {Hempel}, Matthias and {Perego}, Albino and {Liebend{\"o}rfer}, Matthias and {Thielemann}, Friedrich-Karl},
        title = "{PUSHing Core-collapse Supernovae to Explosions in Spherical Symmetry. III. Nucleosynthesis Yields}",
      journal = {\apj},
         year = 2019,
        month = jan,
       volume = {870},
       number = {1},
          eid = {2},
        pages = {2},
          doi = {10.3847/1538-4357/aae7d2},
archivePrefix = {arXiv},
       eprint = {1805.00498},
 primaryClass = {astro-ph.SR},
       adsurl = {https://ui.adsabs.harvard.edu/abs/2019ApJ...870....2C}
}

@ARTICLE{push_4,
       author = {{Ebinger}, Kevin and {Curtis}, Sanjana and {Ghosh}, Somdutta and {Fr{\"o}hlich}, Carla and {Hempel}, Matthias and {Perego}, Albino and {Liebend{\"o}rfer}, Matthias and {Thielemann}, Friedrich-Karl},
        title = "{PUSHing Core-collapse Supernovae to Explosions in Spherical Symmetry. IV. Explodability, Remnant Properties, and Nucleosynthesis Yields of Low-metallicity Stars}",
      journal = {\apj},
         year = 2020,
        month = jan,
       volume = {888},
       number = {2},
          eid = {91},
        pages = {91},
          doi = {10.3847/1538-4357/ab5dcb},
archivePrefix = {arXiv},
       eprint = {1910.08958},
 primaryClass = {astro-ph.SR},
       adsurl = {https://ui.adsabs.harvard.edu/abs/2020ApJ...888...91E}
}

@ARTICLE{push_5,
       author = {{Ghosh}, Somdutta and {Wolfe}, Noah and {Fr{\"o}hlich}, Carla},
        title = "{PUSHing Core-collapse Supernovae to Explosions in Spherical Symmetry. V. Equation of State Dependency of Explosion Properties, Nucleosynthesis Yields, and Compact Remnants}",
      journal = {\apj},
         year = 2022,
        month = apr,
       volume = {929},
       number = {1},
          eid = {43},
        pages = {43},
          doi = {10.3847/1538-4357/ac4d20},
archivePrefix = {arXiv},
       eprint = {2107.13016},
 primaryClass = {astro-ph.HE},
       adsurl = {https://ui.adsabs.harvard.edu/abs/2022ApJ...929...43G}
}

@ARTICLE{harris_2017,
       author = {{Harris}, J. Austin and {Hix}, W. Raphael and {Chertkow}, Merek A. and {Lee}, C.~T. and {Lentz}, Eric J. and {Messer}, O.~E. Bronson},
        title = "{Implications for Post-processing Nucleosynthesis of Core-collapse Supernova Models with Lagrangian Particles}",
      journal = {\apj},
         year = 2017,
        month = jul,
       volume = {843},
       number = {1},
          eid = {2},
        pages = {2},
          doi = {10.3847/1538-4357/aa76de},
archivePrefix = {arXiv},
       eprint = {1701.08876},
 primaryClass = {astro-ph.SR},
       adsurl = {https://ui.adsabs.harvard.edu/abs/2017ApJ...843....2H}
}

@ARTICLE{burrows_vartanyan_2021,
       author = {{Burrows}, A. and {Vartanyan}, D.},
        title = "{Core-collapse supernova explosion theory}",
      journal = {\nat},
         year = 2021,
        month = jan,
       volume = {589},
       number = {7840},
        pages = {29-39},
          doi = {10.1038/s41586-020-03059-w},
archivePrefix = {arXiv},
       eprint = {2009.14157},
 primaryClass = {astro-ph.SR},
       adsurl = {https://ui.adsabs.harvard.edu/abs/2021Natur.589...29B}
}

@ARTICLE{wang_2024a,
       author = {{Wang}, Tianshu and {Burrows}, Adam},
        title = "{Nucleosynthetic Analysis of Three-dimensional Core-collapse Supernova Simulations}",
      journal = {\apj},
         year = 2024,
        month = feb,
       volume = {962},
       number = {1},
          eid = {71},
        pages = {71},
          doi = {10.3847/1538-4357/ad12b8},
archivePrefix = {arXiv},
       eprint = {2311.03446},
 primaryClass = {astro-ph.HE},
       adsurl = {https://ui.adsabs.harvard.edu/abs/2024ApJ...962...71W}
}

@ARTICLE{wang_2024b,
       author = {{Wang}, Tianshu and {Burrows}, Adam},
        title = "{Insights into the Production of $^{44}$Ti and Nickel Isotopes in Core-collapse Supernovae}",
      journal = {\apj},
         year = 2024,
        month = oct,
       volume = {974},
       number = {1},
          eid = {39},
        pages = {39},
          doi = {10.3847/1538-4357/ad6983},
archivePrefix = {arXiv},
       eprint = {2406.13746},
 primaryClass = {astro-ph.HE},
       adsurl = {https://ui.adsabs.harvard.edu/abs/2024ApJ...974...39W}
}

@ARTICLE{wanajo_2018,
       author = {{Wanajo}, Shinya and {M{\"u}ller}, Bernhard and {Janka}, Hans-Thomas and {Heger}, Alexander},
        title = "{Nucleosynthesis in the Innermost Ejecta of Neutrino-driven Supernova Explosions in Two Dimensions}",
      journal = {\apj},
         year = 2018,
        month = jan,
       volume = {852},
       number = {1},
          eid = {40},
        pages = {40},
          doi = {10.3847/1538-4357/aa9d97},
archivePrefix = {arXiv},
       eprint = {1701.06786},
 primaryClass = {astro-ph.SR},
       adsurl = {https://ui.adsabs.harvard.edu/abs/2018ApJ...852...40W}
}

@ARTICLE{wanajo_2006,
       author = {{Wanajo}, Shinya and {Nomoto}, Ken'ichi and {Iwamoto}, Nobuyuki and {Ishimaru}, Yuhri and {Beers}, Timothy C.},
        title = "{Enrichment of Very Metal Poor Stars with Both r-Process and s-Process Elements from 8-10 M$_{solar}$ Stars}",
      journal = {\apj},
         year = 2006,
        month = jan,
       volume = {636},
       number = {2},
        pages = {842-847},
          doi = {10.1086/498293},
archivePrefix = {arXiv},
       eprint = {astro-ph/0509788},
 primaryClass = {astro-ph},
       adsurl = {https://ui.adsabs.harvard.edu/abs/2006ApJ...636..842W}
}

@ARTICLE{frohlich_2006b,
       author = {{Fr{\"o}hlich}, C. and {Hix}, W.~R. and {Mart{\'\i}nez-Pinedo}, G. and {Liebend{\"o}rfer}, M. and {Thielemann}, F.-K. and {Bravo}, E. and {Langanke}, K. and {Zinner}, N.~T.},
        title = "{Nucleosynthesis in neutrino-driven supernovae}",
      journal = {\nar},
         year = 2006,
        month = oct,
       volume = {50},
       number = {7-8},
        pages = {496-499},
          doi = {10.1016/j.newar.2006.06.003},
archivePrefix = {arXiv},
       eprint = {astro-ph/0511584},
 primaryClass = {astro-ph},
       adsurl = {https://ui.adsabs.harvard.edu/abs/2006NewAR..50..496F}
}

@ARTICLE{frohlich_2006a,
       author = {{Fr{\"o}hlich}, C. and {Hauser}, P. and {Liebend{\"o}rfer}, M. and {Mart{\'\i}nez-Pinedo}, G. and {Thielemann}, F.-K. and {Bravo}, E. and {Zinner}, N.~T. and {Hix}, W.~R. and {Langanke}, K. and {Mezzacappa}, A. and {Nomoto}, K.},
        title = "{Composition of the Innermost Core-Collapse Supernova Ejecta}",
      journal = {\apj},
         year = 2006,
        month = jan,
       volume = {637},
       number = {1},
        pages = {415-426},
          doi = {10.1086/498224},
archivePrefix = {arXiv},
       eprint = {astro-ph/0410208},
 primaryClass = {astro-ph},
       adsurl = {https://ui.adsabs.harvard.edu/abs/2006ApJ...637..415F}
}

@ARTICLE{frohlich_2006c,
       author = {{Fr{\"o}hlich}, C. and {Mart{\'\i}nez-Pinedo}, G. and {Liebend{\"o}rfer}, M. and {Thielemann}, F.-K. and {Bravo}, E. and {Hix}, W.~R. and {Langanke}, K. and {Zinner}, N.~T.},
        title = "{Neutrino-Induced Nucleosynthesis of A$>$64 Nuclei: The {\ensuremath{\nu}}p Process}",
      journal = {\prl},
         year = 2006,
        month = apr,
       volume = {96},
       number = {14},
          eid = {142502},
        pages = {142502},
          doi = {10.1103/PhysRevLett.96.142502},
archivePrefix = {arXiv},
       eprint = {astro-ph/0511376},
 primaryClass = {astro-ph},
       adsurl = {https://ui.adsabs.harvard.edu/abs/2006PhRvL..96n2502F}
}

@ARTICLE{thielemann_1986,
       author = {{Thielemann}, F.-K. and {Nomoto}, K. and {Yokoi}, K.},
        title = "{Explosive nucleosynthesis in carbon deflagration models of Type I supernovae}",
      journal = {\aap},
         year = 1986,
        month = apr,
       volume = {158},
       number = {1-2},
        pages = {17-33},
       adsurl = {https://ui.adsabs.harvard.edu/abs/1986A&A...158...17T}
}

@ARTICLE{fischer_2020,
       author = {{Fischer}, Tobias and {Wu}, Meng-Ru and {Wehmeyer}, Benjamin and {Bastian}, Niels-Uwe F. and {Mart{\'\i}nez-Pinedo}, Gabriel and {Thielemann}, Friedrich-Karl},
        title = "{Core-collapse Supernova Explosions Driven by the Hadron-quark Phase Transition as a Rare r-process Site}",
      journal = {\apj},
         year = 2020,
        month = may,
       volume = {894},
       number = {1},
          eid = {9},
        pages = {9},
          doi = {10.3847/1538-4357/ab86b0},
archivePrefix = {arXiv},
       eprint = {2003.00972},
 primaryClass = {astro-ph.HE},
       adsurl = {https://ui.adsabs.harvard.edu/abs/2020ApJ...894....9F}
}

@ARTICLE{nishimura_2015,
       author = {{Nishimura}, Nobuya and {Takiwaki}, Tomoya and {Thielemann}, Friedrich-Karl},
        title = "{The r-process Nucleosynthesis in the Various Jet-like Explosions of Magnetorotational Core-collapse Supernovae}",
      journal = {\apj},
         year = 2015,
        month = sep,
       volume = {810},
       number = {2},
          eid = {109},
        pages = {109},
          doi = {10.1088/0004-637X/810/2/109},
archivePrefix = {arXiv},
       eprint = {1501.06567},
 primaryClass = {astro-ph.SR},
       adsurl = {https://ui.adsabs.harvard.edu/abs/2015ApJ...810..109N}
}

@ARTICLE{pllumbi_2015,
       author = {{Pllumbi}, Else and {Tamborra}, Irene and {Wanajo}, Shinya and {Janka}, Hans-Thomas and {H{\"u}depohl}, Lorenz},
        title = "{Impact of Neutrino Flavor Oscillations on the Neutrino-driven Wind Nucleosynthesis of an Electron-capture Supernova}",
      journal = {\apj},
         year = 2015,
        month = aug,
       volume = {808},
       number = {2},
          eid = {188},
        pages = {188},
          doi = {10.1088/0004-637X/808/2/188},
archivePrefix = {arXiv},
       eprint = {1406.2596},
 primaryClass = {astro-ph.SR},
       adsurl = {https://ui.adsabs.harvard.edu/abs/2015ApJ...808..188P}
}

@ARTICLE{kikuchi_2015,
       author = {{Kikuchi}, Yukihiro and {Hashimoto}, Masa-aki and {Ono}, Masaomi and {Fukuda}, Ryohei},
        title = "{Effects of triple-{\ensuremath{\alpha}} and $^{12}$C({\ensuremath{\alpha}}, {\ensuremath{\gamma}})$^{16}$O reaction rates on the supernova nucleosynthesis in a massive star of 25 M$_{\odot}$}",
      journal = {Progress of Theoretical and Experimental Physics},
         year = 2015,
        month = jun,
       volume = {2015},
       number = {6},
          eid = {063E01},
        pages = {063E01},
          doi = {10.1093/ptep/ptv072},
archivePrefix = {arXiv},
       eprint = {1410.7476},
 primaryClass = {astro-ph.SR},
       adsurl = {https://ui.adsabs.harvard.edu/abs/2015PTEP.2015f3E01K}
}

@ARTICLE{bessell_2015,
       author = {{Bessell}, Michael S. and {Collet}, Remo and {Keller}, Stefan C. and {Frebel}, Anna and {Heger}, Alexander and {Casey}, Andrew R. and {Masseron}, Thomas and {Asplund}, Martin and {Jacobson}, Heather R. and {Lind}, Karin and {Marino}, Anna F. and {Norris}, John E. and {Yong}, David and {Da Costa}, Gary and {Chan}, Conrad and {Magic}, Zazralt and {Schmidt}, Brian and {Tisserand}, Patrick},
        title = "{Nucleosynthesis in a Primordial Supernova: Carbon and Oxygen Abundances in SMSS J031300.36-670839.3}",
      journal = {\apjl},
         year = 2015,
        month = jun,
       volume = {806},
       number = {1},
          eid = {L16},
        pages = {L16},
          doi = {10.1088/2041-8205/806/1/L16},
archivePrefix = {arXiv},
       eprint = {1505.03756},
 primaryClass = {astro-ph.SR},
       adsurl = {https://ui.adsabs.harvard.edu/abs/2015ApJ...806L..16B}
}

@ARTICLE{wu_2015,
       author = {{Wu}, Meng-Ru and {Qian}, Yong-Zhong and {Mart{\'\i}nez-Pinedo}, Gabriel and {Fischer}, Tobias and {Huther}, Lutz},
        title = "{Effects of neutrino oscillations on nucleosynthesis and neutrino signals for an 18 M$_{\odot}$ supernova model}",
      journal = {\prd},
         year = 2015,
        month = mar,
       volume = {91},
       number = {6},
          eid = {065016},
        pages = {065016},
          doi = {10.1103/PhysRevD.91.065016},
archivePrefix = {arXiv},
       eprint = {1412.8587},
 primaryClass = {astro-ph.HE},
       adsurl = {https://ui.adsabs.harvard.edu/abs/2015PhRvD..91f5016W}
}

@ARTICLE{kozyreva_2014,
       author = {{Kozyreva}, A. and {Yoon}, S.-C. and {Langer}, N.},
        title = "{Explosion and nucleosynthesis of low-redshift pair-instability supernovae}",
      journal = {\aap},
         year = 2014,
        month = jun,
       volume = {566},
          eid = {A146},
        pages = {A146},
          doi = {10.1051/0004-6361/201423641},
archivePrefix = {arXiv},
       eprint = {1405.6340},
 primaryClass = {astro-ph.HE},
       adsurl = {https://ui.adsabs.harvard.edu/abs/2014A&A...566A.146K}
}

@ARTICLE{hayakawa_2013,
       author = {{Hayakawa}, T. and {Nakamura}, K. and {Kajino}, T. and {Chiba}, S. and {Iwamoto}, N. and {Cheoun}, M.~K. and {Mathews}, G.~J.},
        title = "{Supernova Neutrino Nucleosynthesis of the Radioactive $^{92}$Nb Observed in Primitive Meteorites}",
      journal = {\apjl},
         year = 2013,
        month = dec,
       volume = {779},
       number = {1},
          eid = {L9},
        pages = {L9},
          doi = {10.1088/2041-8205/779/1/L9},
       adsurl = {https://ui.adsabs.harvard.edu/abs/2013ApJ...779L...9H}
}

@ARTICLE{zha_2019,
       author = {{Zha}, Shuai and {Leung}, Shing-Chi and {Suzuki}, Toshio and {Nomoto}, Ken'ichi},
        title = "{Evolution of ONeMg Core in Super-AGB Stars toward Electron-capture Supernovae: Effects of Updated Electron-capture Rate}",
      journal = {\apj},
         year = 2019,
        month = nov,
       volume = {886},
       number = {1},
          eid = {22},
        pages = {22},
          doi = {10.3847/1538-4357/ab4b4b},
archivePrefix = {arXiv},
       eprint = {1907.04184},
 primaryClass = {astro-ph.HE},
       adsurl = {https://ui.adsabs.harvard.edu/abs/2019ApJ...886...22Z}
}

@ARTICLE{zha_2021,
       author = {{Zha}, Shuai and {O'Connor}, Evan P. and {da Silva Schneider}, Andr{\'e}},
        title = "{Progenitor Dependence of Hadron-quark Phase Transition in Failing Core-collapse Supernovae}",
      journal = {\apj},
         year = 2021,
        month = apr,
       volume = {911},
       number = {2},
          eid = {74},
        pages = {74},
          doi = {10.3847/1538-4357/abec4c},
archivePrefix = {arXiv},
       eprint = {2103.02268},
 primaryClass = {astro-ph.HE},
       adsurl = {https://ui.adsabs.harvard.edu/abs/2021ApJ...911...74Z}
}

@ARTICLE{zha_2022,
       author = {{Zha}, Shuai and {O'Connor}, Evan P. and {Couch}, Sean M. and {Leung}, Shing-Chi and {Nomoto}, Ken'ichi},
        title = "{Hydrodynamic simulations of electron-capture supernovae: progenitor and dimension dependence}",
      journal = {\mnras},
         year = 2022,
        month = jun,
       volume = {513},
       number = {1},
        pages = {1317-1328},
          doi = {10.1093/mnras/stac1035},
archivePrefix = {arXiv},
       eprint = {2112.15257},
 primaryClass = {astro-ph.HE},
       adsurl = {https://ui.adsabs.harvard.edu/abs/2022MNRAS.513.1317Z}
}

@ARTICLE{zha_2024,
       author = {{Zha}, Shuai and {M{\"u}ller}, Bernhard and {Powell}, Jade},
        title = "{Nucleosynthesis in the Innermost Ejecta of Magnetorotational Supernova Explosions in Three Dimensions}",
      journal = {\apj},
         year = 2024,
        month = jul,
       volume = {969},
       number = {2},
          eid = {141},
        pages = {141},
          doi = {10.3847/1538-4357/ad4ae7},
archivePrefix = {arXiv},
       eprint = {2403.02072},
 primaryClass = {astro-ph.HE},
       adsurl = {https://ui.adsabs.harvard.edu/abs/2024ApJ...969..141Z}
}

@ARTICLE{zha_2020,
       author = {{Zha}, Shuai and {O'Connor}, Evan P. and {Chu}, Ming-chung and {Lin}, Lap-Ming and {Couch}, Sean M.},
        title = "{Gravitational-wave Signature of a First-order Quantum Chromodynamics Phase Transition in Core-Collapse Supernovae}",
      journal = {\prl},
         year = 2020,
        month = jul,
       volume = {125},
       number = {5},
          eid = {051102},
        pages = {051102},
          doi = {10.1103/PhysRevLett.125.051102},
archivePrefix = {arXiv},
       eprint = {2007.04716},
 primaryClass = {astro-ph.HE},
       adsurl = {https://ui.adsabs.harvard.edu/abs/2020PhRvL.125e1102Z}
}

@ARTICLE{woosley_2002,
       author = {{Woosley}, S.~E. and {Heger}, A. and {Weaver}, T.~A.},
        title = "{The evolution and explosion of massive stars}",
      journal = {Reviews of Modern Physics},
         year = 2002,
        month = nov,
       volume = {74},
       number = {4},
        pages = {1015-1071},
          doi = {10.1103/RevModPhys.74.1015},
       adsurl = {https://ui.adsabs.harvard.edu/abs/2002RvMP...74.1015W}
}

@ARTICLE{keegans_2023,
       author = {{Keegans}, James D. and {Pignatari}, Marco and {Stancliffe}, Richard J. and {Travaglio}, Claudia and {Jones}, Samuel and {Gibson}, Brad K. and {Townsley}, Dean M. and {Miles}, Broxton J. and {Shen}, Ken J. and {Few}, Gareth},
        title = "{Type Ia Supernova Nucleosynthesis: Metallicity-dependent Yields}",
      journal = {\apjs},
         year = 2023,
        month = sep,
       volume = {268},
       number = {1},
          eid = {8},
        pages = {8},
          doi = {10.3847/1538-4365/ace102},
archivePrefix = {arXiv},
       eprint = {2306.12885},
 primaryClass = {astro-ph.HE},
       adsurl = {https://ui.adsabs.harvard.edu/abs/2023ApJS..268....8K}
}

@ARTICLE{kobayashi_2025,
       author = {{Kobayashi}, Chiaki},
        title = "{Nucleosynthesis and the chemical enrichment of galaxies}",
      journal = {arXiv e-prints},
         year = 2025,
        month = jun,
          eid = {arXiv:2506.20436},
        pages = {arXiv:2506.20436},
          doi = {10.48550/arXiv.2506.20436},
archivePrefix = {arXiv},
       eprint = {2506.20436},
 primaryClass = {astro-ph.GA},
       adsurl = {https://ui.adsabs.harvard.edu/abs/2025arXiv250620436K}
}

@ARTICLE{arcones_thielemann_2023,
       author = {{Arcones}, Almudena and {Thielemann}, Friedrich-Karl},
        title = "{Origin of the elements}",
      journal = {\aapr},
         year = 2023,
        month = dec,
       volume = {31},
       number = {1},
          eid = {1},
        pages = {1},
          doi = {10.1007/s00159-022-00146-x},
       adsurl = {https://ui.adsabs.harvard.edu/abs/2023A&ARv..31....1A}
}

@BOOK{kippenhahn_2013,
       author = {{Kippenhahn}, Rudolf and {Weigert}, Alfred and {Weiss}, Achim},
        title = "{Stellar Structure and Evolution}",
         year = 2013,
          doi = {10.1007/978-3-642-30304-3},
       adsurl = {https://ui.adsabs.harvard.edu/abs/2013sse..book.....K}
}

@ARTICLE{hillebrandt_2000,
       author = {{Hillebrandt}, Wolfgang and {Niemeyer}, Jens C.},
        title = "{Type IA Supernova Explosion Models}",
      journal = {\araa},
         year = 2000,
        month = jan,
       volume = {38},
        pages = {191-230},
          doi = {10.1146/annurev.astro.38.1.191},
archivePrefix = {arXiv},
       eprint = {astro-ph/0006305},
 primaryClass = {astro-ph},
       adsurl = {https://ui.adsabs.harvard.edu/abs/2000ARA&A..38..191H}
}

@ARTICLE{liu_2023,
       author = {{Liu}, Zheng-Wei and {R{\"o}pke}, Friedrich K. and {Han}, Zhanwen},
        title = "{Type Ia Supernova Explosions in Binary Systems: A Review}",
      journal = {Research in Astronomy and Astrophysics},
         year = 2023,
        month = aug,
       volume = {23},
       number = {8},
          eid = {082001},
        pages = {082001},
          doi = {10.1088/1674-4527/acd89e},
archivePrefix = {arXiv},
       eprint = {2305.13305},
 primaryClass = {astro-ph.HE},
       adsurl = {https://ui.adsabs.harvard.edu/abs/2023RAA....23h2001L}
}

@ARTICLE{bethe_1990,
       author = {{Bethe}, H.~A.},
        title = "{Supernova mechanisms}",
      journal = {Reviews of Modern Physics},
         year = 1990,
        month = oct,
       volume = {62},
       number = {4},
        pages = {801-866},
          doi = {10.1103/RevModPhys.62.801},
       adsurl = {https://ui.adsabs.harvard.edu/abs/1990RvMP...62..801B}
}

@INCOLLECTION{ropke_2017,
       author = {{R{\"o}pke}, Friedrich K.},
        title = "{Combustion in Thermonuclear Supernova Explosions}",
    booktitle = {Handbook of Supernovae},
         year = 2017,
       editor = {{Alsabti}, Athem W. and {Murdin}, Paul},
        pages = {1185},
          doi = {10.1007/978-3-319-21846-5_58},
       adsurl = {https://ui.adsabs.harvard.edu/abs/2017hsn..book.1185R}
}

@BOOK{landau_1987,
       author = {{Landau}, L.~D. and {Lifshitz}, E.~M.},
        title = "{Fluid Mechanics}",
         year = 1987,
       adsurl = {https://ui.adsabs.harvard.edu/abs/1987flme.book.....L}
}

@ARTICLE{arnett_1982,
       author = {{Arnett}, W.~D.},
        title = "{Type I supernovae. I - Analytic solutions for the early part of the light curve}",
      journal = {\apj},
         year = 1982,
        month = feb,
       volume = {253},
        pages = {785-797},
          doi = {10.1086/159681},
       adsurl = {https://ui.adsabs.harvard.edu/abs/1982ApJ...253..785A}
}

@ARTICLE{timmes_1992,
       author = {{Timmes}, F.~X. and {Woosley}, S.~E.},
        title = "{The Conductive Propagation of Nuclear Flames. I. Degenerate C + O and O + NE + MG White Dwarfs}",
      journal = {\apj},
         year = 1992,
        month = sep,
       volume = {396},
        pages = {649},
          doi = {10.1086/171746},
       adsurl = {https://ui.adsabs.harvard.edu/abs/1992ApJ...396..649T}
}

@ARTICLE{niemeyer_1995,
       author = {{Niemeyer}, J.~C. and {Hillebrandt}, W.},
        title = "{Turbulent Nuclear Flames in Type IA Supernovae}",
      journal = {\apj},
         year = 1995,
        month = oct,
       volume = {452},
        pages = {769},
          doi = {10.1086/176345},
       adsurl = {https://ui.adsabs.harvard.edu/abs/1995ApJ...452..769N}
}

@ARTICLE{arnett_1969,
       author = {{Arnett}, W. David},
        title = "{A Possible Model of Supernovae: Detonation of $^{12}$C}",
      journal = {\apss},
         year = 1969,
        month = oct,
       volume = {5},
       number = {2},
        pages = {180-212},
          doi = {10.1007/BF00650291},
       adsurl = {https://ui.adsabs.harvard.edu/abs/1969Ap&SS...5..180A}
}

@INCOLLECTION{starrfield_2017,
       author = {{Starrfield}, Sumner},
        title = "{Evolution of Accreting White Dwarfs to the Thermonuclear Runaway}",
    booktitle = {Handbook of Supernovae},
         year = 2017,
       editor = {{Alsabti}, Athem W. and {Murdin}, Paul},
        pages = {1211},
          doi = {10.1007/978-3-319-21846-5_59},
       adsurl = {https://ui.adsabs.harvard.edu/abs/2017hsn..book.1211S}
}

@ARTICLE{whelan_1973,
       author = {{Whelan}, John and {Iben}, Jr., Icko},
        title = "{Binaries and Supernovae of Type I}",
      journal = {\apj},
         year = 1973,
        month = dec,
       volume = {186},
        pages = {1007-1014},
          doi = {10.1086/152565},
       adsurl = {https://ui.adsabs.harvard.edu/abs/1973ApJ...186.1007W}
}

@ARTICLE{nomoto_1982,
       author = {{Nomoto}, K.},
        title = "{Accreting white dwarf models for type I supernovae. I - Presupernova evolution and triggering mechanisms}",
      journal = {\apj},
         year = 1982,
        month = feb,
       volume = {253},
        pages = {798-810},
          doi = {10.1086/159682},
       adsurl = {https://ui.adsabs.harvard.edu/abs/1982ApJ...253..798N}
}

@ARTICLE{canal_1996,
       author = {{Canal}, R. and {Ruiz-Lapuente}, P. and {Burkert}, A.},
        title = "{The Single-Degenerate Scenario for Type IA Supernovae in Cosmic Perspective}",
      journal = {\apjl},
         year = 1996,
        month = jan,
       volume = {456},
        pages = {L101},
          doi = {10.1086/309869},
archivePrefix = {arXiv},
       eprint = {astro-ph/9511138},
 primaryClass = {astro-ph},
       adsurl = {https://ui.adsabs.harvard.edu/abs/1996ApJ...456L.101C}
}

@ARTICLE{han_2004,
       author = {{Han}, Z. and {Podsiadlowski}, Ph.},
        title = "{The single-degenerate channel for the progenitors of Type Ia supernovae}",
      journal = {\mnras},
         year = 2004,
        month = jun,
       volume = {350},
       number = {4},
        pages = {1301-1309},
          doi = {10.1111/j.1365-2966.2004.07713.x},
archivePrefix = {arXiv},
       eprint = {astro-ph/0309618},
 primaryClass = {astro-ph},
       adsurl = {https://ui.adsabs.harvard.edu/abs/2004MNRAS.350.1301H}
}

@ARTICLE{nomoto_1976,
       author = {{Nomoto}, K. and {Sugimoto}, D. and {Neo}, S.},
        title = "{Carbon Deflagration Supernova, an Alternative to Carbon Detonation}",
      journal = {\apss},
         year = 1976,
        month = feb,
       volume = {39},
       number = {2},
        pages = {L37-L42},
          doi = {10.1007/BF00648354},
       adsurl = {https://ui.adsabs.harvard.edu/abs/1976Ap&SS..39L..37N}
}

@ARTICLE{arnett_1971,
       author = {{Arnett}, W. David and {Truran}, J.~W. and {Woosley}, Stanford E.},
        title = "{Nucleosynthesis in Supernova Models. II. The \^\{12\}C Detonation Model}",
      journal = {\apj},
         year = 1971,
        month = apr,
       volume = {165},
        pages = {87},
          doi = {10.1086/150878},
       adsurl = {https://ui.adsabs.harvard.edu/abs/1971ApJ...165...87A}
}

@ARTICLE{khokhlov_1997,
       author = {{Khokhlov}, A.~M. and {Oran}, E.~S. and {Wheeler}, J.~C.},
        title = "{Deflagration-to-Detonation Transition in Thermonuclear Supernovae}",
      journal = {\apj},
         year = 1997,
        month = mar,
       volume = {478},
       number = {2},
        pages = {678-688},
          doi = {10.1086/303815},
archivePrefix = {arXiv},
       eprint = {astro-ph/9612226},
 primaryClass = {astro-ph},
       adsurl = {https://ui.adsabs.harvard.edu/abs/1997ApJ...478..678K}
}

@ARTICLE{ropke_2007,
       author = {{R{\"o}pke}, F.~K.},
        title = "{Flame-driven Deflagration-to-Detonation Transitions in Type Ia Supernovae?}",
      journal = {\apj},
         year = 2007,
        month = oct,
       volume = {668},
       number = {2},
        pages = {1103-1108},
          doi = {10.1086/520830},
archivePrefix = {arXiv},
       eprint = {0709.4095},
 primaryClass = {astro-ph},
       adsurl = {https://ui.adsabs.harvard.edu/abs/2007ApJ...668.1103R}
}

@ARTICLE{plewa_2004,
       author = {{Plewa}, T. and {Calder}, A.~C. and {Lamb}, D.~Q.},
        title = "{Type Ia Supernova Explosion: Gravitationally Confined Detonation}",
      journal = {\apjl},
         year = 2004,
        month = sep,
       volume = {612},
       number = {1},
        pages = {L37-L40},
          doi = {10.1086/424036},
archivePrefix = {arXiv},
       eprint = {astro-ph/0405163},
 primaryClass = {astro-ph},
       adsurl = {https://ui.adsabs.harvard.edu/abs/2004ApJ...612L..37P}
}

@ARTICLE{nomoto_1982b,
       author = {{Nomoto}, K.},
        title = "{Accreting white dwarf models for type I supernovae. II. Off-center detonation supernovae.}",
      journal = {\apj},
         year = 1982,
        month = jun,
       volume = {257},
        pages = {780-792},
          doi = {10.1086/160031},
       adsurl = {https://ui.adsabs.harvard.edu/abs/1982ApJ...257..780N}
}

@ARTICLE{baade_1934,
       author = {{Baade}, W. and {Zwicky}, F.},
        title = "{Cosmic Rays from Super-novae}",
      journal = {Proceedings of the National Academy of Science},
         year = 1934,
        month = may,
       volume = {20},
       number = {5},
        pages = {259-263},
          doi = {10.1073/pnas.20.5.259},
       adsurl = {https://ui.adsabs.harvard.edu/abs/1934PNAS...20..259B}
}

@ARTICLE{ibeling_2013,
       author = {{Ibeling}, Duligur and {Heger}, Alexander},
        title = "{The Metallicity Dependence of the Minimum Mass for Core-collapse Supernovae}",
      journal = {\apjl},
         year = 2013,
        month = mar,
       volume = {765},
       number = {2},
          eid = {L43},
        pages = {L43},
          doi = {10.1088/2041-8205/765/2/L43},
archivePrefix = {arXiv},
       eprint = {1301.5783},
 primaryClass = {astro-ph.SR},
       adsurl = {https://ui.adsabs.harvard.edu/abs/2013ApJ...765L..43I}
}

@article{Aglietta:1987it,
    author = "Aglietta, M. and others",
    title = "{On the event observed in the Mont Blanc Underground Neutrino observatory during the occurrence of Supernova 1987a}",
    doi = "10.1209/0295-5075/3/12/011",
    journal = "EPL",
    volume = "3",
    pages = "1315--1320",
    year = "1987"
}

@article{Bionta:1987qt,
    author = "Bionta, R. M. and others",
    title = "{Observation of a Neutrino Burst in Coincidence with Supernova SN 1987a in the Large Magellanic Cloud}",
    reportNumber = "UCI-NEUTRINO-87-10",
    doi = "10.1103/PhysRevLett.58.1494",
    journal = "Phys. Rev. Lett.",
    volume = "58",
    pages = "1494",
    year = "1987"
}

@article{Kamiokande-II:1987idp,
    author = "Hirata, K. and others",
    editor = "Wali, K. C.",
    collaboration = "Kamiokande-II",
    title = "{Observation of a Neutrino Burst from the Supernova SN 1987a}",
    reportNumber = "UT-ICEPP-87-01, UPR-142E",
    doi = "10.1103/PhysRevLett.58.1490",
    journal = "Phys. Rev. Lett.",
    volume = "58",
    pages = "1490--1493",
    year = "1987"
}

@ARTICLE{hewish_1968,
       author = {{Hewish}, A. and {Bell}, S.~J. and {Pilkington}, J.~D.~H. and {Scott}, P.~F. and {Collins}, R.~A.},
        title = "{Observation of a Rapidly Pulsating Radio Source}",
      journal = {\nat},
         year = 1968,
        month = feb,
       volume = {217},
       number = {5130},
        pages = {709-713},
          doi = {10.1038/217709a0},
       adsurl = {https://ui.adsabs.harvard.edu/abs/1968Natur.217..709H}
}

@ARTICLE{bruenn_1987,
       author = {{Bruenn}, Stephen W.},
        title = "{Neutrinos from SN1987A and current models of stellar-core collapse}",
      journal = {\prl},
         year = 1987,
        month = aug,
       volume = {59},
       number = {8},
        pages = {938-941},
          doi = {10.1103/PhysRevLett.59.938},
       adsurl = {https://ui.adsabs.harvard.edu/abs/1987PhRvL..59..938B}
}

@ARTICLE{arafune_1987,
       author = {{Arafune}, J. and {Fukugita}, M.},
        title = "{Physical implications of the Kamioka observation of neutrinos from supernova 1987A}",
      journal = {\prl},
         year = 1987,
        month = jul,
       volume = {59},
       number = {3},
        pages = {367-369},
          doi = {10.1103/PhysRevLett.59.367},
       adsurl = {https://ui.adsabs.harvard.edu/abs/1987PhRvL..59..367A}
}

@ARTICLE{sato_1987,
       author = {{Sato}, Katsuhiko and {Suzuki}, Hideyuki},
        title = "{Analysis of neutrino burst from the supernova 1987A in the Large Magellanic Cloud}",
      journal = {\prl},
         year = 1987,
        month = jun,
       volume = {58},
       number = {25},
        pages = {2722-2725},
          doi = {10.1103/PhysRevLett.58.2722},
       adsurl = {https://ui.adsabs.harvard.edu/abs/1987PhRvL..58.2722S}
}

@ARTICLE{science_2012,
       author = {{Bhattacharjee}, Yudhijit},
        title = "{How Do Stars Explode?}",
      journal = {Science},
         year = 2012,
        month = jun,
       volume = {336},
       number = {6085},
        pages = {1094},
       adsurl = {https://ui.adsabs.harvard.edu/abs/2012Sci...336.1094B}
}

@ARTICLE{chandrasekhar_1931,
       author = {{Chandrasekhar}, S.},
        title = "{The Maximum Mass of Ideal White Dwarfs}",
      journal = {\apj},
         year = 1931,
        month = jul,
       volume = {74},
        pages = {81},
          doi = {10.1086/143324},
       adsurl = {https://ui.adsabs.harvard.edu/abs/1931ApJ....74...81C}
}

@ARTICLE{baron_1987,
       author = {{Baron}, E. and {Bethe}, H.~A. and {Brown}, G.~E. and {Cooperstein}, J. and {Kahana}, S.},
        title = "{Type II supernovae from prompt explosions}",
      journal = {\prl},
         year = 1987,
        month = aug,
       volume = {59},
       number = {6},
        pages = {736-739},
          doi = {10.1103/PhysRevLett.59.736},
       adsurl = {https://ui.adsabs.harvard.edu/abs/1987PhRvL..59..736B}
}

@ARTICLE{bruenn_1985,
       author = {{Bruenn}, S.~W.},
        title = "{Stellar core collapse - Numerical model and infall epoch}",
      journal = {\apjs},
         year = 1985,
        month = aug,
       volume = {58},
        pages = {771-841},
          doi = {10.1086/191056},
       adsurl = {https://ui.adsabs.harvard.edu/abs/1985ApJS...58..771B}
}

@ARTICLE{bck_1985,
       author = {{Baron}, E. and {Cooperstein}, J. and {Kahana}, S.},
        title = "{Type II supernovae in 12M$_{cirdot}$ and 15M$_{cirdot}$ stars: The equation of state and general relativity}",
      journal = {\prl},
         year = 1985,
        month = jul,
       volume = {55},
       number = {1},
        pages = {126-129},
          doi = {10.1103/PhysRevLett.55.126},
       adsurl = {https://ui.adsabs.harvard.edu/abs/1985PhRvL..55..126B}
}

@ARTICLE{fink_2007,
       author = {{Fink}, M. and {Hillebrandt}, W. and {R{\"o}pke}, F.~K.},
        title = "{Double-detonation supernovae of sub-Chandrasekhar mass white dwarfs}",
      journal = {\aap},
         year = 2007,
        month = dec,
       volume = {476},
       number = {3},
        pages = {1133-1143},
          doi = {10.1051/0004-6361:20078438},
archivePrefix = {arXiv},
       eprint = {0710.5486},
 primaryClass = {astro-ph},
       adsurl = {https://ui.adsabs.harvard.edu/abs/2007A&A...476.1133F}
}

@ARTICLE{zhang_2026,
       author = {{Zhang}, Xiaoyu and {Wang}, Lile and {Gao}, Yang and {Zhou}, Yao},
        title = "{Direct Numerical Simulations of Oxygen-flame-driven Deflagration-to-detonation Transition in Type Ia Supernovae}",
      journal = {\apj},
         year = 2026,
        month = jan,
       volume = {996},
       number = {2},
          eid = {123},
        pages = {123},
          doi = {10.3847/1538-4357/ae28ce},
archivePrefix = {arXiv},
       eprint = {2510.26152},
 primaryClass = {astro-ph.HE},
       adsurl = {https://ui.adsabs.harvard.edu/abs/2026ApJ...996..123Z}
}

@ARTICLE{colgate_1966,
       author = {{Colgate}, Stirling A. and {White}, Richard H.},
        title = "{The Hydrodynamic Behavior of Supernovae Explosions}",
      journal = {\apj},
         year = 1966,
        month = mar,
       volume = {143},
        pages = {626},
          doi = {10.1086/148549},
       adsurl = {https://ui.adsabs.harvard.edu/abs/1966ApJ...143..626C}
}

@ARTICLE{bruenn_1989a,
       author = {{Bruenn}, Stephen W.},
        title = "{The Prompt-Shock Supernova Mechanism. I. The Effect of the Free-Proton Mass Fraction and the Neutrino Transport Algorithm}",
      journal = {\apj},
         year = 1989,
        month = may,
       volume = {340},
        pages = {955},
          doi = {10.1086/167450},
       adsurl = {https://ui.adsabs.harvard.edu/abs/1989ApJ...340..955B}
}

@ARTICLE{bruenn_1989b,
       author = {{Bruenn}, Stephen W.},
        title = "{The Prompt-Shock Supernova Mechanism. II. Supranuclear EOS Behavior and the Precollapse Model}",
      journal = {\apj},
         year = 1989,
        month = jun,
       volume = {341},
        pages = {385},
          doi = {10.1086/167502},
       adsurl = {https://ui.adsabs.harvard.edu/abs/1989ApJ...341..385B}
}

@ARTICLE{bethe_1985,
       author = {{Bethe}, H.~A. and {Wilson}, J.~R.},
        title = "{Revival of a stalled supernova shock by neutrino heating}",
      journal = {\apj},
         year = 1985,
        month = aug,
       volume = {295},
        pages = {14-23},
          doi = {10.1086/163343},
       adsurl = {https://ui.adsabs.harvard.edu/abs/1985ApJ...295...14B}
}

@ARTICLE{janka_1993,
       author = {{Janka}, H.-Th. and {Zwerger}, Th. and {Moenchmeyer}, R.},
        title = "{Does artificial viscosity destroy prompt type-II supernova explosions?}",
      journal = {\aap},
         year = 1993,
        month = feb,
       volume = {268},
       number = {1},
        pages = {360-368},
       adsurl = {https://ui.adsabs.harvard.edu/abs/1993A&A...268..360J}
}

@ARTICLE{rampp_2000,
       author = {{Rampp}, Markus and {Janka}, H.-Thomas},
        title = "{Spherically Symmetric Simulation with Boltzmann Neutrino Transport of Core Collapse and Postbounce Evolution of a 15 M$_{solar}$ Star}",
      journal = {\apjl},
         year = 2000,
        month = aug,
       volume = {539},
       number = {1},
        pages = {L33-L36},
          doi = {10.1086/312837},
archivePrefix = {arXiv},
       eprint = {astro-ph/0005438},
 primaryClass = {astro-ph},
       adsurl = {https://ui.adsabs.harvard.edu/abs/2000ApJ...539L..33R}
}

@ARTICLE{mezzacappa_2001,
       author = {{Mezzacappa}, Anthony and {Liebend{\"o}rfer}, Matthias and {Messer}, O.~E. and {Hix}, W. Raphael and {Thielemann}, Friedrich-Karl and {Bruenn}, Stephen W.},
        title = "{Simulation of the Spherically Symmetric Stellar Core Collapse, Bounce, and Postbounce Evolution of a Star of 13 Solar Masses with Boltzmann Neutrino Transport, and Its Implications for the Supernova Mechanism}",
      journal = {\prl},
         year = 2001,
        month = mar,
       volume = {86},
       number = {10},
        pages = {1935-1938},
          doi = {10.1103/PhysRevLett.86.1935},
archivePrefix = {arXiv},
       eprint = {astro-ph/0005366},
 primaryClass = {astro-ph},
       adsurl = {https://ui.adsabs.harvard.edu/abs/2001PhRvL..86.1935M}
}

@ARTICLE{liebendorfer_2001,
       author = {{Liebend{\"o}rfer}, Matthias and {Mezzacappa}, Anthony and {Thielemann}, Friedrich-Karl and {Messer}, O.~E. and {Hix}, W. Raphael and {Bruenn}, Stephen W.},
        title = "{Probing the gravitational well: No supernova explosion in spherical symmetry with general relativistic Boltzmann neutrino transport}",
      journal = {\prd},
         year = 2001,
        month = may,
       volume = {63},
       number = {10},
          eid = {103004},
        pages = {103004},
          doi = {10.1103/PhysRevD.63.103004},
archivePrefix = {arXiv},
       eprint = {astro-ph/0006418},
 primaryClass = {astro-ph},
       adsurl = {https://ui.adsabs.harvard.edu/abs/2001PhRvD..63j3004L}
}

@ARTICLE{doherty_2017,
       author = {{Doherty}, Carolyn L. and {Gil-Pons}, Pilar and {Siess}, Lionel and {Lattanzio}, John C.},
        title = "{Super-AGB Stars and their Role as Electron Capture Supernova Progenitors}",
      journal = {\pasa},
         year = 2017,
        month = nov,
       volume = {34},
          eid = {e056},
        pages = {e056},
          doi = {10.1017/pasa.2017.52},
archivePrefix = {arXiv},
       eprint = {1703.06895},
 primaryClass = {astro-ph.SR},
       adsurl = {https://ui.adsabs.harvard.edu/abs/2017PASA...34...56D}
}

@ARTICLE{poelarends_2008,
       author = {{Poelarends}, A.~J.~T. and {Herwig}, F. and {Langer}, N. and {Heger}, A.},
        title = "{The Supernova Channel of Super-AGB Stars}",
      journal = {\apj},
         year = 2008,
        month = mar,
       volume = {675},
       number = {1},
        pages = {614-625},
          doi = {10.1086/520872},
archivePrefix = {arXiv},
       eprint = {0705.4643},
 primaryClass = {astro-ph},
       adsurl = {https://ui.adsabs.harvard.edu/abs/2008ApJ...675..614P}
}

@ARTICLE{kitaura_2006,
       author = {{Kitaura}, F.~S. and {Janka}, H.-Th. and {Hillebrandt}, W.},
        title = "{Explosions of O-Ne-Mg cores, the Crab supernova, and subluminous type II-P supernovae}",
      journal = {\aap},
         year = 2006,
        month = apr,
       volume = {450},
       number = {1},
        pages = {345-350},
          doi = {10.1051/0004-6361:20054703},
archivePrefix = {arXiv},
       eprint = {astro-ph/0512065},
 primaryClass = {astro-ph},
       adsurl = {https://ui.adsabs.harvard.edu/abs/2006A&A...450..345K}
}

@ARTICLE{melson_2015,
       author = {{Melson}, Tobias and {Janka}, Hans-Thomas and {Marek}, Andreas},
        title = "{Neutrino-driven Supernova of a Low-mass Iron-core Progenitor Boosted by Three-dimensional Turbulent Convection}",
      journal = {\apjl},
         year = 2015,
        month = mar,
       volume = {801},
       number = {2},
          eid = {L24},
        pages = {L24},
          doi = {10.1088/2041-8205/801/2/L24},
archivePrefix = {arXiv},
       eprint = {1501.01961},
 primaryClass = {astro-ph.SR},
       adsurl = {https://ui.adsabs.harvard.edu/abs/2015ApJ...801L..24M}
}

@ARTICLE{janka_2016,
       author = {{Janka}, Hans-Thomas and {Melson}, Tobias and {Summa}, Alexander},
        title = "{Physics of Core-Collapse Supernovae in Three Dimensions: A Sneak Preview}",
      journal = {Annual Review of Nuclear and Particle Science},
         year = 2016,
        month = oct,
       volume = {66},
       number = {1},
        pages = {341-375},
          doi = {10.1146/annurev-nucl-102115-044747},
archivePrefix = {arXiv},
       eprint = {1602.05576},
 primaryClass = {astro-ph.SR},
       adsurl = {https://ui.adsabs.harvard.edu/abs/2016ARNPS..66..341J}
}

@ARTICLE{janka_2025,
       author = {{Janka}, Hans-Thomas},
        title = "{Long-Term Multidimensional Models of Core-Collapse Supernovae: Progress and Challenges}",
      journal = {Annual Review of Nuclear and Particle Science},
         year = 2025,
        month = sep,
       volume = {75},
       number = {1},
        pages = {425-461},
          doi = {10.1146/annurev-nucl-121423-100945},
archivePrefix = {arXiv},
       eprint = {2502.14836},
 primaryClass = {astro-ph.HE},
       adsurl = {https://ui.adsabs.harvard.edu/abs/2025ARNPS..75..425J}
}

@ARTICLE{radice_2017,
       author = {{Radice}, David and {Burrows}, Adam and {Vartanyan}, David and {Skinner}, M. Aaron and {Dolence}, Joshua C.},
        title = "{Electron-capture and Low-mass Iron-core-collapse Supernovae: New Neutrino-radiation-hydrodynamics Simulations}",
      journal = {\apj},
         year = 2017,
        month = nov,
       volume = {850},
       number = {1},
          eid = {43},
        pages = {43},
          doi = {10.3847/1538-4357/aa92c5},
archivePrefix = {arXiv},
       eprint = {1702.03927},
 primaryClass = {astro-ph.HE},
       adsurl = {https://ui.adsabs.harvard.edu/abs/2017ApJ...850...43R}
}

@INCOLLECTION{hix_2017,
       author = {{Hix}, W. Raphael and {Harris}, J. Austin},
        title = "{The Multidimensional Character of Nucleosynthesis in Core-Collapse Supernovae}",
    booktitle = {Handbook of Supernovae},
         year = 2017,
       editor = {{Alsabti}, Athem W. and {Murdin}, Paul},
        pages = {1771},
          doi = {10.1007/978-3-319-21846-5_77},
       adsurl = {https://ui.adsabs.harvard.edu/abs/2017hsn..book.1771H}
}

@ARTICLE{leblanc_1970,
       author = {{LeBlanc}, J.~M. and {Wilson}, J.~R.},
        title = "{A Numerical Example of the Collapse of a Rotating Magnetized Star}",
      journal = {\apj},
         year = 1970,
        month = aug,
       volume = {161},
        pages = {541},
          doi = {10.1086/150558},
       adsurl = {https://ui.adsabs.harvard.edu/abs/1970ApJ...161..541L}
}

@ARTICLE{meier_1976,
       author = {{Meier}, D.~L. and {Epstein}, R.~I. and {Arnett}, W.~D. and {Schramm}, D.~N.},
        title = "{Magnetohydrodynamic phenomena in collapsing stellar cores}",
      journal = {\apj},
         year = 1976,
        month = mar,
       volume = {204},
        pages = {869-878},
          doi = {10.1086/154235},
       adsurl = {https://ui.adsabs.harvard.edu/abs/1976ApJ...204..869M}
}

@ARTICLE{kuroda_2020,
       author = {{Kuroda}, Takami and {Arcones}, Almudena and {Takiwaki}, Tomoya and {Kotake}, Kei},
        title = "{Magnetorotational Explosion of a Massive Star Supported by Neutrino Heating in General Relativistic Three-dimensional Simulations}",
      journal = {\apj},
         year = 2020,
        month = jun,
       volume = {896},
       number = {2},
          eid = {102},
        pages = {102},
          doi = {10.3847/1538-4357/ab9308},
archivePrefix = {arXiv},
       eprint = {2003.02004},
 primaryClass = {astro-ph.HE},
       adsurl = {https://ui.adsabs.harvard.edu/abs/2020ApJ...896..102K}
}

@ARTICLE{mosta_2014,
       author = {{M{\"o}sta}, Philipp and {Richers}, Sherwood and {Ott}, Christian D. and {Haas}, Roland and {Piro}, Anthony L. and {Boydstun}, Kristen and {Abdikamalov}, Ernazar and {Reisswig}, Christian and {Schnetter}, Erik},
        title = "{Magnetorotational Core-collapse Supernovae in Three Dimensions}",
      journal = {\apjl},
         year = 2014,
        month = apr,
       volume = {785},
       number = {2},
          eid = {L29},
        pages = {L29},
          doi = {10.1088/2041-8205/785/2/L29},
archivePrefix = {arXiv},
       eprint = {1403.1230},
 primaryClass = {astro-ph.HE},
       adsurl = {https://ui.adsabs.harvard.edu/abs/2014ApJ...785L..29M}
}

@ARTICLE{powell_2023,
       author = {{Powell}, Jade and {M{\"u}ller}, Bernhard and {Aguilera-Dena}, David R. and {Langer}, Norbert},
        title = "{Three dimensional magnetorotational core-collapse supernova explosions of a 39 solar mass progenitor star}",
      journal = {\mnras},
         year = 2023,
        month = jul,
       volume = {522},
       number = {4},
        pages = {6070-6086},
          doi = {10.1093/mnras/stad1292},
archivePrefix = {arXiv},
       eprint = {2212.00200},
 primaryClass = {astro-ph.HE},
       adsurl = {https://ui.adsabs.harvard.edu/abs/2023MNRAS.522.6070P}
}

@ARTICLE{obergaulinger_2020,
       author = {{Obergaulinger}, M. and {Aloy}, M. {\'A}.},
        title = "{Magnetorotational core collapse of possible GRB progenitors - I. Explosion mechanisms}",
      journal = {\mnras},
         year = 2020,
        month = mar,
       volume = {492},
       number = {4},
        pages = {4613-4634},
          doi = {10.1093/mnras/staa096},
archivePrefix = {arXiv},
       eprint = {1909.01105},
 primaryClass = {astro-ph.HE},
       adsurl = {https://ui.adsabs.harvard.edu/abs/2020MNRAS.492.4613O}
}

@ARTICLE{sagert_2009,
       author = {{Sagert}, I. and {Fischer}, T. and {Hempel}, M. and {Pagliara}, G. and {Schaffner-Bielich}, J. and {Mezzacappa}, A. and {Thielemann}, F.-K. and {Liebend{\"o}rfer}, M.},
        title = "{Signals of the QCD Phase Transition in Core-Collapse Supernovae}",
      journal = {\prl},
         year = 2009,
        month = feb,
       volume = {102},
       number = {8},
          eid = {081101},
        pages = {081101},
          doi = {10.1103/PhysRevLett.102.081101},
archivePrefix = {arXiv},
       eprint = {0809.4225},
 primaryClass = {astro-ph},
       adsurl = {https://ui.adsabs.harvard.edu/abs/2009PhRvL.102h1101S}
}

@ARTICLE{fischer_2018,
       author = {{Fischer}, Tobias and {Bastian}, Niels-Uwe F. and {Wu}, Meng-Ru and {Baklanov}, Petr and {Sorokina}, Elena and {Blinnikov}, Sergei and {Typel}, Stefan and {Kl{\"a}hn}, Thomas and {Blaschke}, David B.},
        title = "{Quark deconfinement as a supernova explosion engine for massive blue supergiant stars}",
      journal = {Nature Astronomy},
         year = 2018,
        month = oct,
       volume = {2},
        pages = {980-986},
          doi = {10.1038/s41550-018-0583-0},
archivePrefix = {arXiv},
       eprint = {1712.08788},
 primaryClass = {astro-ph.HE},
       adsurl = {https://ui.adsabs.harvard.edu/abs/2018NatAs...2..980F}
}

@ARTICLE{huang_2025,
       author = {{Huang}, Xu-Run and {Zha}, Shuai and {Chu}, Ming-chung and {O'Connor}, Evan P. and {Chen}, Lie-Wen},
        title = "{Phase-transition-induced Collapse of Proto-compact Stars and Its Implication for Supernova Explosions}",
      journal = {\apj},
         year = 2025,
        month = feb,
       volume = {979},
       number = {2},
          eid = {151},
        pages = {151},
          doi = {10.3847/1538-4357/ada146},
archivePrefix = {arXiv},
       eprint = {2409.16189},
 primaryClass = {astro-ph.HE},
       adsurl = {https://ui.adsabs.harvard.edu/abs/2025ApJ...979..151H}
}

@ARTICLE{fischer_2011,
       author = {{Fischer}, T. and {Sagert}, I. and {Pagliara}, G. and {Hempel}, M. and {Schaffner-Bielich}, J. and {Rauscher}, T. and {Thielemann}, F.-K. and {K{\"a}ppeli}, R. and {Mart{\'\i}nez-Pinedo}, G. and {Liebend{\"o}rfer}, M.},
        title = "{Core-collapse Supernova Explosions Triggered by a Quark-Hadron Phase Transition During the Early Post-bounce Phase}",
      journal = {\apjs},
         year = 2011,
        month = jun,
       volume = {194},
       number = {2},
          eid = {39},
        pages = {39},
          doi = {10.1088/0067-0049/194/2/39},
archivePrefix = {arXiv},
       eprint = {1011.3409},
 primaryClass = {astro-ph.HE},
       adsurl = {https://ui.adsabs.harvard.edu/abs/2011ApJS..194...39F}
}

@ARTICLE{papish_2011,
       author = {{Papish}, Oded and {Soker}, Noam},
        title = "{Exploding core collapse supernovae with jittering jets}",
      journal = {\mnras},
         year = 2011,
        month = sep,
       volume = {416},
       number = {3},
        pages = {1697-1702},
          doi = {10.1111/j.1365-2966.2011.18671.x},
archivePrefix = {arXiv},
       eprint = {1103.1554},
 primaryClass = {astro-ph.HE},
       adsurl = {https://ui.adsabs.harvard.edu/abs/2011MNRAS.416.1697P}
}

@ARTICLE{soker_2024,
       author = {{Soker}, Noam},
        title = "{Supernovae in 2023 (review): possible breakthroughs by late observations}",
      journal = {The Open Journal of Astrophysics},
         year = 2024,
        month = apr,
       volume = {7},
          eid = {31},
        pages = {31},
          doi = {10.33232/001c.117147},
archivePrefix = {arXiv},
       eprint = {2311.17732},
 primaryClass = {astro-ph.HE},
       adsurl = {https://ui.adsabs.harvard.edu/abs/2024OJAp....7E..31S}
}

@ARTICLE{webbink_1984,
       author = {{Webbink}, R.~F.},
        title = "{Double white dwarfs as progenitors of R Coronae Borealis stars and type I supernovae.}",
      journal = {\apj},
         year = 1984,
        month = feb,
       volume = {277},
        pages = {355-360},
          doi = {10.1086/161701},
       adsurl = {https://ui.adsabs.harvard.edu/abs/1984ApJ...277..355W}
}

@ARTICLE{iben_1984,
       author = {{Iben}, Jr., I. and {Tutukov}, A.~V.},
        title = "{Supernovae of type I as end products of the evolution of binaries with components of moderate initial mass.}",
      journal = {\apjs},
         year = 1984,
        month = feb,
       volume = {54},
        pages = {335-372},
          doi = {10.1086/190932},
       adsurl = {https://ui.adsabs.harvard.edu/abs/1984ApJS...54..335I}
}

@ARTICLE{limongi_2012,
       author = {{Limongi}, M. and {Chieffi}, A.},
        title = "{Presupernova Evolution and Explosive Nucleosynthesis of Zero Metal Massive Stars}",
      journal = {\apjs},
         year = 2012,
        month = apr,
       volume = {199},
       number = {2},
          eid = {38},
        pages = {38},
          doi = {10.1088/0067-0049/199/2/38},
archivePrefix = {arXiv},
       eprint = {1202.4581},
 primaryClass = {astro-ph.SR},
       adsurl = {https://ui.adsabs.harvard.edu/abs/2012ApJS..199...38L}
}

@ARTICLE{guillochon_2010,
       author = {{Guillochon}, James and {Dan}, Marius and {Ramirez-Ruiz}, Enrico and {Rosswog}, Stephan},
        title = "{Surface Detonations in Double Degenerate Binary Systems Triggered by Accretion Stream Instabilities}",
      journal = {\apjl},
         year = 2010,
        month = jan,
       volume = {709},
       number = {1},
        pages = {L64-L69},
          doi = {10.1088/2041-8205/709/1/L64},
archivePrefix = {arXiv},
       eprint = {0911.0416},
 primaryClass = {astro-ph.HE},
       adsurl = {https://ui.adsabs.harvard.edu/abs/2010ApJ...709L..64G}
}

@ARTICLE{oconnor_2011,
       author = {{O'Connor}, Evan and {Ott}, Christian D.},
        title = "{Black Hole Formation in Failing Core-Collapse Supernovae}",
      journal = {\apj},
         year = 2011,
        month = apr,
       volume = {730},
       number = {2},
          eid = {70},
        pages = {70},
          doi = {10.1088/0004-637X/730/2/70},
archivePrefix = {arXiv},
       eprint = {1010.5550},
 primaryClass = {astro-ph.HE},
       adsurl = {https://ui.adsabs.harvard.edu/abs/2011ApJ...730...70O}
}

@ARTICLE{sumiyoshi_2007,
       author = {{Sumiyoshi}, K. and {Yamada}, S. and {Suzuki}, H.},
        title = "{Dynamics and Neutrino Signal of Black Hole Formation in Nonrotating Failed Supernovae. I. Equation of State Dependence}",
      journal = {\apj},
         year = 2007,
        month = sep,
       volume = {667},
       number = {1},
        pages = {382-394},
          doi = {10.1086/520876},
archivePrefix = {arXiv},
       eprint = {0706.3762},
 primaryClass = {astro-ph},
       adsurl = {https://ui.adsabs.harvard.edu/abs/2007ApJ...667..382S}
}

@ARTICLE{kochanek_2008,
       author = {{Kochanek}, Christopher S. and {Beacom}, John F. and {Kistler}, Matthew D. and {Prieto}, Jos{\'e} L. and {Stanek}, Krzysztof Z. and {Thompson}, Todd A. and {Y{\"u}ksel}, Hasan},
        title = "{A Survey About Nothing: Monitoring a Million Supergiants for Failed Supernovae}",
      journal = {\apj},
         year = 2008,
        month = sep,
       volume = {684},
       number = {2},
        pages = {1336-1342},
          doi = {10.1086/590053},
archivePrefix = {arXiv},
       eprint = {0802.0456},
 primaryClass = {astro-ph},
       adsurl = {https://ui.adsabs.harvard.edu/abs/2008ApJ...684.1336K}
}

@ARTICLE{macfadyen_1999,
       author = {{MacFadyen}, A.~I. and {Woosley}, S.~E.},
        title = "{Collapsars: Gamma-Ray Bursts and Explosions in ``Failed Supernovae''}",
      journal = {\apj},
         year = 1999,
        month = oct,
       volume = {524},
       number = {1},
        pages = {262-289},
          doi = {10.1086/307790},
archivePrefix = {arXiv},
       eprint = {astro-ph/9810274},
 primaryClass = {astro-ph},
       adsurl = {https://ui.adsabs.harvard.edu/abs/1999ApJ...524..262M}
}

@ARTICLE{arnett_1973,
       author = {{Arnett}, W. David},
        title = "{Explosive Nucleosynthesis in Stars}",
      journal = {\araa},
         year = 1973,
        month = jan,
       volume = {11},
        pages = {73},
          doi = {10.1146/annurev.aa.11.090173.000445},
       adsurl = {https://ui.adsabs.harvard.edu/abs/1973ARA&A..11...73A}
}

@ARTICLE{clifford_1965,
       author = {{Clifford}, F.~E. and {Tayler}, R.~F.},
        title = "{The equilibrium distribution of nuclides in matter at high temperatures (Summary of paper in Memoris, 69, 2,1965)}",
      journal = {\mnras},
         year = 1965,
        month = jan,
       volume = {129},
        pages = {104},
          doi = {10.1093/mnras/129.1.104},
       adsurl = {https://ui.adsabs.harvard.edu/abs/1965MNRAS.129..104C}
}

@ARTICLE{smith_2014,
       author = {{Smith}, Nathan},
        title = "{Mass Loss: Its Effect on the Evolution and Fate of High-Mass Stars}",
      journal = {\araa},
         year = 2014,
        month = aug,
       volume = {52},
        pages = {487-528},
          doi = {10.1146/annurev-astro-081913-040025},
archivePrefix = {arXiv},
       eprint = {1402.1237},
 primaryClass = {astro-ph.SR},
       adsurl = {https://ui.adsabs.harvard.edu/abs/2014ARA&A..52..487S}
}

@ARTICLE{heger_2010,
       author = {{Heger}, Alexander and {Woosley}, S.~E.},
        title = "{Nucleosynthesis and Evolution of Massive Metal-free Stars}",
      journal = {\apj},
         year = 2010,
        month = nov,
       volume = {724},
       number = {1},
        pages = {341-373},
          doi = {10.1088/0004-637X/724/1/341},
archivePrefix = {arXiv},
       eprint = {0803.3161},
 primaryClass = {astro-ph},
       adsurl = {https://ui.adsabs.harvard.edu/abs/2010ApJ...724..341H}
}

@ARTICLE{woosley_1978,
       author = {{Woosley}, S.~E. and {Howard}, W.~M.},
        title = "{The p-processes in supernovae.}",
      journal = {\apjs},
         year = 1978,
        month = feb,
       volume = {36},
        pages = {285-304},
          doi = {10.1086/190501},
       adsurl = {https://ui.adsabs.harvard.edu/abs/1978ApJS...36..285W}
}
\bibliographystyle{unsrt}

\end{multicols}
\end{document}